\documentclass[prb,twocolumn,showpacs]{revtex4}

\usepackage{graphicx}
\usepackage{dcolumn}
\usepackage{amsmath}
\usepackage{amssymb}
\usepackage{xspace}

\newcommand{\rmi}{{\mathrm i}}
\newcommand{\rmd}{{\mathrm d}}
\newcommand{\rmc}{{\mathrm c}}
\newcommand{\rms}{{\mathrm s}}
\newcommand{\rmf}{{\mathrm F}}
\newcommand{\rmr}{{\mathrm R}}
\newcommand{\rml}{{\mathrm L}}
\newcommand{\rmb}{{\mathrm B}}
\newcommand{\rT}{\mathrm{T}}
\newcommand{\pp}{\mathrm{pp}}
\newcommand{\ph}{\mathrm{ph}}
\newcommand{\tpar}{t_\|}
\newcommand{\tperp}{t_\perp}
\newcommand{\anc}{c^{}}
\newcommand{\cdag}{c^\dagger}
\newcommand{\anp}{\psi^{}}
\newcommand{\pdag}{\psi^\dagger}
\newcommand{\ket}[1]{| #1 \rangle}
\newcommand{\bra}[1]{\langle #1 |}
\newcommand{\ve}{\varepsilon}
\newcommand{\vp}{\varphi}
\newcommand{\sgn}{\mathrm{sgn}}
\newcommand{\gc}{G^\rmc}
\newcommand{\gs}{G^\rms}
\newcommand{\gcs}{G^{\rmc(\rms)}}

\newcommand{\Fcs}{F^{\rmc(\rms)}}
\newcommand{\mac}{\mathcal{A}^\rmc}
\newcommand{\mas}{\mathcal{A}^\rms}
\newcommand{\macs}{\mathcal{A}^{\rmc(\rms)}}
\newcommand{\mcc}{\mathcal{C}^\rmc}
\newcommand{\mcs}{\mathcal{C}^\rms}
\newcommand{\mccs}{\mathcal{C}^{\rmc(\rms)}}
\newcommand{\mdc}{\mathcal{D}^\rmc}
\newcommand{\mds}{\mathcal{D}^\rms}
\newcommand{\mdcs}{\mathcal{D}^{\rmc(\rms)}}
\newcommand{\mcu}{\mathcal{U}}
\newcommand{\mcv}{\mathcal{V}}
\newcommand{\mcw}{\mathcal{W}}
\newcommand{\mco}{\mathcal{O}}
\newcommand{\nus}{\frac{\nu}{2}}
\newcommand{\tnus}{\frac{3\nu}{2}}

\newcommand{\vk}{\ensuremath{\boldsymbol{k}}}
\newcommand{\vq}{\ensuremath{\boldsymbol{q}}}
\newcommand{\vkp}{\ensuremath{\boldsymbol{k'}}}
\newcommand{\vkf}{\ensuremath{\boldsymbol{k}_{\rmf}}}
\newcommand{\vkfz}{\ensuremath{\boldsymbol{k}_{\rmf,0}}}
\newcommand{\vu}{\ensuremath{\boldsymbol{u}}}
\newcommand{\vup}{\ensuremath{\boldsymbol{u'}}}

\newcommand{\ie}{i.e.\@\xspace}
\newcommand{\fs}{Fermi surface\xspace}
\newcommand{\fss}{Fermi surfaces\xspace}
\newcommand{\se}{self-energy\xspace}
\newcommand{\ses}{self-energies\xspace}

\begin{document}
\graphicspath{{/EPSF/}{figures/}{./}} 

\title{Interaction-induced Fermi surface deformations in quasi one-dimensional electronic systems}

\author{S\'ebastien Dusuel$^{1}$, Beno\^{\i}t Dou\c{c}ot$^{1}$}

\address{
$^1$ Laboratoire de Physique Th\'eorique et Hautes
 Energies, CNRS UMR 7589, Universit\'e Paris VII-Denis Diderot,\\
4, Place Jussieu, 75252 Paris Cedex 05, France.}


\begin{abstract}

We consider serious conceptual problems with the application of standard perturbation theory, in its zero temperature version, to the computation of the dressed \fs for an interacting electronic system. In order to overcome these difficulties, we set up a variational approach which is shown to be equivalent to the renormalized perturbation theory where the dressed \fs is fixed by recursively computed counterterms. The physical picture that emerges is that couplings that are irrelevant tend to deform the \fs in order to become more relevant (irrelevant couplings being those that do not exist at vanishing excitation energy because of kinematical constraints attached to the \fs). These insights are incorporated in a renormalization group approach, which allows for a simple approximate computation of \fs deformation in quasi one-dimensional electronic conductors. We also analyze flow equations for the effective couplings and quasiparticle weights. For systems away from half-filling, the flows show three regimes corresponding to a Luttinger liquid at high energies, a Fermi liquid, and a low-energy incommensurate spin-density wave. At half-filling Umklapp processes allow for a Mott insulator regime where the dressed \fs is flat, implying a confined phase with vanishing effective transverse single-particle coherence. The boundary between the confined and Fermi liquid phases is found to occur for a bare transverse hopping amplitude of the order of the Mott charge gap of a single chain.

\end{abstract}



\maketitle

\section{Introduction}
\label{sec:intro}

One of the striking results obtained in the last decade on strongly correlated electronic systems is the coexistence of a notion of \fs and of strong deviations from the predictions of Fermi liquid theory for many low-energy properties. This has been extensively studied experimentally for high-temperature superconducting cuprates, where angular resolved photoemission spectroscopy (ARPES) has revealed the presence of \fs arcs, even in the underdoped regime which is characterized by the pseudo-gap seen with most low-energy probes.\cite{Timusk99} Although these systems exhibit intermediate or even strong electron interactions, they have triggered many theoretical works using perturbative tools.\cite{Zanchi96,Halboth00,Honerkamp01} 

At the beginning of any perturbative analysis, the shape of the \fs is crucial in determining which couplings survive in an effective low-energy description.\cite{Shankar94} For most crystalline materials the absence of continuous rotational invariance allows for a deformation of the \fs away from the bare free electron \fs, as interactions are switched on. In many metallic systems this effect is not expected to play much role beyond usual renormalizations of effective parameters of band theory. But in some situations, like the vicinity of a Van-Hove singularity, the presence of a nesting vector, or for strongly anisotropic conductors, it seems essential to understand how to compute the dressed \fs, since it is the relevant object for the construction of an effective low-energy theory. 

In the case of quasi one-dimensional (quasi 1D) systems, this \fs deformation is intimately connected to the widely studied notion of transverse coherence. Experimental and theoretical investigations converge towards a description in terms of almost uncoupled Luttinger liquids along the chains, at high enough energies.\cite{Jerome82_dans_articles,Bourbonnais91} At low energies, optical conductivity measurements\cite{Vescoli98} have shown the existence of two types of behaviors: either the system remains confined in a Mott-insulator phase (in the TMTTF compounds) or the transverse hopping of electrons takes over and establishes a long-ranged transverse phase coherence, leading to a two-dimensional (2D) Fermi liquid phase (for the TMTSF). In the latter case the dressed \fs remains warped while in the former it becomes completely flat under the effect of sufficiently strong interactions.\cite{Prigodin79,Bourbonnais85}

Because of their difficulty, precise computations of \fs deformations for model systems have been undertaken only recently. A direct numerical evaluation of the electron propagator to second order in interaction has been performed for the 2D Hubbard model.\cite{Zlatic95,Halboth97} Similar studies have also been carried for more phenomenological models where electrons are scattered by dynamical spin fluctuations.\cite{Yanase99,Morita00} Although these computations yield valuable physical understanding of the processes involved in the \fs deformation, they suffer from at least two serious problems. First, they identify the dressed \fs with the locus of points in $k$-space for which the dressed quasiparticle energy is equal to the (interacting) chemical potential, which is of course correct. But this does not imply that the imaginary part of the \se vanishes on this surface and for frequencies equal to the chemical potential. Therefore this procedure does not lead to a picture of asymptotically stable quasiparticles at low energies. This remark is valid in the zero temperature approach, which is the only one we are using in this paper, because of its conceptual simplicity. Second, this problem is not cured while going to higher orders in perturbation theory. Furthermore, some new problems arise (namely infrared divergences) at these higher orders for both zero and finite temperature formalisms.

The underlying assumption of the standard perturbation scheme as used above is that one can generate the interacting ground-state by adiabatically switching on the interactions, starting from the non-interacting ground-state. This has to be questioned for large systems for which the ground-state lies at the edge of an energy continuum. Because of this, the perturbation algorithm acting on various excited states of the original systems, associated to different shapes of the \fs, has the possibility to generate energy levels' crossings. This implies that the seed state to be used in perturbation theory is not known a priori, when interactions do deform the \fs. This difficulty has been pointed out in the sixties by Kohn and Luttinger,\cite{Kohn60} and also Nozi\`eres.\cite{Nozieres_anglais} These ideas have been revived recently in a mathematically rigorous framework.\cite{Feldman96} The conclusion of all these works is that a sound formalism is obtained when one works with a bare propagator which singularities are pinned to the {\em dressed} \fs. This is achieved in practice by the introduction of counterterms, which have to be computed order by order in perturbation
theory. The main difficulty in practical implementations of this philosophy (which may be called renormalized perturbation theory) is that it provides only an implicit determination of the dressed \fs, since this algorithm expresses the bare \fs as a function of the dressed one. Although formally this connection has been proved to be invertible,\cite{Feldman98} this remains a formidable task which has never been, to our knowledge, practically undertaken. Note that the necessity to use these counterterms is not a pathology of the zero temperature approach. It also appears in the Matsubara formalism at finite temperature which is the one used in the rigorous works just described.

As a first step towards the realization of this program, several groups have performed self-consistent computations. Their basic principle is to start with a trial \fs, which is adjusted so that it matches with the calculated \fs. A first example follows directly the standard Hartree-Fock method.\cite{Valenzuela01}
It has been applied to the 2D Hubbard model in the presence of second-neighbor hopping and nearest neighbor interaction, and the possibility of a change in \fs topology (from hole-like to electron-like) has been observed. A rather sophisticated scheme has also been developed by Nojiri,\cite{Nojiri99} in which the \se is self-consistently computed from the corresponding second order Feynman diagram. This work addressed the simplest 2D Hubbard model with on-site interaction for which the \fs deformation was found to be very small and to preserve the \fs topology. Note that the quantitative difference between this self-consistent scheme and a standard perturbation theory\cite{Zlatic95,Halboth97} appears to be small.

In spite of their merits, these approaches lack the ability to keep track of the growth of some effective couplings, as the typical energy scale is lowered. These effects play a crucial role for the 2D Hubbard model near half-filling, or for quasi 1D conductors. A natural way of handling these trends is to use a renormalization group (RG) approach. Several groups have incorporated the RG methodology in the computation of the dressed \fs.\cite{Prigodin79,Bourbonnais85,Kishine98,Honerkamp01} Similar studies have also been carried for two coupled chains where the \fs reduces to four Fermi points.\cite{Fabrizio93,Tsuchiizu99,LeHur01} Our understanding of these works is that they always begin with a known bare \fs and compute the evolution of the effective \fs, as the high-energy cut-off is gradually decreased. 
Although this is very reasonable on physical grounds, we may wonder whether this fits with the general rigorous analysis described in the last but one paragraph. We believe there are two ways to combine the corresponding requirements with a RG approach. The first one uses the renormalized perturbation theory described above, with a running energy cut-off. After the usual mode integration in a small energy shell, the kinetic term in the effective action is corrected to preserve the shape of the dressed \fs. In the process of integrating the RG flow, one has to keep track of and sum all these counterterms to obtain the bare \fs as a function of the dressed one. Alternatively, one would fix the bare high-energy theory, and perform the mode integration is such a way that modes being integrated out always remain at a finite distance from the flowing \fs. But then one has to ensure that all modes are integrated over exactly once with a uniform weight. This is indeed possible but requires some slight modifications of the Wilson-Polchinski usual RG equations.\cite{Drazen} We believe the practical implementation of either approach remains to be attempted.

The bulk of this paper is composed of three sections. 
Sec.~\ref{sec:csfs} begins with a general discussion of some difficulties with the standard perturbation theory. We then develop a physical understanding of the driving force that deforms the \fs on the basis of a simple variational calculation for a system of two spinless chains. The main insight gained here is that the couplings which tend to deform the \fs are those for which external momenta of in and out going particles can not be simultaneously taken on the \fs, because of momentum conservation. In the RG language, these interactions are usually called irrelevant. We finally establish the equivalence between this procedure and a standard renormalized perturbation theory where the dressed \fs is fixed by counterterms. The reader interested in more technical aspects is referred to Appendices \ref{app:equi}, \ref{app:diff_re} and \ref{app:diff_im} (the first two begin with some simple first order calculations on the system of two spinless chains, whose results can be compared to the ones obtained in Sec.~\ref{sec:csfs}).
In Secs.~\ref{sec:RG_formalism} and \ref{sec:RG_numerical} we show how the
RG can be implemented in the study of quasi 1D systems. We want to emphasise
that we have not made use of a single RG scheme, but of two coupled RG schemes.
We describe our motivations for performing such a study in 
Secs.~\ref{sec:sub:sub:gen_th_an} and \ref{sec:sub:motiv_use_two_RG}, but let 
us very briefly explain what they are, before coming to a more detailed 
description of Secs.~\ref{sec:RG_formalism} and \ref{sec:RG_numerical}. The 
field-theoretical RG in the spirit of Gell-Mann and Low\cite{Gell-Mann54} is 
a simple but powerful way of computing low-energy properties of systems
described by a renormalizable field-theory. This is why we adopted it for 
this purpose (this method is discussed in detail in Appendix 
\ref{app:field_th}). However, it cannot be used to compute the dressed \fs, 
for the
simple reason that the \fs is defined as the locus of the zeros, in $k$-space 
of the inverse propagator evaluated at {\em zero frequency}. There is thus no
low-energy scale $\nu$ that can be varied to get RG equations as is done for 
example for the low-energy vertices, when relating the values of these vertices
at two different scales $\nu$ and $\nu'$. However, one can use the approach 
known under the name cut-off scaling, and developed by 
S\'olyom.\cite{Solyom79_dans_articles} This RG does not suffer from the 
limitation just described, because it is the high-energy cut-off and not the
low-energy scale that is varied, and we have used it for the computation of 
the dressed \fs. The high-energy part of the flows, in which the \fs 
deformation takes place, is thus described by the cut-off scaling. The dressed
\fs that one obtains in this way then serves as an input parameter for the 
field-theoretical RG which governs the low-energy part of the flows. Let us
say that RG flow equations appear neither in Sec.~\ref{sec:RG_formalism} nor 
in Sec.~\ref{sec:RG_numerical}, but they all have been gathered in Appendix 
\ref{app:flow_eq}. 
In Sec.~\ref{sec:RG_formalism} we set up the cut-off scaling approach for the 
study of \fs deformations in a quasi 1D system of weakly coupled electronic 
chains. In order to make the ideas more concrete, this method is then 
applied to the simplest possible example, and we end the section with a 
comparison to other methods that can be found in the literature.
We then turn to numerical investigations, that are presented in 
Sec.~\ref{sec:RG_numerical}, for short range, Hubbard-like, repulsive electron 
interactions.
Sec.~\ref{sec:sub:first_num_std} deals with considerations about systems away from half-filling which exhibit an incommensurate nesting vector for their \fs. The flow pattern involves a high-energy Luttinger liquid regime, followed by a Fermi liquid at intermediate energy, and finally a long-range ordered spin-density wave (SDW) phase is the stable low-energy attractor. Special attention has been given to the scale and transverse size dependence of the quasiparticle weight. We then focus on the half-filled (and nearly half-filled) case in Sec.~\ref{sec:sub:umklapps_limitations}, where Umklapp processes may drive the system into a confined low-energy phase and pin the SDW on the crystal lattice. In particular we study the cross-over between the confined and the Fermi liquid regimes. It is shown to occur for bare values of the inter-chain hopping of the order of the 1D Mott charge gap.


\section{Computing the shape of the \fs: various difficulties and their resolution}
\label{sec:csfs}
\subsection{General considerations}
\label{sec:sub:gen_cons}

\begin{figure}[t]
\includegraphics[width=8cm]{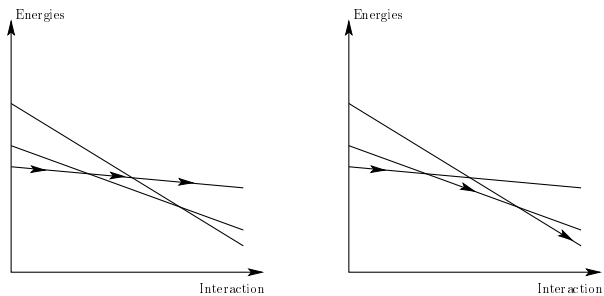}
\caption{Schematic energy level pattern as a function of interaction strength for a conducting Fermi system. Different levels correspond to different choices for the \fs of the non-interacting system. The left figure represents what happens in the standard perturbation theory, where the level repulsion at avoided crossings can not be resolved, so that one obtains a non-adiabatic evolution of the system wave-function as interactions are increased, therefore generating an excited state. On the right we represent the effect of applying the standard perturbation theory in a finite size system. In this case an adiabatic generation of the interacting ground-state is possible.}
\label{fig:branch_adiab_et_cr_niv}
\end{figure}

As emphasized in the Introduction, the computation of the dressed
Fermi surface in an interacting metallic state encounters some
obstacles because of the presence of a continuum of low-lying energy
states in the immediate vicinity of the non-interacting ground-state.
This has been already discussed in a very inspiring paper by
Kohn and Luttinger.\cite{Kohn60} There, they have shown that the standard Brueckner-Goldstone
perturbation theory for the ground-state energy is not consistent 
with a careful procedure of taking the zero-temperature limit of the
total energy computed in the grand-canonical ensemble. They interpret
this failure in terms of the pattern of energy levels of an interacting
Fermi system as a function of the interaction strength. When the shape
of the Fermi surface changes, a deep reshuffling of the spectrum takes
place, leading to a huge number of level crossings. A simple illustration
for this is given on Fig.~\ref{fig:branch_adiab_et_cr_niv}. 
At this stage, it is important to distinguish
between two situations, which have both interesting physical realizations.
For some simple models, such as a ladder of interacting spinless fermions,
or a single chain of spin $1/2$ electrons, the total number 
of particles of a given species 
(transverse momentum in the ladder case, or the $z$ component of
the spin for spin $1/2$ electrons) may be conserved. As a result of this
symmetry, the level crossings just mentioned are an essential feature
of the exact many-body spectrum. In more general situations, these
level crossings appear at any {\em finite} order in a perturbative computation
of the spectrum as a function of interaction strength, although they are
expected to disappear in an exact treatment for a finite-size system. 
Let us first concentrate on the former case for a while, since it
shows dramatically why and where difficulties arise. In such situations,
the conventional assumption often made in many-body computations
does not hold. It states that one can get the interacting many-body 
ground-state by adiabatically switching on the interactions, starting from 
the non-interacting ground-state. A trivial example where the
adiabatic switching procedure most often generates an excited state 
is provided in the case of the free Hamiltonian:
\begin{equation}
H_\lambda = \sum_k \ve_{\lambda}(k)\cdag(k)c(k),
\end{equation}
where we arbitrarily split $\ve_{\lambda}(k)$ in two parts:
\begin{equation}
\ve_{\lambda}(k) = \ve_{0}(k)+\lambda \ve_{1}(k).
\end{equation}
This induces a decomposition of $H_\lambda$ as a sum $H_\lambda=H_{0}+\lambda H_{1}$,
where $H_{0}$ is the ``unperturbed'' Hamiltonian, and
$\lambda H_{1}$ the perturbation. Since $H_{0}$ and $H_{1}$
commute, the eigenstates of $H_\lambda$ do not depend on the strength $\lambda$
of the perturbation. But energy levels as functions of $\lambda$ are
free to cross, so the initial ground-state (\ie for $\lambda = 0$)
becomes in general an excited state for finite $\lambda$. This is
reflected on the computation of the single particle Green's function
in the zero temperature formalism. Starting with the ``bare''
propagator $G^{(0)}(k,\omega)^{-1}=\omega-\ve_{0}(k)+\rmi\eta\sgn\left(\ve_{0}(k)-\mu_{0}\right)$, the conventional algorithm yields 
a ``dressed'' propagator 
$\tilde{G}^{(\lambda)}(k,\omega)^{-1}=\omega-\ve_{\lambda}(k)+\rmi\eta\sgn\left(\ve_{0}(k)-\mu_{0}\right)$
instead of the correct result:
$G^{(\lambda)}(k,\omega)^{-1}=\omega-\ve_{\lambda}(k)+\rmi\eta\sgn\left(\ve_{\lambda}(k)-\mu_{\lambda}\right)$,
where $\mu_{0}$ and $\mu_{\lambda}$ denote the bare and the dressed
chemical potentials respectively.
Note that the problem would apparently disappear in a finite temperature
approach using the Matsubara formalism. However, Kohn and Luttinger
have shown that special care is needed in taking the zero temperature
limit, since they have found a class of diagrams (they have called
them anomalous diagrams) for which the zero temperature limit and the
infinite volume limit do not commute. Taking the former limit first
yields a vanishing contribution for those diagrams, and therefore the
wrong result of the standard zero temperature formalism is obtained.
The correct result for an infinite system is obtained by taking the
other order of limits, where anomalous diagrams do provide
finite contributions.

\begin{figure}[t]
\includegraphics[width=8cm]{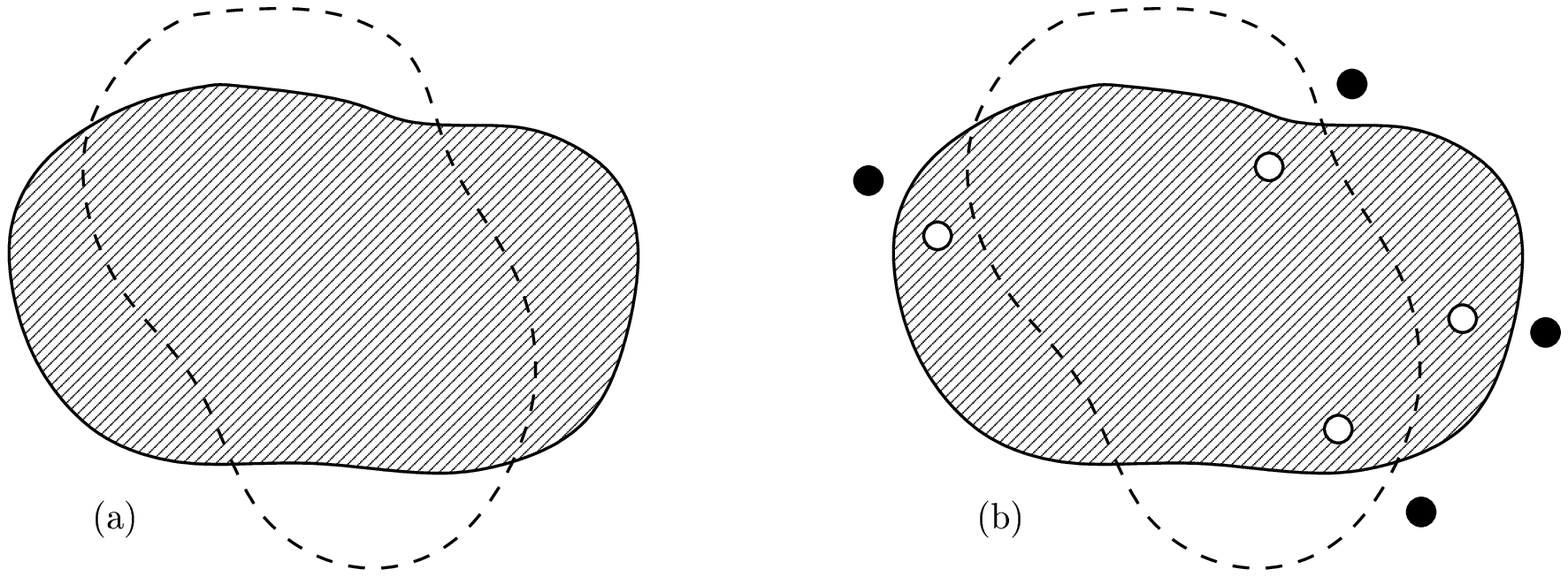}
\caption{Examples of possible initial states for the perturbation algorithm. These states are Slater determinants with occupied single-particle states depicted by the dashed areas in $k$-space. The dashed line denotes the non-interacting \fs. In state (a), the \fs is deformed, but no additional particle-hole excitations are present, unlike in state (b).}
\label{fig:sf_lib_hab}
\end{figure}

For this reason, and also given the conceptual interest of this
problem, we shall use only the zero temperature formalism throughout
this paper. In this framework, a natural way to circumvent this
problem with level crossings is to start the standard perturbation
algorithm with any arbitrary eigenstate of the non-interacting Hamiltonian
$H_{0}$. Intuitively, we believe in most cases it is sufficient
to choose an initial state where the locus of occupied single particle states
is singly connected (\ie it has no isolated particle-hole excitations
from the viewpoint of $H_{0}$), but with a deformed Fermi surface, as
shown on Fig.~\ref{fig:sf_lib_hab}. 
The selection of the correct initial state
is performed by minimizing the total energy of the dressed state it generates,
after switching on the interactions. An example of this procedure is
given below (Sec.~\ref{sec:sub:2chaines_ordre1}) for a simple two-chain model.

For practical purposes, it is important to note that this approach 
may also be implemented through a perturbative computation of the
single particle Green's function. Instead of using the free propagator
$G^{(0)}(k,\omega)^{-1}=\omega-\ve_{0}(k)+\rmi\eta\sgn\left(\ve_{0}(k)-\mu_{0}\right)$, we should first make a guess for the
dressed Fermi surface. This allows us to define a function $\Phi (k)$
such that $\Phi (k)=1$ if $k$ does not belong to the trial Fermi sea,
and $\Phi (k)=-1$ if $k$ belongs to it. 
The locus of points in $k$-space
where  $\Phi (k)$ jumps from $-1$ to $+1$ is our trial Fermi surface,
and points on this set will be generically denoted as $k_{F}$ in the
present discussion. The corresponding bare propagator
to be used in Feynman graph expansions is:
\begin{equation}
G^{(0)}_{\Phi}(k,\omega)^{-1}=\omega-\ve_{0}(k)+\rmi\eta \Phi (k).
\end{equation}
As usual, the dressed propagator is obtained as
$G_{\Phi}(k,\omega)^{-1}=\omega-\ve_{0}(k)-\Sigma_{\Phi}(k,\omega)$,
where the subscript $\Phi$ in $\Sigma_{\Phi}(k,\omega)$ is to stress
the influence of the choice of a trial Fermi surface encoded in the
function $\Phi$. If this trial Fermi surface is the correct one for the
interacting Fermi system, we expect the \se satisfies the following
well-known conditions:

i) There exists a well defined chemical potential $\mu$ so that for any
$k_{F}$ belonging to the trial Fermi surface, we have:
\begin{equation}
  \label{eq:conditioon_re_fs}
  \mu-\ve_{0}(k_{F})-\Re\Sigma_{\Phi}(k_{F},\mu)=0.
\end{equation}

ii) The inverse life-time of ``quasiparticles'' vanishes on the trial
Fermi surface so that:
\begin{equation}
\Im \Sigma_{\Phi}(k_{F},\mu)=0.
\end{equation}
Of course, these conditions are not satisfied for most trial
Fermi surfaces, as the reader will immediately notice on simple
examples. We have checked on several examples that both procedures
(\ie minimizing the total energy, or satisfying conditions i) and
ii) on the dressed single particle propagator) yield the same
dressed Fermi surface. In Appendix \ref{app:equi}, we provide a formal proof
of this equivalence, first in the finite volume case, and then
in the case of an infinite volume. When the choice of $\Phi$ is not
the correct one, it is impossible to satisfy both conditions i) and ii)
simultaneously. In the case of standard perturbation theory $\Phi$ is taken to be $\Phi^{(0)}$ corresponding to
the bare Fermi surface, obtained from $H_{0}$.\cite{Zlatic95,Halboth97}
The dressed Fermi surface is assumed to be determined 
from an equation which resembles condition i), namely:
\begin{equation}
\mu-\ve_{0}(k_{F})-\Re\Sigma_{\Phi^{(0)}}(k_{F},\mu)=0.
\end{equation}
But doing this yields two severe flaws: 
as shown in Appendix \ref{app:diff_re}, this does not generate the same dressed Fermi surface 
as the two procedures presented above and argued to be the correct
ones do. Furthermore, in perturbation theory, 
$\Im\Sigma_{\Phi^{(0)}}(k_{F},\mu)$ changes sign on the
non-interacting Fermi surface, and for $\omega$ equal to the non-interacting
chemical potential $\mu_{0}$ as shown in Appendix \ref{app:diff_im}.

This discussion holds clearly in the case where energy level crossings
associated to various initial shapes of the Fermi surface are protected
by some symmetries of the full Hamiltonian, as stated at the beginning of
this section. Here, we would like to emphasize that a similar qualitative
picture also holds in a more generic situation. On general grounds, 
we expect that energy levels of a finite system do not cross 
as the interaction strength is increased. This is the famous 
phenomenon of energy level repulsion which plays a key role in the field
of ``quantum chaos'' (see for instance the book by Gutzwiller\cite{Gutzwiller}).
So, standard perturbation theory starting from the unperturbed ground-state
is expected to generate the correct interacting ground-state for a {\em finite}
system. However, to get the full single energy level resolution
in the spectrum with all the avoided level crossings clearly requires
going to very high orders in perturbation theory. Instead, in most many-body
computations, we first get formal expressions for various quantities
such as Green's functions for a chosen {\em finite} order 
in powers of the interaction,
and we most often take the thermodynamical limit {\em before} summing the
perturbation series. We believe this procedure is most likely to generate
in the end an excited state of the interacting system, although
the seed of the perturbation series is the non-interacting ground-state.
This belief is confirmed by the simple computations in Appendix 
\ref{app:diff_re}, which do not require any special symmetry of the 
full Hamiltonian.


\subsection{Two chains of spinless fermions: Energy minimization}
\label{sec:sub:2chaines_ordre1}

\subsubsection{Model and notations}
\label{sec:sub:sub:model_notations}

Let us first focus on the simplest possible model exhibiting the features described previously: a system of two chains of interacting spinless fermions. We will assume this system to be anisotropic, described by a tight-binding Hamiltonian, with a hopping $\tpar$ along the chain much larger than the transverse hopping $\tperp$. Hence, we have two bands, named by the transverse momentum they correspond to, \ie 0 (bonding) and $\pi$ (anti-bonding). We suppose the filling is such that both bands are partially filled. We will furthermore focus on the low-energy properties, so that we can linearize the spectrum around the four Fermi points, giving rise to four types of fermions: $(\rmr,0)$, $(\rmr,\pi)$, $(\rml,0)$ and $(\rml,\pi)$. As usual, we extend the spectrum for arbitrary momenta. The low-energy free Hamiltonian is thus given by: 
\begin{eqnarray}
  \label{eq:ham_libre_lin}
  &&H_0=\sum_k \sum_{I=0,\pi}\\
  &&\hspace{0.5cm}\Bigg\lbrace \left[\mu^{(0)}+v^{(0)}_{\rmf,I} (k-k^{(0)}_{\rmf,I})\right] \cdag_{\rmr,I}(k) \anc_{\rmr,I}(k) + \nonumber\\
  &&\hspace{1cm}\left[\mu^{(0)}-v^{(0)}_{\rmf,I}(k+k^{(0)}_{\rmf,I})\right] \cdag_{\rml,I}(k) \anc_{\rml,I}(k) \Bigg\rbrace.\nonumber
\end{eqnarray}
In the above expression, all the superscripts $^{(0)}$ denote free quantities. $\mu^{(0)}$ is the chemical potential, $v^{(0)}_{\rmf,I}$ and $k^{(0)}_{\rmf,I}$ the Fermi velocity and momentum on chain $I$. $\cdag_{\rmr,I}(k)$ is the creation operator of a right fermion on chain $I$, with parallel momentum $k$. The sum over $k$ is to be understood as an integral for a system in the thermodynamic limit. In all that follows, we will simplify the problem and suppose that the Fermi velocities for both branches are equal, and they will simply be denoted as $v_\rmf^{(0)}$.

\begin{figure}[t]
\includegraphics[width=8cm]{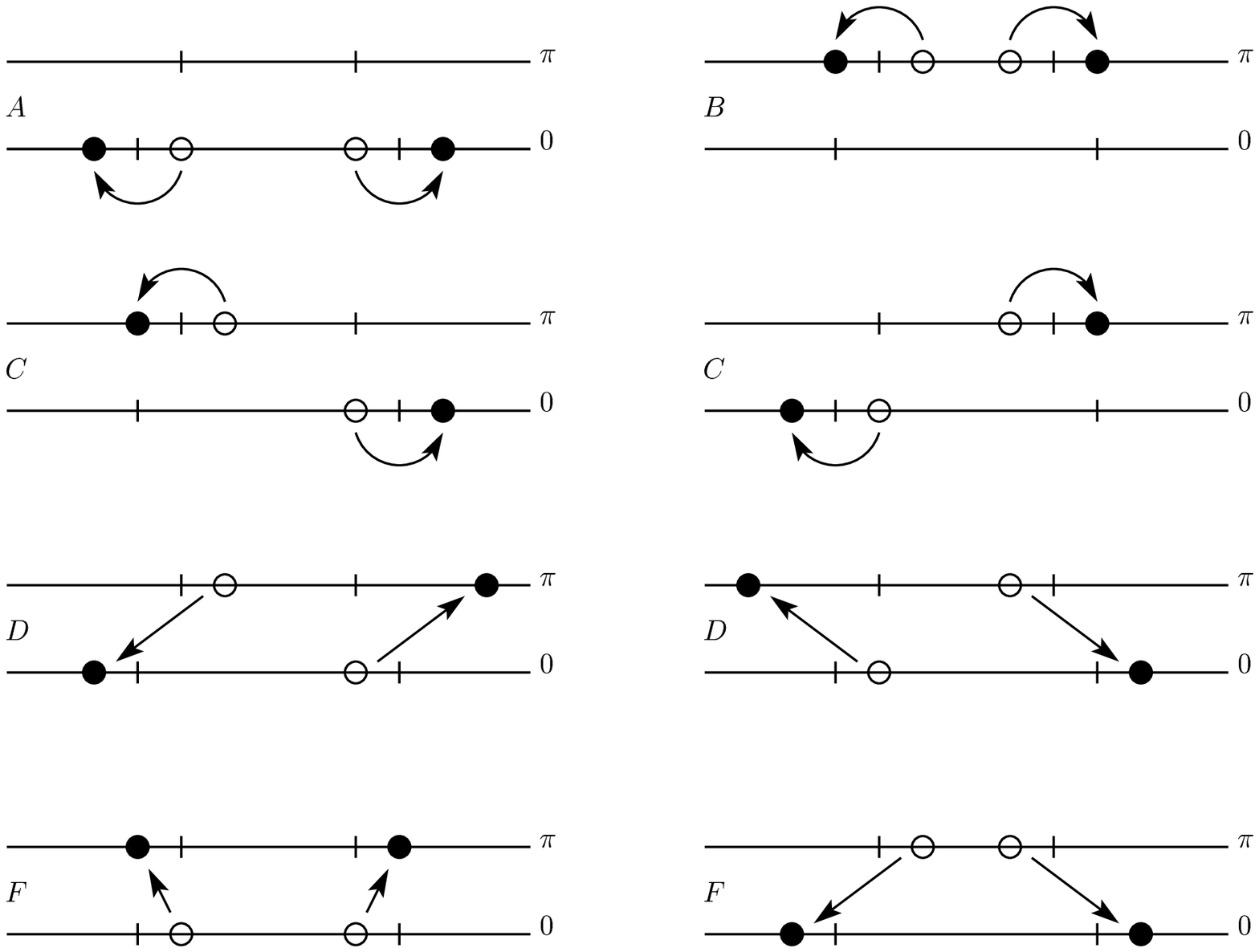}
\caption{Selected low-energy interactions for the two-chain model.}
\label{fig:interactions_2chaines}
\end{figure}

We shall also make simplifying assumptions about the interactions. Thus, the only low-energy interaction processes we will be interested in, are of the forward scattering type ($g_2$), classified as $A$, $B$, $C$, $D$, and $F$. They are represented on Fig.~\ref{fig:interactions_2chaines}. 
We shall neglect the Umklapps, assuming the filling is not commensurate. $g_4$ interactions, involving four right or four left fermions, are also discarded, because we shall restrict ourselves to first and second order effects, to which these interactions give no contribution. In order to save space, we only give the $D$ type interaction Hamiltonian:
\begin{eqnarray}
  &&H_\mathrm{int}^{(D)}=\frac{D}{L}\sum_{k,k',q}\\
  &&\hspace{0.5cm}\Big\lbrace\cdag_{\rmr,\pi}(k+q) \cdag_{\rml,0}(k'-q) \anc_{\rml,\pi}(k') \anc_{\rmr,0}(k) +\mathrm{h.c.} \Big\rbrace,\nonumber
\end{eqnarray}
where h.c. means the hermitic conjugate.


\subsubsection{First order}
\label{sec:sub:sub:min_energy}

We will here compute the energy to order one in the usual quantum mechanical perturbation theory, of eigenstates obtained from two types of free eigenstates. The first ones, denoted as $\ket{0;k_{\rmf,0},k_{\rmf,\pi}}_0$, are free states for which the bonding (respectively anti-bonding) band is filled up to $k_{\rmf,0}$ (respectively $k_{\rmf,\pi}$). The ground-state of the free system is thus obviously $\ket{0;k_{\rmf,0}^{(0)},k_{\rmf,\pi}^{(0)}}_0$. Of course, as the number of particles is fixed, the condition $k_{\rmf,0}+k_{\rmf,\pi}=k_{\rmf,0}^{(0)}+k_{\rmf,\pi}^{(0)}$ must be satisfied. As we wish to understand what happens if one adds a particle to the system, we will also consider states that are simply obtained from the first ones by adding a particle of momentum $q$ on branch 0 or $\pi$ (with $q\geqslant k_{\rmf,0}$ or $q\geqslant k_{\rmf,\pi}$). We will refer to these states as $\ket{1,q,0(\pi);k_{\rmf,0},k_{\rmf,\pi}}_0$. We shall neither consider states with one hole, nor states with an arbitrary number of particles or holes. 

\begin{figure}[t]
\includegraphics[width=7cm]{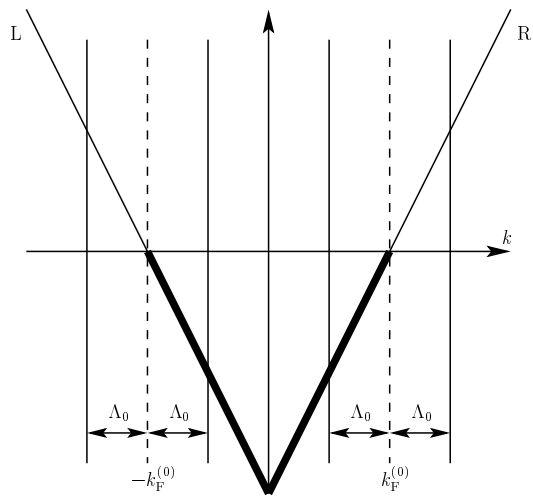}
\caption{Here we show how the ultra-violet cut-off is chosen around the free \fs, for one band.}
\label{fig:cut-off}
\end{figure}

First of all we can compute the energies of these states, in the non-interacting case. Of course, because
our linearized dispersion relations have been extended to include infinitely many single-particle states,
there is strictly speaking an infinite particle density in these Dirac seas, which yields divergent
expressions for the total energy.
We will regularize these divergences by putting an ultra-violet cut-off $\Lambda_0$ on the momenta, around the four {\em free} Fermi momenta, as shown on Fig.~\ref{fig:cut-off} for one band.
For the sake of simplicity, we work in the thermodynamic limit, and after a bit of algebra we find:
\begin{eqnarray}
  &&E^{(0)}(0;k_{\rmf,0},k_{\rmf,\pi})=\frac{L}{\pi}(2\mu^{(0)}\Lambda_0-v_\rmf^{(0)}\Lambda_0^2)\nonumber\\
  &&\hspace{2cm}+\frac{v_\rmf^{(0)} L}{\pi}\left(k_{\rmf,0}-k_{\rmf,0}^{(0)}\right)^2, \mbox{ and }\\
  &&E^{(0)}(1,q,0(\pi);k_{\rmf,0},k_{\rmf,\pi})=E^{(0)}(0;k_{\rmf,0},k_{\rmf,\pi})\nonumber\\
  &&\hspace{2cm}+\mu^{(0)}+v_\rmf^{(0)}(q-k_{\rmf,0(\pi)}^{(0)}).
\end{eqnarray}
It is obvious that the minimum of the energy is obtained for the free \fs. The value of $\mu^{(0)}$ does not play a role here since we have fixed the total particle number.

To order one in the couplings, it is well known that the energy of a free state is simply shifted by the mean value of the interaction for this state. As a consequence, the $D$ and $F$ couplings do not give any contribution. They will only start playing a role to second order. It is a very simple matter to check that:
\begin{eqnarray}
  &&\Delta E^{(1)}(0;k_{\rmf,0},k_{\rmf,\pi})=\frac{L}{(2\pi)^2}\Bigg[ A \left(\Lambda_0+ (k_{\rmf,0}-k_{\rmf,0}^{(0)}) \right)^2\nonumber\\
  \label{eq:en_1_sf}
  &&\hspace{2cm}+ B \left(\Lambda_0-( k_{\rmf,0}-k_{\rmf,0}^{(0)}) \right)^2\\
  &&\hspace{0.3cm}+2C \left(\Lambda_0+ (k_{\rmf,0}-k_{\rmf,0}^{(0)}) \right) \left(\Lambda_0-( k_{\rmf,0}-k_{\rmf,0}^{(0)}) \right)\Bigg],\nonumber\\
  &&\Delta E^{(1)}(1,q,0;k_{\rmf,0},k_{\rmf,\pi})=\Delta E^{(1)}(0;k_{\rmf,0},k_{\rmf,\pi})\nonumber\\
  \label{eq:en_1_1part0}
  &&\hspace{1cm}+\frac{1}{2\pi}\Bigg[ A\left(\Lambda_0+ (k_{\rmf,0}-k_{\rmf,0}^{(0)}) \right)\\
  &&\hspace{3cm}+ C \left(\Lambda_0-( k_{\rmf,0}-k_{\rmf,0}^{(0)}) \right)\Bigg],\nonumber\\
  &&\Delta E^{(1)}(1,q,\pi;k_{\rmf,0},k_{\rmf,\pi})=\Delta E^{(1)}(0;k_{\rmf,0},k_{\rmf,\pi})\nonumber\\
  \label{eq:en_1_1partpi}
  &&\hspace{1cm}+\frac{1}{2\pi}\Bigg[ B\left(\Lambda_0- (k_{\rmf,0}-k_{\rmf,0}^{(0)}) \right)\\
  &&\hspace{3cm}+ C \left(\Lambda_0+( k_{\rmf,0}-k_{\rmf,0}^{(0)}) \right)\Bigg].\nonumber
\end{eqnarray}
We have used the conservation of the number of particles so that the above expressions are expressed only in terms of the Fermi momenta on branch 0. Thus we minimize the energy $E^{(1)}=E^{(0)}+\Delta E^{(1)}$ simply by requiring for its derivative with respect to $k_{\rmf,0}$ to vanish. This yields:
\begin{equation}
  \label{eq:dk0_vrai}
  k_{\rmf,0}^{(1)}-k_{\rmf,0}^{(0)}=(B-A)\frac{\Lambda_0}{4\pi v_\rmf^{(0)}}\left(1+\frac{A+B-2C}{4\pi v_\rmf^{(0)}}\right)^{-1}.
\end{equation}

Let us show how the chemical potential can be computed, using the energies of the states with one added particle. First of all, we notice that the expressions $\Delta E^{(1)}(q)$ are independent of $q$. It implies the energy for adding a particle to the system on branch 0 ($\pi$) will be minimal if $q$ is as small as possible, \ie $q=k_{\rmf,0}$ ($q=k_{\rmf,\pi}$). This confirms that $k_{\rmf,0}$ and $k_{\rmf,\pi}$ are the actual Fermi momenta. Now if we require this minimal energy to be the same on the two branches, equal to the renormalized chemical potential, we obtain the two following conditions:
\begin{eqnarray}
  &&\mu^{(1)}=\mu^{(0)}+v_\rmf^{(0)}(k_{\rmf,0}-k_{\rmf,0}^{(0)})\nonumber\\
  &&\hspace{1cm}+\Delta E^{(1)}(1,q=k_{\rmf,0},0;k_{\rmf,0},k_{\rmf,\pi})\\
  &&\hspace{2cm}-\Delta E^{(1)}(0;k_{\rmf,0},k_{\rmf,\pi}),\nonumber\\
  &&\mu^{(1)}=\mu^{(0)}+v_\rmf^{(0)}(k_{\rmf,\pi}-k_{\rmf,\pi}^{(0)})\nonumber\\
  &&\hspace{1cm}+\Delta E^{(1)}(1,q=k_{\rmf,\pi},\pi;k_{\rmf,0},k_{\rmf,\pi})\\
  &&\hspace{2cm}-\Delta E^{(1)}(0;k_{\rmf,0},k_{\rmf,\pi}).\nonumber
\end{eqnarray}
One can check these equations give the deformation (\ref{eq:dk0_vrai}) of the \fs. This is physically desirable. Indeed, imposing that the minimum energies to add one particle on one branch or the other are identical, should be equivalent to the requirement that taking two particles at the \fs on one branch and putting them at the \fs on the other branch costs nothing (in the thermodynamical limit). Finally we find the chemical potential:
\begin{eqnarray}
  \label{eq:dmu_vrai}
  &&\mu^{(1)}=\mu^{(0)}+(A+B+2C)\frac{\Lambda_0}{4\pi}\\
  &&\hspace{1cm}-(B-A)^2\frac{\Lambda_0}{4\pi v_\rmf^{(0)}}\left(1+\frac{A+B-2C}{4\pi v_\rmf^{(0)}}\right)^{-1}.\nonumber
\end{eqnarray}

To conclude this section about first order computations, we show two figures of what would happen for a total energy of the following simplified form: $E^{(1)}=(k_{\rmf,0}-k_{\rmf,0}^{(0)})^2+(A-B)(k_{\rmf,0}-k_{\rmf,0}^{(0)})$. Fig.~\ref{fig:croisements_1} illustrates the level crossings: we represent the energy as a function of $(B-A)$ (assumed positive), for various values of $(k_{\rmf,0}-k_{\rmf,0}^{(0)})$. 
\begin{figure}[t]
  \includegraphics[width=8cm]{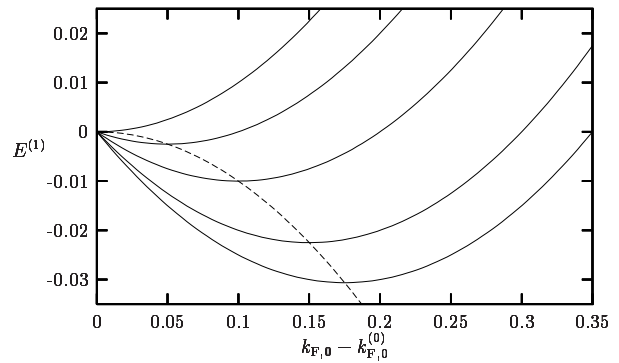}
  \caption{Energies as functions of $(k_{\rmf,0}-k_{\rmf,0}^{(0)})$ for different values of $B-A$ (0, 0.1, 0.2, 0.3, 0.35). The dashed curve gives the energy of the ground-state, and thus goes through the minima of all the different curves.}
  \label{fig:croisements_2}
\end{figure}
Fig.~\ref{fig:croisements_2} proposes an alternative vision of the same thing (see the caption).
\begin{figure}[t]
  \includegraphics[width=8cm]{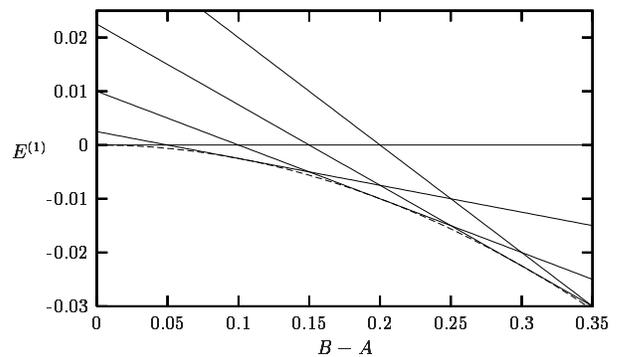}
  \caption{Energies as functions of $B-A$, for states with different values of $(k_{\rmf,0}-k_{\rmf,0}^{(0)})$ (0, 0.05, 0.1, 0.15 et 0.2). The crossings show us that the \fs will be deformed. The dashed curve is the envelope of all these curves. It is thus the energy of the interacting ground-state, as a function of the interaction.}
  \label{fig:croisements_1}
\end{figure}

Eq.~(\ref{eq:dk0_vrai}) shows us that the deformation of the \fs at first order is due to the difference between the couplings on the branches: if $A=B$, no deformation takes place. We can understand the sign of the deformation very simply. Suppose the fermions repel each other (\ie the couplings are positive), but that the repulsion is bigger on chain $\pi$ for example: $B-A>0$. It is then natural, in order to lower the energy of the system, that some fermions of chain $\pi$ go to chain $0$, so that $(k_{\rmf,0}-k_{\rmf,0}^{(0)})$ should be positive. This is indeed what we find. We will now see that things are different at second order: the $D$ couplings tend to flatten the \fs, whatever their sign, and without having to invoke a difference between two couplings.


\subsubsection{Second order and further}
\label{sec:sub:2chaines_ordre2}

We shall now discuss in detail perturbation theory to second order (this notion of order being simply the number of vertices in the corresponding Feynman graphs), and we will also see that some problems arise to third order and beyond, showing that another perturbation scheme is needed.

We have already seen the effects of $A$, $B$ and $C$ interactions on the \fs's shape to first order. We let the reader check that these three interactions play no essential role in the deformation of the \fs at second order. Indeed, when we compute the energy of the states $\ket{0;k_{\rmf,0},k_{\rmf,\pi}}_0$, if we only keep contributions that diverge when $\Lambda_0\to\infty$, we get a quantity that is proportional to $\Lambda_0^2$, and {\em independent} of the dressed Fermi momenta. Only finite terms do depend on the dressed Fermi momenta. We shall thus neglect these contributions, and focus on $D$ and $F$ interactions.

Let us begin with the effect of $D$ interaction, which is the only one that does not exist at zero energy if the \fs is not strictly flat. This is due to the constraint of momentum conservation, and is easily visualized from Fig.~\ref{fig:interactions_2chaines}. Second order perturbation theory tells us that the eigenenergy of an eigenstate $\ket{n}$, obtained from the free eigenstate $\ket{n}_0$, is obtained by shifting the free eigenenergy of a quantity: 
\begin{equation}
  \label{eq:formule_th_pert_2}
  \sum_{k\neq n} \frac{V_{nk}V_{kn}}{E^{(0)}_n-E^{(0)}_k}, \mbox{ where } V_{nk}=\,_0\bra{n} V \ket{k}_0,
\end{equation}
and where $V$ is the interaction potential. This formula involves energy denominators. If these become smaller, the energy will decrease. When $D$ interactions are considered, we understand from these considerations that they will tend to flatten the \fs, because this will allow for smaller energy denominators. That this is true can be checked by explicitly computing the energy shift, which is found to be:
\begin{equation}
  \Delta E^{(2)}_D=\frac{L D^2}{v_\rmf^{(0)} (2\pi)^3} (k_{\rmf,0}-k_{\rmf,\pi})^2\ln\left(\frac{\Lambda_0}{|k_{\rmf,0}-k_{\rmf,\pi}|}\right).
\end{equation}
We stress that this result has been found computing (\ref{eq:formule_th_pert_2}), keeping only terms that are divergent when $\Lambda_0\to\infty$ and that depend on the Fermi momenta. If only $D$ terms are considered, it is now easy to show that the free and renormalized Fermi momenta are linked by the following formula:
\begin{equation}
  \label{eq:def_sf_D2}
  \Delta k_\rmf^{(0)}=\Delta k_\rmf\left[ 1+2\left(\frac{D}{2\pi v_\rmf^{(0)}}\right)^2 \ln\left( \frac{\Lambda_0}{|\Delta k_\rmf|} \right)\right],
\end{equation}
where we have set $\Delta k_\rmf=k_{\rmf,0}-k_{\rmf,\pi}$ (and the same for free quantities). This clearly shows the tendency towards the flattening of the \fs, induced by $D$ terms.

\begin{figure}[t]
  \includegraphics[width=7cm]{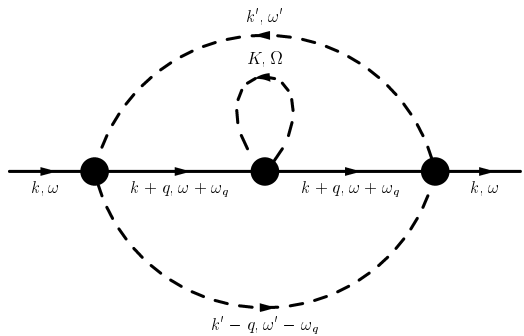}
  \caption{Example of non-skeleton diagram giving rise to infrared divergences.}
  \label{fig:diag_squelette_3_spinless}
\end{figure}
What about $F$ interactions? Perturbation theory at second order is divergent in the low-energy limit. Indeed $\ket{0;k_{\rmf,0},k_{\rmf,\pi}}_0$ states that are not the free ground-states, are coupled to a continuum of excited states composed of two particles and two holes, which have kinetic energies arbitrarily close to the one of the seed state $\ket{0;k_{\rmf,0},k_{\rmf,\pi}}_0$. This yields energy denominators that are very small in absolute value, and even zero.
In the \se formalism (constructed from an excited state, see Appendix \ref{app:equi} for details), this problem is regularized by the imaginary parts $\pm\rmi\eta$ in the \se approach. For the minimization of energy, one can similarly define the divergent integrals with a principal part, and one finds the same results as in this \se version. But this infrared divergence is only the first one, and is not the most problematic. Things become worse and worse for higher orders. This has already been discussed by Feldman, Salmhofer and Trubowitz\cite{Feldman96} so that we shall be brief. In the language of Feynman diagrams, the divergences come from repeated \se insertions, or to say it differently, with non-skeleton diagrams. An example of the lowest order diagrams of this type (apart from the Kohn-Luttinger diagram we have already discussed at second order, and which is zero) is given in Fig.~\ref{fig:diag_squelette_3_spinless}.
The problem with such a diagram is the following. Because of the inserted first order \se in the internal right propagator, we now have two right internal propagators. This gives a bad behavior of the integral over $q$ around $q=0$ for $\omega=0$ and $k=0$, once all other variables have been integrated out. It is clear that things get even worse if two or more such first or higher order \ses are inserted. We thus have to find a way of getting rid of these infrared problems that plague our perturbation theory. This is achieved by the use of counterterms, that we will expose now.


\subsection{The use of counterterms in the two-chain model}
\label{sec:sub:use_ct}

\subsubsection{Notations and first order calculation}

In order to simplify the notations, we will denote the Fermi velocity by $v_\rmf$ instead of $v_\rmf^{(0)}$. We will again suppose this velocity to be independent of the chain index and its renormalization will be neglected throughout this paper to simplify the discussion. The use of counterterms in interacting fermionic systems, for which the \fs gets deformed, is quite old, and can for example be found in the beautiful discussion by Nozi\`eres,\cite{Nozieres_anglais} where the reader will find more details. The main idea, which has been illustrated very recently,\cite{Neumayr02,Ledowski02}, is to take the interacting Fermi sea as the starting point of perturbation theory. As it is a priori unknown, we must ensure in the end of the calculation, that the ``guessed'' \fs is indeed the dressed one. In order to have a good starting point, the most natural idea is to split the free Hamiltonian into two bits: one that is a modified free Hamiltonian with the correct interacting \fs, and another that will be the difference between the true free Hamiltonian, and the modified one. We will thus write:
\begin{eqnarray}
  &&H_0=\sum_{I=0,\pi} \sum_k \Bigg\lbrace \left[\mu+v_\rmf (k-k_{\rmf,I})\right] \cdag_{\rmr,I}(k) \anc_{\rmr,I}(k)\nonumber\\
  &&\hspace{2cm}+\left[\mu-v_\rmf(k+k_{\rmf,I})\right] \cdag_{\rml,I}(k) \anc_{\rml,I}(k) \Bigg\rbrace\nonumber\\
  &&\hspace{3cm} + H_{0,\mathrm{ct}}^{(\mu)}+H_{0,\mathrm{ct}}^{(k)},
\end{eqnarray}
with
\begin{equation}
  H_{0,\mathrm{ct}}^{(\mu)}=\delta\mu \sum_{I,k} \cdag_{\rmr,I}(k) \anc_{\rmr,I}(k) + \delta\mu \sum_{I,k} \cdag_{\rml,I}(k) \anc_{\rml,I}(k),
\end{equation}
and
\begin{eqnarray}
  H_{0,\mathrm{ct}}^{(k)}&=&-v_\rmf \delta k_I \sum_{I,k} \cdag_{\rmr,I}(k) \anc_{\rmr,I}(k)\nonumber\\
  &&- v_\rmf \delta k_I \sum_{I,k} \cdag_{\rml,I}(k) \anc_{\rml,I}(k).
\end{eqnarray}
\begin{figure}[t]
  \includegraphics[width=5cm]{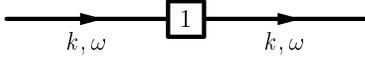}
  \caption{Graphical representation of the chemical potential counterterm, at first order.}
  \label{fig:dmu1_spinless}
\end{figure}
The counterterms $\delta\mu$ and $\delta k_I$ are found by the requirement that $H_0$ remains the true free Hamiltonian:
\begin{eqnarray}
  &&\mu^{(0)}=\mu+\delta\mu, \mbox{ with }\\
  &&\delta\mu=\delta\mu^{(1)}+\delta\mu^{(2)}+\ldots=G^\alpha \delta \mu_1^\alpha+\ldots\\
  &&k_{\rmf,I}^{(0)}=k_{\rmf,I}+\delta k_I, \mbox{ with }\\
  &&\delta k_I=\delta k_I^{(1)}+\delta k_I^{(2)}+\ldots=G^\alpha \delta k_{I,1}^\alpha+\ldots.
\end{eqnarray}
Note that we have used a symbolic notation $G^\alpha$ for the couplings $A$ to $F$, and the sum over $\alpha$ is implicit. We stress that there is not only one counterterm for the chemical potential (or for the Fermi momentum of each chain), but an infinity, which are all the $\delta\mu^{(n)}$'s, for $n=1,2,\ldots$. The number $n$ gives the power in the couplings of the considered counterterm. Counterterms have to be computed order by order, one after the other, in a perturbation theory. When using the counterterms, the Luttinger theorem simply says that: $\sum_I \delta k_I=0$, or for each order $j$: $\sum_I \delta k_I^{(j)}=0$. Now the free (R,0) propagator with which Feynman diagrams are computed is:
\begin{equation}
  G^*_{\rmr,0}(k,\omega)=\frac{1}{\omega-\left[\mu+v_\rmf(k-k_{\rmf,I})\right]+\rmi \eta \,\sgn(k-k_{\rmf,I})},
\end{equation}
and similarly for other types of fermions. Both real and imaginary parts of these propagators refer to the interacting \fs.

\begin{figure}[t]
  \includegraphics[width=5cm]{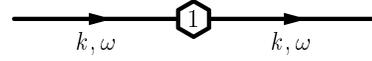}
  \caption{Graphical representation of the Fermi momenta counterterms, at first order.}
  \label{fig:dk1_spinless}
\end{figure}

We shall now see how to implement the use of counterterms in the perturbation theory of the two-chain model. For this, it is useful to associate a graphical representation to the counterterms. This is illustrated at first order in Fig.~\ref{fig:dmu1_spinless} for the chemical potential, and in Fig.~\ref{fig:dk1_spinless} for the Fermi momenta. 
The chemical potential counterterm is represented by a square, whereas the Fermi momenta counterterms are denoted by hexagons. In both cases, the number written inside the symbol is the order $n$ mentioned previously. Notice that for Fermi momenta, we do not need to explicitly write down the chain index $I$, because it would be redundant with the chain index of the propagators. The reader should also remark that counterterms for right or left fermions are exactly identical.

\begin{figure}[b]
\includegraphics[width=6cm]{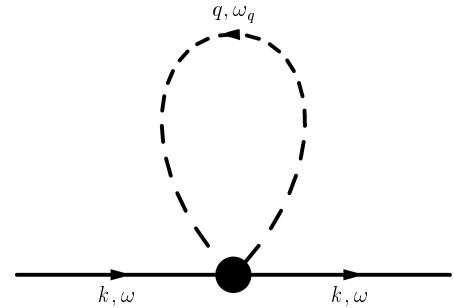}
\caption{First order contribution to the \se: the tadpole graph.}
\label{fig:tadpole_spinless}
\end{figure}

Now that the general notations have been given, let us see what the counterterm approach gives to first order, for the two chains. In all that follows, we will not use an ultra-violet cut-off around the free \fs, but {\em around the interacting \fs}. This will slightly alter the results, but it makes the computation simpler, without involving a qualitatively different physics. The tadpole diagram of Fig.~\ref{fig:tadpole_spinless}, computed with the new free propagator $G_*^{(0)}$ and the new cut-off, gives the following contribution to the \se:
\begin{eqnarray}
  &&\Sigma_{\rmr,0}^{(1)}(k,\omega)=(A+C)\frac{\Lambda_0}{2\pi}, \mbox{ and }\\
  &&\Sigma_{\rmr,\pi}^{(1)}(k,\omega)=(B+C)\frac{\Lambda_0}{2\pi}.
\end{eqnarray}
But we also have the counterterm contributions (diagrams of Figs.~\ref{fig:dmu1_spinless} and \ref{fig:dk1_spinless}):
\begin{eqnarray}
  &&\Sigma_{\mathrm{ct};\rmr,0}^{(\mu;1)}(k,\omega)=\Sigma_{\mathrm{ct};\rmr,\pi}^{(\mu;1)}(k,\omega)=\delta\mu^{(1)},\\
  &&\Sigma_{\mathrm{ct};\rmr,0}^{(k;1)}(k,\omega)=-v_\rmf\delta k_0^{(1)},\\
  &&\Sigma_{\mathrm{ct};\rmr,\pi}^{(k;1)}(k,\omega)=-v_\rmf\delta k_\pi^{(1)}.
\end{eqnarray}
\begin{figure}[t]
  \includegraphics[width=6cm]{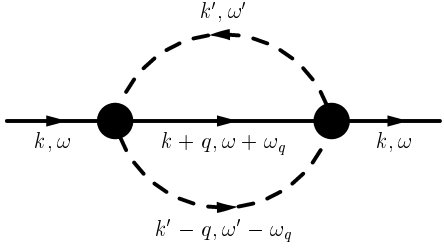}
  \caption{Sunrise diagram contributing to the second order \se.}
  \label{fig:sunrise_spinless}
\end{figure}
\begin{figure}[b]
  \includegraphics[width=6cm]{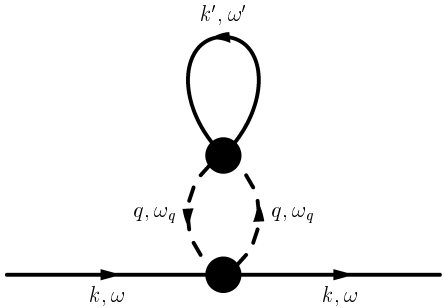}
  \caption{Kohn-Luttinger diagram contributing to the second order \se.}
  \label{fig:KL_spinless}
\end{figure}
\begin{figure}[b]
  \includegraphics[width=6cm]{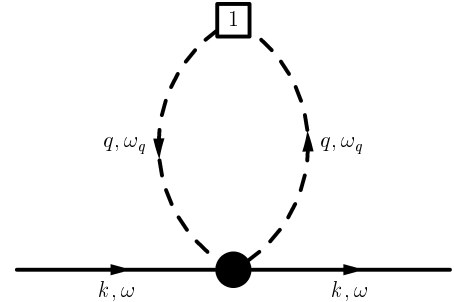}
  \caption{``Mixed'' contribution to the second order \se. This graph is a tadpole, with an insertion of the first order chemical potential counterterm.}
  \label{fig:tadpole_dmu1_spinless}
\end{figure}
The dressed propagators $G$ are such that they satisfy the Dyson equation: $G^{-1}={G_*^{(0)}}^{-1}-\Sigma-\Sigma_{\mathrm{ct}}$. The chemical potential and Fermi momenta are found by requiring that they vanish for $\omega=\mu$ and for $k$ on the interacting \fs, and that the Luttinger theorem is satisfied:
\begin{eqnarray}
  &&G_{\rmr,0}^{-1}(k=k_{\rmf,0},\omega=\mu)=0,\\
  &&G_{\rmr,\pi}^{-1}(k=k_{\rmf,\pi},\omega=\mu)=0,\\
  &&\sum_I \delta k_I^{(j)}=0,
\end{eqnarray} 
with $j=1$ here, because we're working at first order for the moment. It is very easy to check that one finds:
\begin{eqnarray}
  &&\delta\mu^{(1)}=-(A+B+2C)\frac{\Lambda_0}{4\pi},\\
  &&\delta k_0^{(1)}=(A-B)\frac{\Lambda_0}{4\pi v_\rmf}.
\end{eqnarray}
This is fully compatible with equations (\ref{eq:dk0_vrai}) and (\ref{eq:dmu_vrai}), except for second order terms that we do not find here, because we have changed the way we choose the cut-off. 


\subsubsection{Second order calculation with counterterms and next-order considerations}

\begin{figure}[t]
  \includegraphics[width=6cm]{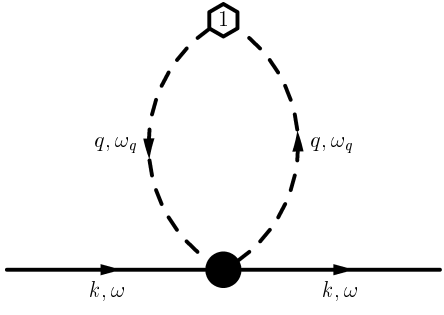}
  \caption{``Mixed'' contribution to the second order \se. This graph is a tadpole, with an insertion of the first order Fermi momentum counterterm.}
  \label{fig:tadpole_dk1_spinless}
\end{figure}

As for the first order calculation, we have ``usual'' contributions to the \se, namely the sunrise and Kohn-Luttinger diagrams of Figs.~\ref{fig:sunrise_spinless} and \ref{fig:KL_spinless}. We also have ``pure'' counterterms contributions, as in Figs.~\ref{fig:dmu1_spinless} and \ref{fig:dk1_spinless}, with the index 1 replaced by an index 2. But now we also have two ``mixed'' contributions, involving counterterms of the previous order, shown in Figs.~\ref{fig:tadpole_dmu1_spinless} and \ref{fig:tadpole_dk1_spinless}. 
In fact, these two graphs vanish, for the same reason the Kohn-Luttinger graph vanishes. This is consistent, because there is no divergence from the Kohn-Luttinger graph to cancel. 

Before studying the sunrise graph, let us see how useful the counterterms are for third order graphs, on the example of Fig.~\ref{fig:diag_squelette_3_spinless}. It is now obvious that it will be completely canceled by the same graph, with the inserted tadpole replaced by the two first order counterterms. Notice that the fourth order graph, consisting of still the same graph, with an inserted sunrise instead of a tadpole, would not be canceled by the graph with inserted second-order counterterms. The reason is that the sunrise is frequency and momentum dependent, but the counterterms are not. However, the counterterms allow for the infrared divergence cancellation obtained at zero external momentum and frequency.

The sunrise is easily computed, and one gets the following contributions (the interaction index $G$ refers to the interaction associated to the two black dots in the sunrise):
\begin{eqnarray}
  &&\Sigma_{\rmr,0;G}^{(2)}(k=k_{\rmf,0}+\kappa,\omega=\mu+\nu)\\
  &&\hspace{0.5cm}=\frac{1}{4}\left(\frac{G}{2\pi v_\rmf}\right)^2 (\nu-v_\rmf \kappa)\ln\left(\frac{| \nu^2-(v_\rmf \kappa)^2 |}{(2v_\rmf\Lambda_0)^2} \right),\nonumber\\
  &&\hspace{3cm}\mbox{ for } G=A,C,F;\nonumber\\
  \nonumber\\
  &&\Sigma_{\rmr,\pi;G}^{(2)}(k=k_{\rmf,\pi}+\kappa,\omega=\mu+\nu)\\
  &&\hspace{0.5cm}=\frac{1}{4}\left(\frac{G}{2\pi v_\rmf}\right)^2 (\nu-v_\rmf \kappa)\ln\left(\frac{| \nu^2-(v_\rmf \kappa)^2 |}{(2v_\rmf\Lambda_0)^2} \right),\nonumber\\
  &&\hspace{3cm}\mbox{ for } G=B,C,F;\nonumber\\
  \nonumber\\
  &&\Sigma_{\rmr,0;D}^{(2)}(k=k_{\rmf,0}+\kappa,\omega=\mu+\nu)\nonumber\\
  &&\hspace{0.5cm}=\frac{1}{4}\left(\frac{D}{2\pi v_\rmf}\right)^2 [\nu-v_\rmf (\kappa+2\Delta k_\rmf)]\\
  &&\hspace{1.5cm}\times\ln\left(\frac{| \nu^2-[v_\rmf (\kappa+2\Delta k_\rmf)]^2 |}{(2v_\rmf\Lambda_0)^2} \right),\nonumber\\
  \label{eq:contrib_se_DR0}
  \nonumber\\
  &&\Sigma_{\rmr,\pi;D}^{(2)}(k=k_{\rmf,\pi}+\kappa,\omega=\mu+\nu)\nonumber\\
  \label{eq:contrib_se_fin}
  &&\hspace{0.5cm}=\frac{1}{4}\left(\frac{D}{2\pi v_\rmf}\right)^2 [\nu-v_\rmf (\kappa-2\Delta k_\rmf)]\\
  &&\hspace{1.5cm}\times\ln\left(\frac{| \nu^2-[v_\rmf (\kappa-2\Delta k_\rmf)]^2 |}{(2v_\rmf\Lambda_0)^2} \right)\nonumber.
\end{eqnarray}
The second-order conditions ensuring that the trial \fs is indeed the interacting one read:
\begin{eqnarray}
  &&-\frac{1}{2}\left(\frac{D}{2\pi v_\rmf}\right)^2 \left( -2v_\rmf\Delta k_\rmf\right)\ln\left(\frac{|\Delta k_\rmf|}{\Lambda_0}\right)\nonumber\\
  &&\hspace{3cm}-\delta\mu^{(2)}+v_\rmf \delta k_0^{(2)}=0,\\
  &&-\frac{1}{2}\left(\frac{D}{2\pi v_\rmf}\right)^2 \left( 2v_\rmf\Delta k_\rmf\right)\ln\left(\frac{|\Delta k_\rmf|}{\Lambda_0}\right)\nonumber\\
  &&\hspace{3cm}-\delta\mu^{(2)}+v_\rmf \delta k_\pi^{(2)}=0,\\
  &&\delta k_0^{(2)}+\delta k_\pi^{(2)}=0.
\end{eqnarray}
These equations lead to $\delta\mu^{(2)}=0$, and to $\delta k_0^{(2)}=-\left(\frac{D}{2\pi v_\rmf}\right)^2 \Delta k_\rmf \ln\left(|\Delta k_\rmf|/\Lambda_0\right)$, which is nothing but (\ref{eq:def_sf_D2}). The dressed (R,0) propagator (and others as well) can finally be deduced from all this:
\begin{eqnarray}
  &&{G_{\rmr,0}^{(2)}}^{-1}(k=k_{\rmf,0}+\kappa,\omega=\mu+\nu)=\nu-v_\rmf\kappa\nonumber\\
  &&\hspace{1cm}-\frac{1}{4}\left[\left(\frac{A}{2\pi v_\rmf}\right)^2+\left(\frac{C}{2\pi v_\rmf}\right)^2+\left(\frac{F}{2\pi v_\rmf}\right)^2\right]\nonumber\\
  \label{eq:dressed_prop_2c}
  &&\times(\nu-v_\rmf \kappa)\ln\left(\frac{| \nu^2-(v_\rmf \kappa)^2 |}{(2v_\rmf\Lambda_0)^2} \right)\\
  &&-\frac{1}{4}\left(\frac{D}{2\pi v_\rmf}\right)^2 (\nu-v_\rmf \kappa)\ln\left(\frac{| \nu^2-[v_\rmf (\kappa+2\Delta k_\rmf)]^2 |}{(2v_\rmf\Lambda_0)^2} \right)\nonumber\\
  &&+\frac{1}{4}\left(\frac{D}{2\pi v_\rmf}\right)^2 (2v_\rmf\Delta k_\rmf)\ln\left(\frac{| \nu^2-[v_\rmf (\kappa+2\Delta k_\rmf)]^2 |}{(2v_\rmf\Delta k_\rmf)^2} \right).\nonumber
\end{eqnarray}
We could now define renormalized propagators, introducing a wave function renormalization, and show how to implement a RG calculation of the dressed \fs. In order not to be too redundant, we will do this for the more general case of $N$ chains of spin 1/2 electrons, which is anyway physically motivated by the case of quasi 1D systems.


\section{Cut-off scaling RG calculation for a system of $N$ chains of spin 1/2 electrons: formalism}
\label{sec:RG_formalism}

\subsection{Setting of the model}

\label{sec:setting_model}

The free Hamiltonian is much like the one of Eq.~(\ref{eq:ham_libre_lin}), except that there are now $N$ chains instead of 2, and that the fermions carry a spin index $\sigma$:
\begin{eqnarray}
  \label{eq:ham_libre_lin_N}
  &&H_0=\sum_k \sum_{I=1}^N \sum_{\sigma=\uparrow,\downarrow}\nonumber\\
  &&\Bigg\lbrace \left[\mu^{(0)}+v^{(0)}_{\rmf,I} (k-k^{(0)}_{\rmf,I})\right] \cdag_{\rmr,I,\sigma}(k) \anc_{\rmr,I,\sigma}(k)\\
  &&+\left[\mu^{(0)}-v^{(0)}_{\rmf,I}(k+k^{(0)}_{\rmf,I})\right] \cdag_{\rml,I,\sigma}(k) \anc_{\rml,I,\sigma}(k) \Bigg\rbrace.\nonumber
\end{eqnarray}
We will in fact assume, as we did previously, that the Fermi velocity is independent of the chain index $I$, and that it remains unrenormalized. We will thus simply use the notation $v_\rmf$. 

As in the two-chain model, we select low-energy interaction processes. Those are of two types. The first one denoted by $G$, generalizes the interactions $A$ to $F$ of the two-chain model. They are forward or backward scattering interactions. We shall only be interested in interactions that are invariant under spin rotations. Thus, we will use the charge and spin couplings $\gc$ and $\gs$. We refer the reader to our previous paper\cite{Dusuel02} for more details about this parametrization. There is however one major difference between the situation described in this article, and the one we are interested in here. Because of periodic boundary conditions in the transverse direction, all indices $I$, $J$ and $\delta$ are defined modulo the number of chains $N$. This was not the case in our previous article, where the chains were obtained after considering patches on a nearly square \fs, thus the $N$ chains had boundaries, and as a consequence the chains were not all equivalent. 

Furthermore, if the filling is not too far from one half, we have to consider Umklapp scatterings. These will be denoted by $U$. It is easy to convince oneself that due to the Pauli principle, there is no need to consider exchange couplings for the Umklapps. The interaction Hamiltonian is thus:
\begin{equation}
  \label{eq:H_int_basse_en}
  H_\mathrm{int}=H_\mathrm{int}^{(G)}+H_\mathrm{int}^{(U)}, \mbox{ with:}
\end{equation}
\begin{eqnarray}
  &&H_\mathrm{int}^{(G)}=\frac{2\pi v_\rmf}{NL}\sum_{I,J,\delta}~ \sum_{k,k',q}~ \sum_{\tau,\tau'}~ \sum_{\rho,\rho'}\nonumber\\
  \label{eq:H_int_basse_en_G}
  &&\Bigg\lbrace \Big[ \gc_\delta(I,J) \mathbb{I}_{\tau,\tau'} \mathbb{I}_{\rho,\rho'} + \gs_\delta(I,J) \boldsymbol{\sigma}_{\tau,\tau'} \cdot \boldsymbol{\sigma}_{\rho,\rho'} \Big]\\
  &&\times\cdag_{\rmr,I+\delta,\tau}(k+q) \cdag_{\rml,J-\delta,\rho}(k'-q) \anc_{\rml,J,\rho'}(k') \anc_{\rmr,I,\tau'}(k)~ \Bigg\rbrace,\nonumber
\end{eqnarray}
\begin{figure}[t]
  \includegraphics[width=6cm]{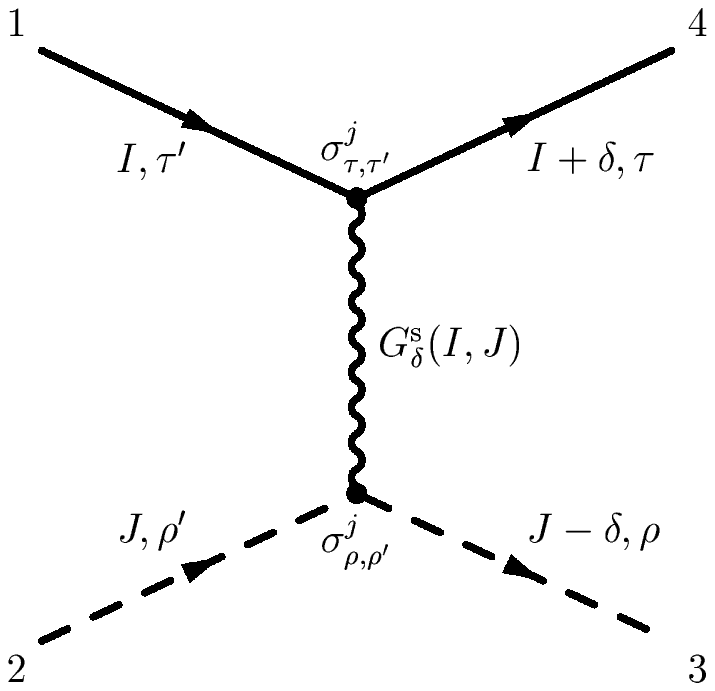}
  \caption{Graphical representation of the spin interaction $\gs_\delta(I,J)$.}
  \label{fig:gs}
\end{figure}
and
\begin{eqnarray}
  &&H_\mathrm{int}^{(U)}=\frac{\pi v_\rmf}{NL}\sum_{I,J,\delta}~ \sum_{k,k',q}~ \sum_{\tau,\tau'}~ \sum_{\rho,\rho'} \nonumber\\
  &&\Bigg\lbrace U_\delta(I,J) \mathbb{I}_{\tau,\tau'} \mathbb{I}_{\rho,\rho'}\\
  \label{eq:H_int_basse_en_U}
  &&\times\cdag_{\rmr,I+\delta,\tau}(k+q) \cdag_{\rmr,J-\delta,\rho}(k'-q) \anc_{\rml,J,\rho'}(k') \anc_{\rml,I,\tau'}(k)\nonumber\\
  &&\hspace{5cm} + \mbox{ h. c.}~ \Bigg\rbrace. \nonumber
\end{eqnarray}

The factors $1/N$ are required to yield a good thermodynamical limit. The $2\pi v_\rmf$ terms have been factorized, so that the couplings are dimensionless, and this will suppress many $2\pi v_\rmf$ denominators in the following. In the case of $G$ couplings, the left-right symmetry requires $\gcs_\delta(I,J)=\gcs_{-\delta}(J,I)$, and the hermiticity of $H_\mathrm{int}$ yields $\gcs_\delta(I,J)=\gcs_{-\delta}(I+\delta,J-\delta)$. The first of these relations, \ie $U_\delta(I,J)=U_{-\delta}(J,I)$, naturally holds for the Umklapps because of the Pauli principle, so that the interaction that destroys two left fermions on chains $I$ and $J$, and creates two right fermions on chains $I+\delta$ and $J-\delta$ is present twice. The difference of a 1/2 factor between the Umklapps and the $G$ interactions, is here to compensate this. We let the reader check that in the case of the Hubbard model with an interaction Hamiltonian $\mcu\sum_i n_{i,\uparrow} n_{i,\downarrow}$, one has (up to $2\pi v_\rmf$ factors) $\gc=\mcu/2$, $\gs=-\mcu/2$ and $U=\mcu$. Because of this last equality, we will simply give the value of $U$ when referring to Hubbard couplings. Of course the Hubbard model, in terms of right and left fermions, also contains $g_4$ interactions, but these have been set to zero, for the reasons already given in Sec.~\ref{sec:sub:sub:model_notations}. 
\begin{figure}[t]
  \includegraphics[width=6cm]{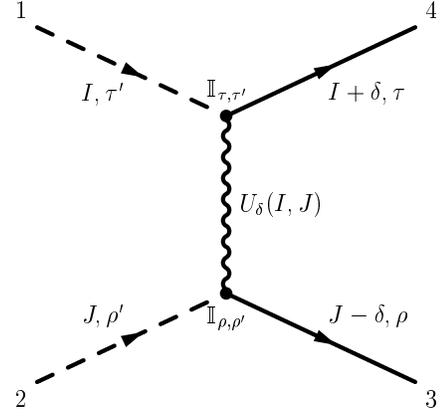}
  \caption{Graphical representation of the Umklapp interaction $U_\delta(I,J)$.}
  \label{fig:u}
\end{figure}

In order to make our notations for the interactions a bit more concrete, we show two Feynman graphs in Figs.~\ref{fig:gs} and \ref{fig:u}, associated respectively with $\gs$ and $U$ terms. The representation for $\gc$ is the same as the one for $\gs$, except it involves $\mathbb{I}$ matrices instead of $\sigma$ matrices. Notice we do not use the single dot notation as we did previously, because it is not suited for the Umklapps, but we have adopted the wiggly line instead. In these graphs we also show which external legs are numbered 1, 2, 3 and 4.


\subsection{Cut-off scaling calculation of the \fs}

\subsubsection{General considerations}
\label{sec:sub:sub:gen_th_an}

One of the main conclusions of Sec.~\ref{sec:csfs} is the necessity to use a renormalized perturbation theory in situations where the \fs changes as a function of interaction strength. In the standard many-body formalism, this is achieved by the introduction of counterterms which pin the dressed \fs. The outcome is a precise connection such as Eq.~(\ref{eq:def_sf_D2}) between the bare and dressed \fss which may in principle be computed to any order in perturbation theory. As noticed already long ago by Gell-Mann and Low,\cite{Gell-Mann54} it is possible to sum infinite classes of contributions using a renormalization group procedure. This idea has played a crucial role in building a consistent physical picture of quasi 1D conductors for instance.\cite{Solyom79_dans_articles}

The most general and flexible way to implement a renormalization approach
is based on Wilson's idea of gradual mode elimination. Several groups have recently implemented Wilson's approach to the RG, expressed via the Polchinski equation,\cite{Zanchi96,Halboth00} or its one-particle irreducible version.\cite{Jungnickel96,Honerkamp01} Although these equations are exact, they are quite complicated, since effective interactions involving an arbitrary number of particles are generated along the RG flow. Any numerical computation requires therefore drastic truncations in the effective action. For this reason, we have prefered
to use a simplified version of RG which is known as ``cut-off'' scaling.
This procedure has been initiated in the pioneering work by Anderson et al.
\cite{Anderson70} for the Kondo problem, and put in a more mathematical
form by Abrikosov and Migdal \cite{Abrikosov70} and Fowler and
Zawadowski.\cite{Fowler71} A very extensive review on this method has been
written by S\'olyom.\cite{Solyom79_dans_articles} 

This scheme amounts to constructing a one parameter family of ``bare''
Hamiltonians. These are defined on the single particle states whose momentum 
lies in a strip of width $\Lambda$ away from the \fs.
It is therefore natural to parametrize these Hamiltonians as a function of
$\Lambda$. Note that by contrast to Wilson's effective action, which includes
all the possible types of interactions (relevant, marginal {\em and}
irrelevant ones), the cut-off scaling procedure only considers relevant and
marginal couplings. So unlike what is achieved in Wilson's RG, it is no
longer possible to preserve invariance of the {\em full} set of low-energy
correlation functions as $\Lambda$ is gradually decreased. The cut-off
scaling approach only allows to preserve a restricted set of low-energy
observables, for instance the first derivatives of the two-point function
with respect to external momentum and frequency, and the value of the
four-point function for external legs taken on-shell at the Fermi level.   

Actual computations within this scheme encounter a new difficulty when
the \fs is sensitive to the strength of interactions. As explained in detail
in Sec.~\ref{sec:sub:use_ct} the bare propagators used in Feynman graphs
are required to be singular on the {\em dressed} \fs. But clearly, this is
not known until the whole computation has been performed. This calls for
an iterative procedure. For a given microscopic model 
(defined with an initial cut-off $\Lambda_{0}$), 
the one-particle part of the corresponding Hamiltonian defines a \fs 
which will be called the $\Lambda_{0}$-\fs. This is a natural
first choice for a trial dressed \fs in the iterative computation.
One can then construct the flow equations for the running bare \fs 
(the $\Lambda$-\fs) and the effective couplings using cut-off scaling. 
In the limit where $\Lambda$ goes to zero (or at least to its minimal value 
before a phase transition occurs), the $\Lambda$-\fs goes towards a new 
dressed \fs, which is to be used as the new trial dressed \fs in the next 
step of the iteration.

On physical grounds, such a computation is expected to converge although we
have not embarked yet in checking this statement. Instead we have tried to
bypass the intrinsic complication of an iterative procedure by appealing to
the physical insights gained in Sec.~\ref{sec:sub:2chaines_ordre1} devoted 
to the energetic approach. The main ideas are the following: firstly, in 
the high-energy regime, and in the logarithmic approximation, the dressed 
Fermi momenta do not appear in the flow couplings' equations; secondly, in the low-energy 
part of the flow, the \fs should not move too much, because the couplings 
that deform it are irrelevant in this regime. 
The first point will be checked on the RG equations. 
The second one has already been checked in the two-chain model, 
where only the $D$ coupling, that does not exist at very low energies because 
of the non-flatness of the \fs, deforms the \fs. We will furthermore check 
it remains true in the case of $N$ chains. Assuming what happens in the 
intermediate energy regime (defined by the curvature of the \fs) is not 
essential, the computation of the dressed \fs is now possible in a {\em single}
 step. Indeed, we do not need a priori knowledge of the dressed \fs anymore, 
since it disappears from the RG couplings' flow equations in the logarithmic approximation 
of the high-energy regime. The flow of the $\Lambda$-\fs, will then be stopped 
when the cut-off $\Lambda$ becomes comparable to the maximal momentum scale 
defined by the running $\Lambda$-\fs. 
Note that this is the least controlled step of this approximated scheme, 
because the \fs does not define one single 
momentum scale, but rather a continuum of scales. 
This will for example prevent us from using this scheme when the Umklapp 
couplings are taken into account, in a system too far from half-filling. 

The couplings' flow equations in the cut-off scaling are given in Appendix 
\ref{app:flow_eq}, since their derivation is standard. 
We shall now focus on the RG computation of the \fs. 
The basic equation is Eq.~(\ref{eq:fs_counterterm}). The \se that appears in
this equation can be found from Eq.~(\ref{eq:se_Nchaines}), where, since we work in the cut-off
scaling scheme, the cut-off $\Lambda_0$ should now be replaced by the running
cut-off $\Lambda$, and where the couplings are to be understood as running bare
couplings.
 Given that $\delta k_I=k^{(0)}_I(\Lambda)-k_I$, Eq.~(\ref{eq:fs_counterterm}) 
allows us to express $k^{(0)}_I(\Lambda)$ as a function of $\Lambda$, the 
running bare couplings and the dressed Fermi momenta $k_I$. But it is the 
latter who are fixed independently of the value of $\Lambda$, so that it is 
more convenient to invert this relation, working only at a second order 
accuracy, to get $k_I$ as a function of $\Lambda$, the running bare couplings 
and the running Fermi momenta $k^{(0)}_I(\Lambda)$. Asking for the invariance 
of $k_I$ as $\Lambda$ is changed yields the \fs flow equation that we give in 
Appendix \ref{app:flow_eq}, Eq.~(\ref{eq:flot_fs}). 

Only the couplings for which the curvature of the \fs is felt, \ie for which $\Delta k_\alpha(I,J)\neq 0$ or $\Delta k_\alpha^{U}(I,J)\neq 0$ contribute to the flow of $k_I^{(0)}(\Lambda)$. This is what we had already noticed in the two-chain system, where the $D$ coupling was the only one to give a deformation of the \fs. This confirms that only couplings that will be irrelevant in the low-energy regime contribute to the deformation of the \fs. 

Finally, notice that we have not taken the first order contribution into account. This is not justified in the general case, but for the initial condition we are interested in, \ie with all charge, spin and Umklapp couplings equal to $G_\rmb^\rmc$, $G_\rmb^\rms$ and $U_\rmb$, the first order contribution vanishes. We emphasize this is true only in the high-energy regime, where the couplings will have a purely one-dimensional (1D) flow (because of the logarithmic approximation), and thus will remain equal (with one value for each of the three types of couplings). Once the low-energy regime is reached the couplings become different, so that one should take the first-order deformation into account. 

The \se at one loop is given by the contribution of the tadpole diagram. It is easy to see that out of the three couplings $\gc$, $\gs$ and $U$, only the $\gc$ couplings contribute. Compared to the spinless case, there will be a factor of 2, because of the two possible spin states of the propagator in the loop. We let the reader check that:
\begin{eqnarray}
  &&\Sigma_{\rmr,I}^{(1)}(k=k_{\rmf,I}+\kappa,\omega=\mu+\nu)=\nonumber\\
  &&\hspace{2cm}\left[\frac{1}{N}\sum_J \gc_0(I,J)\right] (2 v_\rmf\Lambda),
\end{eqnarray}
and that the corresponding first-order Fermi momenta counterterms read:
\begin{equation}
  \delta k_I^{(1)}=\frac{2v_\rmf\Lambda}{N}\left( \sum_J \gc_0(I,J)-\frac{1}{N}\sum_{I,J} \gc_0(I,J)\right).
\end{equation}

However, because we are interested in systems for which $\tperp$ is small, \ie for systems that are nearly 1D, we know that all the chains are nearly equivalent, so that the RHS of the previous equation will be nearly independent of $I$, and thus, very small. That this is true can be checked on Fig.~\ref{fig:evolution_coupC_charge}, that will be described later, and on which the couplings $\gc_0(I,J)$ are represented after running the flow, into the low-energy regime. It is clear on this figure that the term $\sum_J \gc_0(I,J)$ is nearly independent of $I$.


\subsubsection{Analytical study of the simplest example}

Let us illustrate all this on the simplest possible case, for which $G_\rmb^\rms=0$ and $U_\rmb=0$. This physically corresponds to a system away from half filling, so that it is justified to neglect the Umklapps. Furthermore we have set the spin couplings to zero, which corresponds to the Luttinger liquid fixed point. This simplifies the flow, because the charge couplings then remain constant along it. Furthermore, this will allow us to compare our results to those obtained in the literature, starting from decoupled Luttinger liquids, coupled by a hopping term. It is easy to check that the general flow equation of the \fs (\ref{eq:flot_fs}) can be simplified in:
\begin{eqnarray}
  \partial_t k_{\rmf,I}^{(0)}&=&\frac{{\gc_\rmb}^2}{N^2}\sum_{J,\alpha} \left[(k_{\rmf,I+\alpha}^{(0)}+k_{\rmf,J}^{(0)})-(k_{\rmf,I}^{(0)}+k_{\rmf,J-\alpha}^{(0)})\right]\nonumber\\
  &=&-{\gc_\rmb}^2 k_{\rmf,I}^{(0)}+\frac{{\gc_\rmb}^2}{N}\sum_\alpha k_{\rmf,I+\alpha}^{(0)}.
\end{eqnarray}
If we denote by $\overline{k}=(\sum_I k_{\rmf,I})/N=(\sum_I k_{\rmf,I}^{(0)})/N$ the mean value of the Fermi momenta, and if we write $k_{\rmf,I}^{(0)}(t)=\overline{k}+\delta k_{\rmf,I}^{(0)}(t)$, the differential equation is easily solved and the solution is:
\begin{equation}
  \label{eq:lien_mult_kf_nu_hab}
  \delta k_{\rmf,I}^{(0)}(\Lambda)=\delta k_{\rmf,I}^{(0)}(\Lambda_0)\left( \frac{\Lambda}{\Lambda_0}\right)^{{\gc_\rmb}^2}.
\end{equation}

The question that now arises, is how to determine at what scale $\Lambda^*$ the flow should be stopped. This scale cannot be determined precisely in the cut-off scaling scheme, which is only a very simple version of the Wilsonian approach. Only the latter approach could precisely describe the transition between the two regimes, which would not even occur at a single scale (because of the large number of different scales $K$ appearing in the RHS of the RG flow equations). We will thus adopt the simple and pragmatic following point of view: the flow of the \fs will be stopped, when the biggest of these scales is reached, \ie when the scale given by the difference between the biggest and the smallest Fermi momenta (denoted by $\Delta k_\rmf^{\max}(\Lambda)$) is reached:
\begin{equation}
  \label{eq:def_lambda_star}
  \Lambda^*=\Delta k_\rmf^{\max}(\Lambda^*).
\end{equation}
Notice that if the \fs flattens more quickly than the RG time decreases, this scale will never be reached.

According to Eq.~(\ref{eq:lien_mult_kf_nu_hab}), the differences between the Fermi momenta and the mean value are all multiplied by the same factor. The biggest (respectively smallest) momentum will remain the biggest (respectively smallest) along the flow. We thus have $\Delta k_\rmf^{\max}(\Lambda)=\Delta k_\rmf^{\max}(\Lambda_0) (\Lambda/\Lambda_0)^{{\gc_\rmb}^2}$. We will stop the flow at the scale $\Lambda^*$ such that $\Lambda^*\simeq\Delta k_\rmf^{\max}(\Lambda_0) (\Lambda^*/\Lambda_0)^{{\gc_\rmb}^2}$. Finally we get the following link between high-energy and low-energy momenta:
\begin{eqnarray}
  \delta k_{\rmf,I}&\simeq&\delta k_{\rmf,I}^{(0)}(\Lambda_0) \left( \frac{\Delta k_\rmf^{\max}(\Lambda_0)}{\Lambda_0}\right)^{\frac{{\gc_\rmb}^2}{1-{\gc_\rmb}^2}} \\ 
  \label{eq:lien_DkfR_Dkfnu}
  \Rightarrow \Delta k_\rmf^{\max}&\simeq&\Delta k_\rmf^{\max}(\Lambda_0) \left( \frac{\Delta k_\rmf^{\max}(\Lambda_0)}{\Lambda_0}\right)^{\frac{{\gc_\rmb}^2}{1-{\gc_\rmb}^2}}.
\end{eqnarray}
As $\Delta k_\rmf^{\max}(\Lambda_0)=2\tperp/\tpar$, we obtain the result already found in the literature:\cite{Prigodin79,Bourbonnais85}
\begin{equation}
  \label{eq:def_sf_litterature}
  \tperp^\mathrm{eff}\sim\tperp \left( \frac{\tperp}{\tpar}\right)^\frac{\alpha}{1-\alpha},
\end{equation}
where $\alpha$ is the single-particle Green's function's exponent: $\alpha=(K_\rho+1/K_\rho)/4-1/2$, with $K_\rho=\sqrt{(1-2G^\rmc_\rmb)/(1+2G^\rmc_\rmb)}$. Perturbatively, $\alpha=\left.{G^\rmc_\rmb}\right.^2$, so that our result is indeed the same as Eq.~(\ref{eq:def_sf_litterature}), to lowest order. Let us mention that the power-law behavior of Eq.~(\ref{eq:def_sf_litterature}) has been confirmed numerically, for a two-chain system, using exact diagonalization techniques.\cite{Capponi98}

Notice that according to Eq.~(\ref{eq:lien_DkfR_Dkfnu}), if ${\gc_\rmb}^2\geqslant 1$, the effective transverse hopping vanishes, and the dressed \fs is flat. Although this result is confirmed by the non-perturbative (in the coupling) result Eq.~(\ref{eq:def_sf_litterature}) after replacing ${G^\rmc_\rmb}^2$ by $\alpha$, we shall not use the perturbative RG in such situations that lay outside the validity range of this approach.


\subsubsection{Comparison to other previous results of the literature}
\label{sec:sub:sub:comparison}

Before turning to numerical calculations, we shall compare our equations describing the deformation of the \fs to some more results of the recent literature. Let us begin with the article by Kishine and Yonemitsu,\cite{Kishine98} which treats exactly the same problem as ours. We shall compare our equations (\ref{eq:flot_fs}) to their flow equation for the effective transverse hopping (Eq.~(4)). Note that they obtained this equation from the previous works by Bourbonnais and co-workers\cite{Bourbonnais84,Bourbonnais91,Bourbonnais93} and Kimura\cite{Kimura75} (see also the review article by Firsov, Prigodin and Seidel\cite{Firsov85}). We however choose the paper by Kishine and Yonemitsu because their formalism is the closest to ours. 

The comparison is easily achieved in two steps: first notice that our Fermi momenta do not flow when no interactions are turned on (which is desirable, since the \fs should not get deformed in this case), while their $\tperp$ flows in this case, because of the first term of their equation coming from the rescaling they have performed, so that we should simply forget this term if we want to compare our results. Second, we have to take the particular set of couplings they have chosen, namely local couplings. This is obtained when setting all charge couplings to the same value, and doing the same for spin couplings and Umklapps. Repeating exactly what we have done in the previous section, we find:
\begin{equation}
  \partial_t \delta k_{\rmf,I}^{(0)}=-\delta k_{\rmf,I}^{(0)} \left({\gc_\rmb}^2+3{\gs_\rmb}^2+\frac{U^2}{2}\right).
\end{equation}
This in particular means that all Fermi momenta $\delta k_{\rmf,I}^{(0)}$ will be scaled by the same factor. Next, we have to link our charge and spin couplings to the g-ology notation. It is easily checked that one simply has: $\gc=g_2-g_1/2$ and $\gs=-g_1/2$ (here all $g_\|$ and $g_\perp$ couplings of the g-ology are equal because we have restricted ourselves to spin-rotation invariant couplings). We also have the trivial identification $U=g_3$. The difference in the numerical factor 4 simply comes from a different normalization of the dimensionless couplings (we divided the couplings by $2\pi v_\rmf$ and they divided them by $\pi v_\rmf$). Finally, we have $\Delta k_\rmf^{\max}=2(\tperp/\tpar)$ which shows the equivalence between the two approaches.

We want to stress that this equivalence relies on our simple approximation 
according to which the \fs is deformed only in the high energy regime, since
this deformation is driven by irrelevant couplings. It would be possible to
go beyond this approximation by implementing the iterative procedure outlined
in Sec.~\ref{sec:sub:sub:gen_th_an}. To estimate the residual deformation
of the \fs induced by these irrelevant couplings in the low-energy regime 
remains an interesting open question, that could be addressed within the 
general framework discussed in this paper. 
Furthermore, such a calculation would enable us to take
into account the deformation of the \fs induced by the Hartree terms which are
effective only when the forward scattering amplitudes significantly vary along
the Fermi line. Such a dependence is only generated when the running cut-off 
becomes comparable to or smaller than the natural scale associated to the 
transverse dispersion.

For the sake of completeness, we shall give a simple and quick derivation of the RG equation, which emphasizes the role of the 1D chains (see note 31 of Ref.~\cite{Bourbonnais84}), and explains why the exponent obtained in Eq.~(\ref{eq:def_sf_litterature}) is the 1D propagator's exponent. The idea is to assume that the full propagator at scale $\Lambda$ can be obtained by taking the corresponding purely 1D propagator at scale $\Lambda$, and correcting it with the dispersion relation induced by the bare $\tperp$: $G^{-1}(\Lambda)=G^{-1}_{\mathrm{1D}}(\Lambda)+2\tperp\cos(k_\perp)$. (In other words, this amounts to assume that when computing the effective action at scale $\Lambda$, one puts the inter-chain hopping aside, so that the flow is purely 1D, and the (unrenormalized) inter-chain hopping is reintroduced in the effective action at the end of the computation). But we can write $G^{-1}_{\mathrm{1D}}=Z^{-1}_{\mathrm{1D}}(\Lambda) [\omega-\tilde{\ve}_\Lambda(k_\|)]$. Note that in the previous two formulas, we have denoted by $k_\perp$ and $k_\|$ the transverse and longitudinal momenta. 
$Z_{\mathrm{1D}}$ is the 1D wave-function renormalization, and $\tilde{\ve}_\Lambda(k_\|)$ is the renormalized 1D dispersion relation. We thus get $G^{-1}(\Lambda)=Z^{-1}_{\mathrm{1D}}(\Lambda) [\omega-\tilde{\ve}_\Lambda(k_\|)+2Z_{\mathrm{1D}}(\Lambda)\tperp\cos(k_\perp)]$, showing that the effective inter-chain hopping at scale $\Lambda$ reads: $\tperp(\Lambda)=Z_{\mathrm{1D}}(\Lambda)\tperp$. The effective $\tperp$ at two different scales are thus proportionally related by the 1D $Z$ function, whose flow equation can easily be deduced from Eq.~(\ref{eq:flot_phi}) specialized to the 1D case. This yields the correct flow equation for $\tperp$. Let us also mention that this way of taking into account the inter-chain tunneling has been recently adopted by Essler and Tsvelik,\cite{Essler02} except that they use the exact 1D Green's function instead of the result of a perturbative RG computation.

Let us now compare our results with those of Fabrizio,\cite{Fabrizio93} whose work is devoted to the two-chain model without longitudinal Umklapps, but with Fermi velocity renormalization. Fabrizio used a Wilsonian RG (at two loops) for the calculation of the deformation of the \fs. As one can expect, this formalism allows to cross the energy scales coming from the non-flatness of the \fs (see how the flows are defined piece-wise in his appendix A, and for which the various RHS never diverge). Our equations coincide with those of Fabrizio in the high-energy regime (when his function $C_2$ is expanded to lowest order in $h$, the dimensionless $\Delta k_\rmf$), and in the low-energy regime where the flow of the \fs vanishes. The intermediate regime is of course different. Note for the comparison that Fabrizio's coupling $g_b$ is our coupling $D$.

Finally, we would like to note that our results at one loop are consistent with the article of Louis, Alvarez and Gros,\cite{Louis01} (see their Eqs.~(9) and (13)) if we specialize these to the case of uniform Fermi velocities.


\section{Coupled cut-off scaling and field-theoretical RG calculations for a system of $N$ chains of spin 1/2 electrons: numerical results}
\label{sec:RG_numerical}

\subsection{Motivation of the use of two RG schemes}

\label{sec:sub:motiv_use_two_RG}

Before we present our results for specific models, we wish to emphasize that
one of our motivations besides the \fs deformation was to describe precisely
the connection between the essentially 1D high-energy regime, and the 2D 
low-energy physics where the system is sensitive to the warping of the \fs.
This can be viewed as a complement to the RG analysis of Lin et al.\cite{Lin97}
who have focused exclusively on the low-energy side where the cut-off is much
smaller than the scale associated to the transverse dispersion. In this work,
they could relate the high and low energy regimes without actually solving
RG flow equations for the former since they assumed very weak bare couplings
(so that these couplings were barely renormalized in the high energy part of
the flow). Although working with the full Wilsonian effective action allows
one to get through such intermediate energy scales, it is not clear to us 
that this can be achieved in a reliable way with the cut-off scaling. Indeed,
our view of this procedure is that it provides a simple approximation of the
full Wilsonian RG, which is certainly well controlled when the running cut-off
is much larger than the intrinsic low-energy scales of the system's dynamics. 
Because of this we have decided to study the low-energy part of the flow in
the field-theoretical framework. This latter scheme heavily relies on 
the existence of an
infinite cut-off limit (continuum limit), or in other words the corresponding
theory of 1D fermions with linear dispersion and point-like interactions is 
renormalizable. This statement is {\em independent} of the existence 
of intrinsic low-energy
scales such as a mass term, or variations in Fermi wave vectors with the chain
index. In this context RG equations are obtained by relating physical 
properties measured at different running energy scales. To avoid confusion 
with cut-off scaling this running scale has been denoted by $\nu$ in Appendix
\ref{app:field_th} which presents some details on this approach. 
Since we have used a logarithmic approximation, the high-energy flows of the 
couplings in both
cut-off scaling and field-theoretical RG are identical. It is an interesting
question whether the two schemes give the same physical low-energy results or 
not. We plan to study this in more detail in a forthcoming work.

\subsection{A first numerical study: incommensurate nesting}

\label{sec:sub:first_num_std}

We will now show what information can be deduced from numerical computations. 
For this we choose to focus on a simple example, where the Umklapps are set 
to zero, but we still assume a perfect \fs nesting. This situation is realized 
in several interesting systems as for instance in two dimensional molybdenum
and tungsten bronzes. For a review, see for example the paper by Foury and 
Pouget.\cite{Foury93}

\begin{figure}[t]
  \includegraphics[width=8cm]{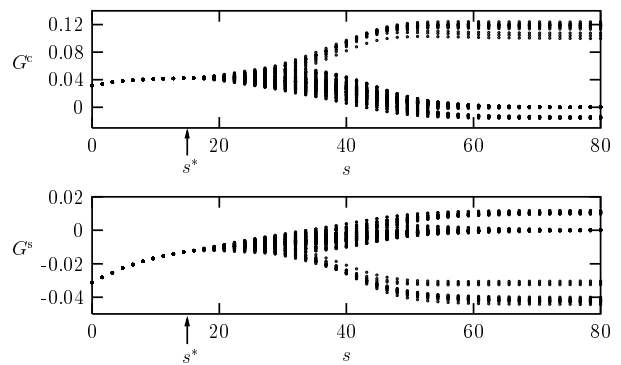}
  \caption{Flow of the normalized charge (top) and spin (bottom) couplings, as a function of the ``good'' time $s$, for $N=8$. Initially the true couplings (not normalized) are $\gc=0.3=-\gs$, and the bare hopping is $\tperp/\tpar=0.1$. We have indicated the value $s^*$ corresponding to the value $\Lambda^*$ (see Eq.~(\ref{eq:def_lambda_star})).}
  \label{fig:flot_gcs}
\end{figure}
\begin{figure}[b]
  \includegraphics[width=8cm]{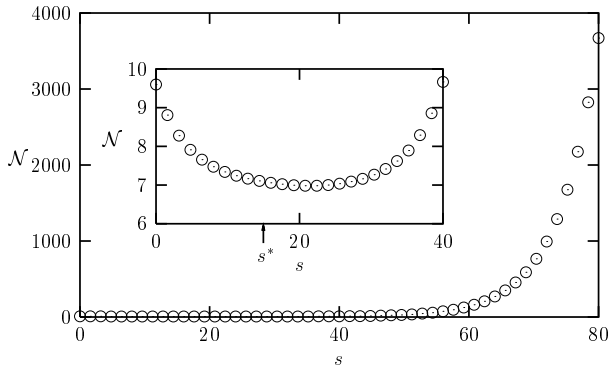}
  \caption{Flow of the norm $\mathcal{N}$ as $s$ grows, corresponding to Fig.~\ref{fig:flot_gcs}. The inserted flow is a zoom on short times.}
  \label{fig:norme}
\end{figure}
\begin{figure}[t]
  \includegraphics[width=8cm]{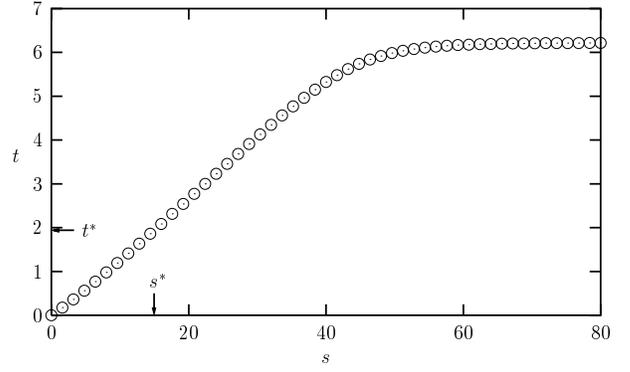}
  \caption{Link between the ``good'' RG time $s$ and the ``true'' RG time $t$, corresponding to Fig.~\ref{fig:flot_gcs}. We have indicated the values of $s^*$ and of $t^*=\ln(\Lambda_0/\Lambda^*)$ corresponding to $\Lambda^*$}
  \label{fig:temps}
\end{figure}
\begin{figure}[b]
  \includegraphics[width=8cm]{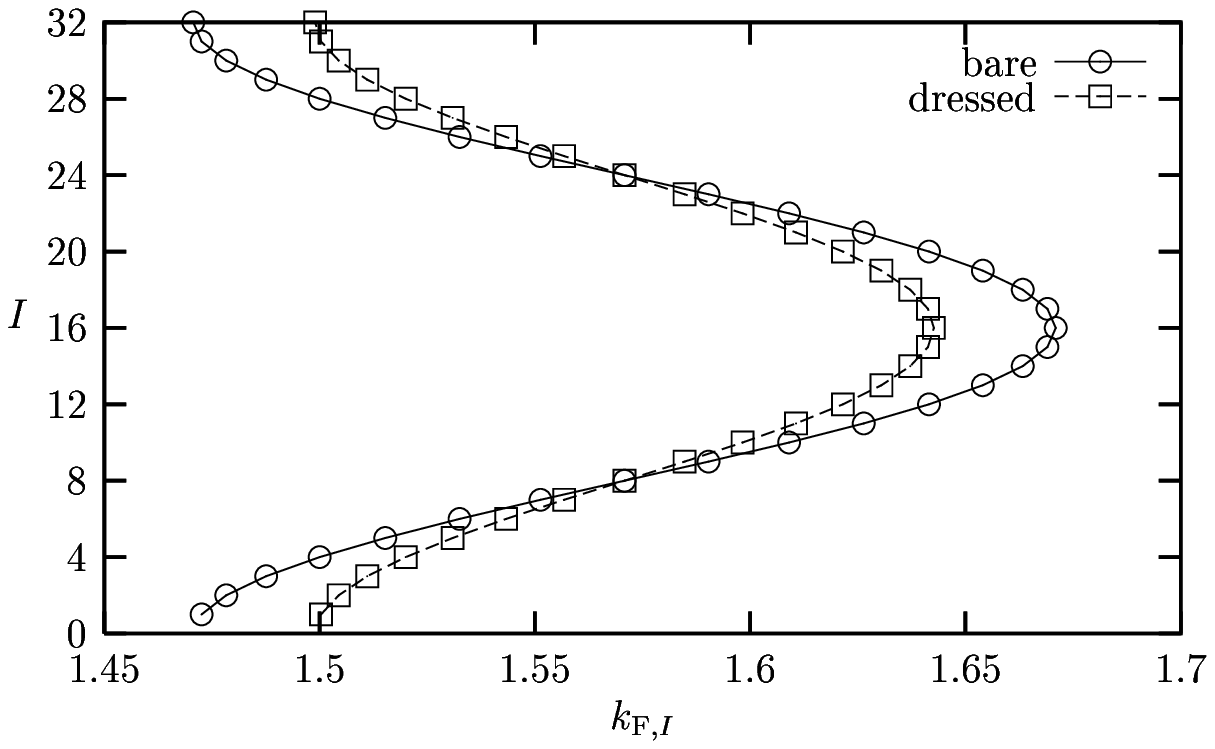}
  \caption{Bare and dressed \fs for $N=32$.}
  \label{fig:defSF32}
\end{figure}
\begin{figure}[t]
  \includegraphics[width=8cm]{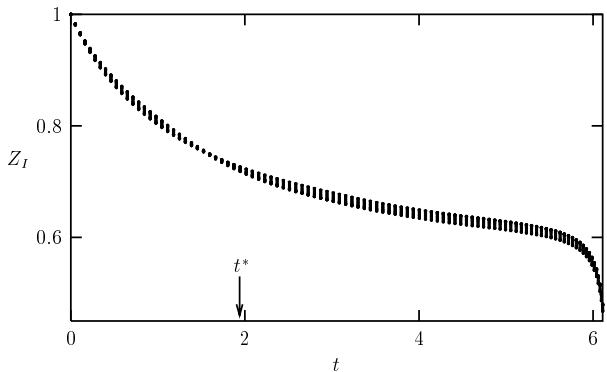}
  \caption{Flow of the quasiparticle weights $Z_I=1/\vp_I$, for $N=32$. We have represented these for $I=16$ to $I=32$, as the ones for smaller values of $I$ can be found using the top-bottom ($I\leftrightarrow N-I$ because $N$ is even) symmetry. Note that we have indicated the value of $t^*=\ln(\Lambda_0/\Lambda^*)$ on the time axis.}
  \label{fig:Z32}
\end{figure}
\begin{figure}[b]
  \includegraphics[width=8cm]{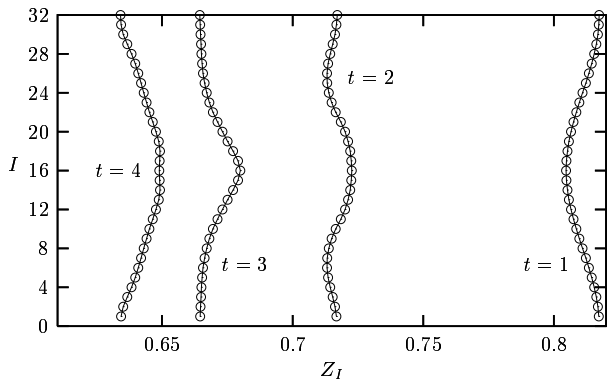}
  \caption{Here we show the transverse momentum (\ie chain number $I$) dependence of the quasiparticle weights $Z_I$, at four RG times $t=1$, $2$, $3$ and $4$ (see Fig.~\ref{fig:Z32} for a time reference).}
  \label{fig:Z32_t}
\end{figure}

We have chosen an initial condition for which all charge couplings are equal, and all spin couplings too, with $\gc=0.3=-\gs$. The bare hopping is $\tperp/\tpar=0.1$. The couplings are quite large so that the deformation of the \fs will be visible. The \fs could be deduced analytically, but we have computed it numerically as all other quantities. All the results are contained in Figs.~\ref{fig:flot_gcs} to \ref{fig:Z32_t}. The first three (respectively last three) of these figures have been computed with $N=8$ (respectively $N=32$). The reasons for these choices are that we could not represent all the couplings (the first of the six figures) for a too high value of $N$, because the number of couplings grows like $N^3$. But this was no problem for the \fs and the quasiparticle weights, except for a longer computation time.

In these figures, that we shall comment one after the other, we have made use of some notions such as the norm of the couplings, the normalized couplings, and the adapted RG time $s$. All these notions, and some others (such as fixed directions, etc.) have been dealt with extensively in our previous paper,\cite{Dusuel02} so we shall simply give the few basic definitions. The norm is the Euclidean norm of the coupling vector, and the normalized couplings are the usual couplings divided by the norm (we give an explicit formula for the charge couplings only): 
\begin{eqnarray}
  &&\mathcal{N}=\sqrt{\sum_{I,J,\delta} {\gc_\delta(I,J)}^2 + {\gs_\delta(I,J)}^2 + {U_\delta(I,J)}^2},\quad\\
  &&\widetilde{\gc}_\delta(I,J)=\frac{\gc_\delta(I,J)}{\mathcal{N}}.
\end{eqnarray}
In all the figures in which flows will be represented (such as Fig.~\ref{fig:flot_gcs}), we adopt the following convention: in the caption, we give the initial condition for {\em usual} couplings, while in the figures themselves we draw the {\em normalized} couplings, and suppress the $\tilde{\:}$ in the $y$-axis legends. The reader should not get confused by this abuse of notation. The time $s$ that we have used in the numerical simulation is defined by $\rmd s=\mathcal{N}(t)\rmd t$, and is the time adapted for zooming on the flow singularities. 

The first of the six figures, Fig.~\ref{fig:flot_gcs}, represents the ``field-theoretical'' RG flow of the normalized charge and spin couplings, as functions of the RG time $s$. This flow is divided into three regions. In the first one ($0\leqslant s\leqslant s^*\simeq 15$), corresponding to the high-energy regime, all charge couplings and all spin couplings remain equal. Indeed, in this regime, the curvature of the \fs is not felt at all, in the logarithmic approximation we use. All chains are thus identical (remember we use periodic boundary conditions in the transverse direction), the system is purely one-dimensional, so that the symmetry between the chains cannot be broken. 

In this high-energy regime, the cut-off scaling flow is exactly the same as the ``field-theoretical'' one, so that we could use the latter for the computation of the deformation of the \fs. After $s\simeq 15$, the flow of the \fs is stopped, because the scale $\Lambda^*$ (as previously defined by $\Lambda^*=\Delta k_\rmf^{\max}(\Lambda^*)$) is reached. The dressed \fs is the one obtained at that scale, and is then used in the ``field-theoretical'' flow of the couplings. This dressed \fs, and the bare one, are represented on Fig.~\ref{fig:defSF32}. As discussed in Sec.~\ref{sec:sub:sub:comparison}, within the approximation we use, the dressed \fs is still given by $\tperp \cos(k_\perp)$ but with the dressed value of the interchain hopping. Higher harmonics for $\Delta k_\rmf$ as a function of $k_\perp$, corresponding to longer range transverse hoppings, are expected to be generated only in the low-energy part of the \fs flow. Indeed it is only in this regime that effective couplings acquire a dramatic dependence with respect to transverse momenta. But this goes beyond the scope of the simple (\ie non iterative) procedure used for the numerical computations presented here.

We emphasize that {\em by contrast} to the flow of the couplings, for the quasiparticle weights $Z_I$ functions associated with the renormalized propagator, {\em the whole flow must be computed with the fixed dressed \fs}, because the flow equations do depend on the \fs in the high-energy regime (see Eq.~(\ref{eq:flot_phi})). As a consequence it is necessary to first compute the dressed \fs, and then use it to compute the flow of the $Z_I$'s. These flows are represented on Fig.~\ref{fig:Z32}. We furthermore show the variation of the $Z_I$'s along the Fermi line at four different RG times on Fig.~\ref{fig:Z32_t}. We note that the dispersion in the $Z_I$'s is small, in qualitative agreement with the dynamical mean field theory results obtained by Biermann et al,\cite{Biermann01} and this provides a consistency check of Bourbonnais' computation (see Sec.~\ref{sec:sub:sub:comparison}). Fig.~\ref{fig:Z32_t} shows that the evolution of the quasiparticle weights with typical energy scale exhibits some similarity with the results of Kishine and Yonemitsu:\cite{Kishine99} in the early stages of the flow, the quasiparticle weight is larger for $k_y=\pm\pi$ ($I=0$ or $N$), than for $k_y=\pi/2$ ($I=N/2$), and this ordering is reversed at later stages. However, we stress that the two models are different since Kishine and Yonemitsu have considered a 2D model with flat \fs segments, and it is not obvious that the end points of these segments should exhibit the same properties as the extremal points $k_y=\pm\pi$ in our quasi 1D model. The variation of the quasiparticle weight along the \fs has also been investigated by D. Zanchi\cite{Zanchi01} for the 2D Hubbard model, where he found a much stronger reduction of the $Z$ factor in the vicinity of the Van-Hove singularities than for typical \fs points. We believe this effect requires to take into account the variation of the Fermi velocity along the \fs which we have not done here. The influence of these variations on the flow of couplings for a $N$-leg Hubbard ladder has been recently studied\cite{Ledermann00,Ledermann01} (for $\tperp<t$), with the conclusion that they play a dramatic role only below a cross-over scale which is extremely small as $N$ becomes large.

The second regime ($15\lesssim s\lesssim 60$) is a transient between the 1D high-energy flow, and the low-energy regime, where the shape of the \fs is felt, and where the differentiation between the couplings takes place. In more physical terms, it corresponds to a Fermi liquid regime, located between a Luttinger liquid state at higher energies, and an ordered phase at lower energies. One might have expected that in this Fermi liquid regime, the $Z_I$'s would remain constant, so that if this Fermi liquid regime was the final one, the quasiparticle residue would be finite. Instead of this, we see on Fig.~\ref{fig:Z32}, that the time derivative of the $Z_I$'s decreases (in absolute value) but does not vanish. 
\begin{figure}[t]
  \includegraphics[width=8cm]{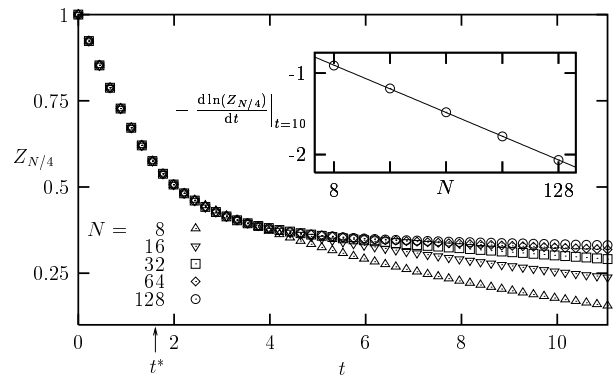}
  \caption{Flow of $Z_{N/4}$ as a function of the ``true'' RG time $t$, for different values of $N$, and for $\tperp/\tpar=0.1$. These flows were obtained assuming that all coupling remain constant to their bare values ($\gc=-\gs=0.3$). Note that we chose $I=N/4$ because $Z_{N/4}$ is about the mean value of the $Z_I$'s. The inserted graph is a base 10 log-log representation of the value of the time derivative of $-\ln(Z_{N/4})$ at time $t=10$ as a function of $N$. The solid curve is the numerical fit, which shows a $1/N$ behavior.}
  \label{fig:verif_FL_Z}
\end{figure}
This is in fact a finite $N$ effect, as can be seen on the flows of the $Z_I$'s one obtains (see Fig.~\ref{fig:verif_FL_Z}), assuming that the couplings are constant, equal to their bare value, all along the flow. The inserted graph shows that the growth rate of the logarithm of the $Z_I$'s behaves like $1/N$ in the Fermi liquid regime. This is indeed consistent with the flow equations (\ref{eq:flot_phi}) in the low-energy regime, where only the terms with a vanishing $\Delta k_\alpha(I,J)$ contribute, and whose number is of order $N$. This factor $N$ combined with the $1/N^2$ denominator explains the numerical result. The relevance of these considerations with constant couplings is demonstrated by Figs.~\ref{fig:norme} and \ref{fig:temps}. Indeed they show that the norm is almost constant for the interval $t\in[0,5]$ in which we are interested (see Fig.~\ref{fig:Z32}). (Note that Figs.~\ref{fig:norme} and \ref{fig:temps} were obtained for $N=8$ not for $N=32$, but we have checked that on this interval $t\in[0,5]$, the flows of the two norms are identical, apart from a multiplicative factor of $8=(32/8)^{3/2}$, whose origin is the number $N^3$ of couplings for a given $N$, and which is irrelevant for our discussion).

The final regime is one of a fixed direction, for which some normalized couplings are zero, and the others are gathered around specific values. It is in this final phase of the flow that the norm of the couplings explodes, as can be seen on Fig.~\ref{fig:norme}. In order to be complete, we have also represented the link between the two RG times $s$ and $t$ on Fig.~\ref{fig:temps}. Notice that because of the definition of $s$ (see the comment after Eq.~(10) of our previous paper), and because the norm explodes in the end of the flow, the time $t$ ``saturates'', as a function of $s$, to a value which is roughly the critical temperature at which the final phase sets in.

\begin{figure}[t]
  \includegraphics[width=8cm]{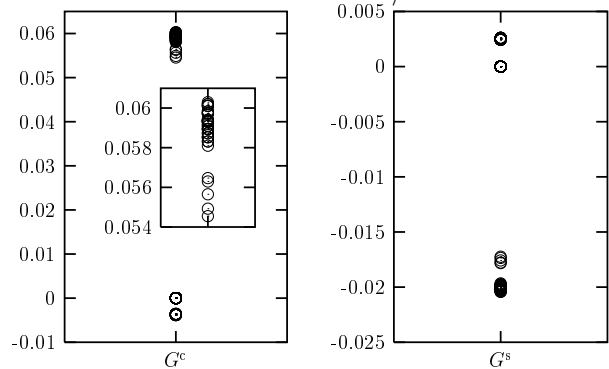}
  \caption{Values of the charge and spin couplings, for the fixed direction that is finally reached on Fig.~\ref{fig:flot_gcs}, but for $N=16$ here. The inserted graph is a zoom on the values taken by the positive charge couplings, and shows that these can be grouped into two sets.}
  \label{fig:analyse_df_finale}
\end{figure}

Let us now study more precisely the final fixed direction, in the spirit of our previous paper. First of all, let us have a more precise look at the values of
the couplings, on the fixed direction that is reached. These values are shown on Fig.~\ref{fig:analyse_df_finale}, for the $N=16$ case. 
We did not choose $N=8$ as on Fig.~\ref{fig:flot_gcs}, because we wanted to have more values (which was manageable here since we represent the whole set of values only once). 

When briefly looking at these values, one can deduce that the couplings seem to be grouped into a few sets of similar values (with lots of couplings being equal to zero). Furthermore, forgetting about the zero value, it seems the three values (for charge or spin couplings) are not independent, but one is the sum of the other two. Finally, the values of the charge couplings seem to differ by a factor of three (and a minus sign) from the spin couplings. In fact, if we also look at Fig.~\ref{fig:flot_gcs}, we see that this will probably not be an exact statement for all values of $N$, but only in the limit of infinite $N$. 

It is then interesting to study what types of couplings take non-zero values. For the system we study, the notation $G_\delta(I,J)$ which is best adapted to superconductivity, can favorably be changed for $\Fcs_\delta(I,J)=\gcs_{J-I-\delta}(I,J)$. $\delta$ is then the transferred transverse momentum, between the R particle that is destroyed, and the L particle that is created. In this notation, only $F$ couplings with $\delta=J-I$ or with $\delta=N/2$ are numerically found to have non-zero values. Notice that here, as $N$ is even, $N/2$ is an integer (we will discuss the odd $N$ case a bit further). The couplings for which $\delta=J-I$ and $J-I\neq N/2$ will be denoted as $\mccs$ couplings, and correspond to the charge (respectively spin) couplings that are negative (respectively positive) on the fixed direction. They are the usual forward scattering couplings. The couplings that satisfy $\delta=N/2$ and $J-I\neq N/2$ will be denoted as $\mdcs$, whereas the ones for which $\delta=N/2$ and $J-I=N/2$ will be denoted as $\macs$ couplings. Both have a transferred transverse momentum $\delta$ which is half the number of chains (\ie $\pi$ if we use the usual momentum units). The vector linking a point of the \fs on the R side, to the one on the L side, and $N/2$ chains further is a {\em nesting} vector, which explains why these couplings are present in the final low-energy fixed direction.

\begin{figure}[b]
  \includegraphics[width=8cm]{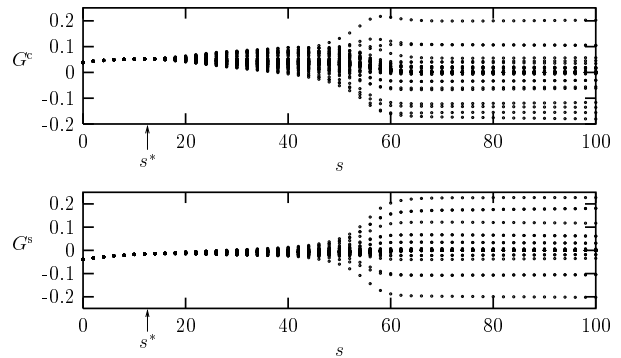}
  \caption{Flow of the charge and spin couplings, for the same values of the parameters as in Fig.~\ref{fig:flot_gcs}, but with $N=7$.}
  \label{fig:flot_Nimpair}
\end{figure}
We let the reader write down the RG equations satisfied by the $F$ couplings, specialize these for the three types of couplings above, and deduce the equations satisfied for the final fixed direction, in the spirit of our previous paper.\cite{Dusuel02} The resulting equations are:
\begin{widetext}
  \begin{equation}
    \label{eq:fd}
    \left\lbrace
      \begin{array}{l}
        N\mac=(N-1)({\mdc}^2+3{\mds}^2)\\
        N\mas=4{\mas}^2 +2(N-1)({\mds}^2+\mds\mdc)\\
        N\mcc=-({\mdc}^2+3{\mds}^2)\\
        N\mcs=4{\mcs}^2 +2({\mds}^2-\mds\mdc)\\
        N\mdc=(N-2)({\mdc}^2+3{\mds}^2)+2(\mac-\mcc)\mdc+6(\mas-\mcs)\mds\\
        N\mds=2(N-2)({\mds}^2+\mds\mdc)+2(\mas-\mcs)\mdc+2(\mac-\mcc)\mds+4(\mas+\mcs)\mds
      \end{array}
    \right..
  \end{equation}
\end{widetext}

This set of coupled equations is nothing but Eq.~(46) of our previous paper, with usual letters replaced by calligraphic letters. The condition $A^{\rm c,s}=C^{\rm c,s}+D^{\rm c,s}$ was satisfied, and ensured the SU$(N)$ symmetry of the interaction Hamiltonian. Here we will thus also be able to fulfill the relation $\macs=\mccs+\mdcs$, which was previously guessed when looking at the fixed direction obtained numerically. The values of the couplings can be found in Table II of our previous paper. It is clear that in the present situation, it is the so called $(+,-)$ fixed direction that is selected since in the infinite $N$ limit, it is the one for which the charge couplings equal minus three times the spin couplings.

The effective low-energy interaction Hamiltonian has the following schematic form (we drop the charge and spin structure):
\begin{eqnarray}
  H_\mathrm{int}^\mathrm{eff} &\sim &\mathcal{C}\sum_{q} \Big[ \sum_{I,k} \cdag_{\rmr,I} (k+q)\anc_{\rmr,I} (k) \Big]\nonumber\\
  \label{eq:Hint_schematic}
  &&\hspace{1cm}\times\Big[ \sum_{J,k'} \cdag_{\rml,J} (k'-q)\anc_{\rml,J} (k') \Big]\\
  &&- \mathcal{D} \sum_{q,k,k'} \Big[ \sum_{J} \cdag_{\rmr,J-N/2} (k+q)\anc_{\rml,J} (k') \Big]\nonumber\\
  &&\hspace{1cm}\times\Big[ \sum_{I} \cdag_{\rml,I+N/2} (k'-q)\anc_{\rmr,I} (k) \Big]\;.\nonumber
\end{eqnarray}
Let us describe the physics associated with such an effective Hamiltonian. We will assume that $N$ is large, so that we can neglect all finite $N$ corrections. Thus, for example, only the $\mathcal{D}$ terms (Peierls couplings) survive, as the forward couplings $\mathcal{C}$ are a correction of order $1/N$. Furthermore, in the infinite $N$ limit, the couplings take the values $\mdc=3/4$ and $\mds=-1/4$, so that $\mdc=-3\mds$. This relation implies that the interaction exists only in the triplet channel, and the effective Hamiltonian can be written as ($g>0$):
\begin{eqnarray}
  \label{eq:Hint_eff_esp_reel}
  &&H_\mathrm{int}^\mathrm{eff} = -\frac{g}{N}\int_0^L \rmd x \nonumber\\
  &&\hspace{1cm}: \left[ \sum_J \pdag_{\rmr,J-N/2,\rho}(x) \boldsymbol{\sigma}_{\rho,\rho'} \anp_{\rml,J,\rho'}(x) \right]\\
  &&\hspace{1.7cm}\times \left[ \sum_I \pdag_{\rml,I+N/2,\tau}(x) \boldsymbol{\sigma}_{\tau,\tau'}\anp_{\rmr,I,\tau'}(x) \right]:.\nonumber
\end{eqnarray}

The only difference (except for changes of notation) between this effective Hamiltonian and the one we arrived at in Eq.~(70) of our previous paper, is in the shift of the creation operators' chain number by an amount of $N/2$. We thus expect the physics to be essentially the same as we had discussed in our previous paper, apart from a different SDW's wave vector that will now be $(2\overline{k},\pi)$ (remember $\overline{k}$ is the average Fermi momentum). We refer the reader to our previous paper for details.

We thus have shown that after a high-energy 1D regime where the \fs's shape is not felt in the flow of the couplings, and after the crossing of the typical energy-scale given by the curvature of the \fs, the system goes to a strong coupling phase of the SDW type, with the above effective low-energy Hamiltonian. The $(2\overline{k},\pi)$ nesting vector naturally arises from the RG flow, and there is no need to artificially introduce it.

Let us make a final remark, about the odd $N$ case. The flow of the charge and spin couplings for $N=7$ is shown on Fig.~\ref{fig:flot_Nimpair}. 
This figure is obviously different from Fig.~\ref{fig:flot_gcs}. The reason for this is that there is no {\em exact} nesting vector anymore when $N$ is odd. We have analysed which couplings are non-zero in the low-energy phase of Fig.~\ref{fig:flot_Nimpair}, and these turn out to be BCS type couplings, indicating a superconducting low-energy phase. This is not in contradiction with what has been said before in the even $N$ case. When $N$ grows, the nesting is better and better in the odd $N$ case, so that the RG flow will first be towards the same fixed direction as in the even $N$ case. Then, there will be a shift from this fixed direction to another one, corresponding to superconductivity. But, this will take place at very low energies, and in regimes where the norm of the couplings has exploded. The conclusion is that the low-energy phase, in the thermodynamical limit, is always the one we have observed in the even $N$ case. This discussion has been quite brief, but we refer the reader to our previous paper where we had analysed in detail how the observed shift from one fixed direction to another one, in a finite $N$ situation, slows down as $N$ increases and finally disappears in the infinite $N$ limit. We have checked all this numerically, but unfortunately it requires quite a large value of $N$ (more than 30) to be visible, so that we could not depict it in this paper.


\subsection{Taking account of the Umklapps and limitations of the method}

\label{sec:sub:umklapps_limitations}

In the above two simple examples we have studied, we encountered no real limitation of the computation scheme we have proposed. Of course, we had to use a non rigorous (but plausible) argument to define the scale at which we had to stop the flow of the couplings. We will see that there are some cases where it is not possible to use such a simple point of view. The well known main limitation of a RG approach is its perturbative nature. We cannot fully trust the RG flows when they go to strong couplings, even if the fixed direction that is reached in this regime gives an insight of what physics takes place. For the two computations of the \fs we have given previously, it is clear that this was no limitation, because the deformation of the \fs occurred in a weak coupling regime (remember Figs.~\ref{fig:flot_gcs} and \ref{fig:norme}, where the norm diminishes between $0\leqslant s \lesssim 20$, which is the time interval where the deformation of the \fs takes place).

The half-filled system is interesting, because it exhibits a variety of behaviors, depending on the strength of the bare couplings (compared to the value of the bare transverse hopping). We will discuss the strong, intermediate and weak coupling situations, which do not give the same low-energy physics. After this, we will consider the case of a nearly half-filled system.


\subsubsection{Half-filled system in strong coupling}

\begin{figure}[t]
  \includegraphics[width=8cm]{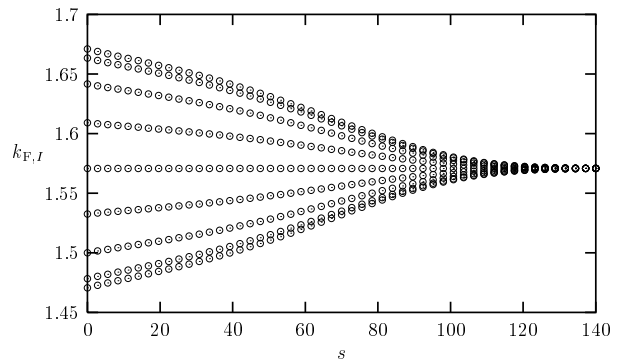}
  \caption{Flow of the Fermi momenta, for $N=16$ chains, $\tperp/\tpar=0.1$ and Hubbard initial condition $\gc=0.2=-\gs$ and $U=0.4$.}
  \label{fig:defSF_um_coup_fort}
\end{figure}

By strong coupling, we mean the initial couplings are large enough for the behavior of the system to remain purely one-dimensional. To make this statement more precise, let us study the flows for one of these strong coupling initial conditions. We will assume, as we always did, that $\tperp/\tpar=0.1$. In this case, the Hubbard-condition $\gc=0.2=-\gs$ and $U=0.4$ is a strong coupling condition, for which the RG leads to a fully flat \fs. In fact, as the couplings are large and grow quickly (because of the Umklapps), the \fs flattens quickly, so that the decreasing cut-off never catches the scale of the \fs, and there is no non-zero value of $\Lambda^*$. The flow of the couplings is thus purely 1D all along the flow and is well known, so that we do not show it. However, to be concrete, we show the evolution of the Fermi momenta on Fig.~\ref{fig:defSF_um_coup_fort}. 
About the norm, let us say that its value at the beginning of the flow is about 30, and at $s=140$, it is about 1400, so that it is around 50 times bigger. This means the couplings have grown a lot. For example for the Umklapps, $U\simeq 19$ which is very big, so that the RG is not valid anymore. However, if we believe the RG is qualitatively valid, the flow of the Fermi momenta seems to show the existence of a confined phase (the effective $\tperp$ is zero). As the system goes to strong coupling, and remains purely 1D, it would thus be natural to directly start from a system of decoupled (no hopping) Luttinger liquids or 1D Mott insulators. We shall simply direct the reader to some papers on this line of approach.\cite{Boies95,Arrigoni00,Essler02}


\subsubsection{Half-filled system in weak coupling}
\label{sec:sub:sub:hfwc}

\begin{figure}[t]
  \includegraphics[width=8cm]{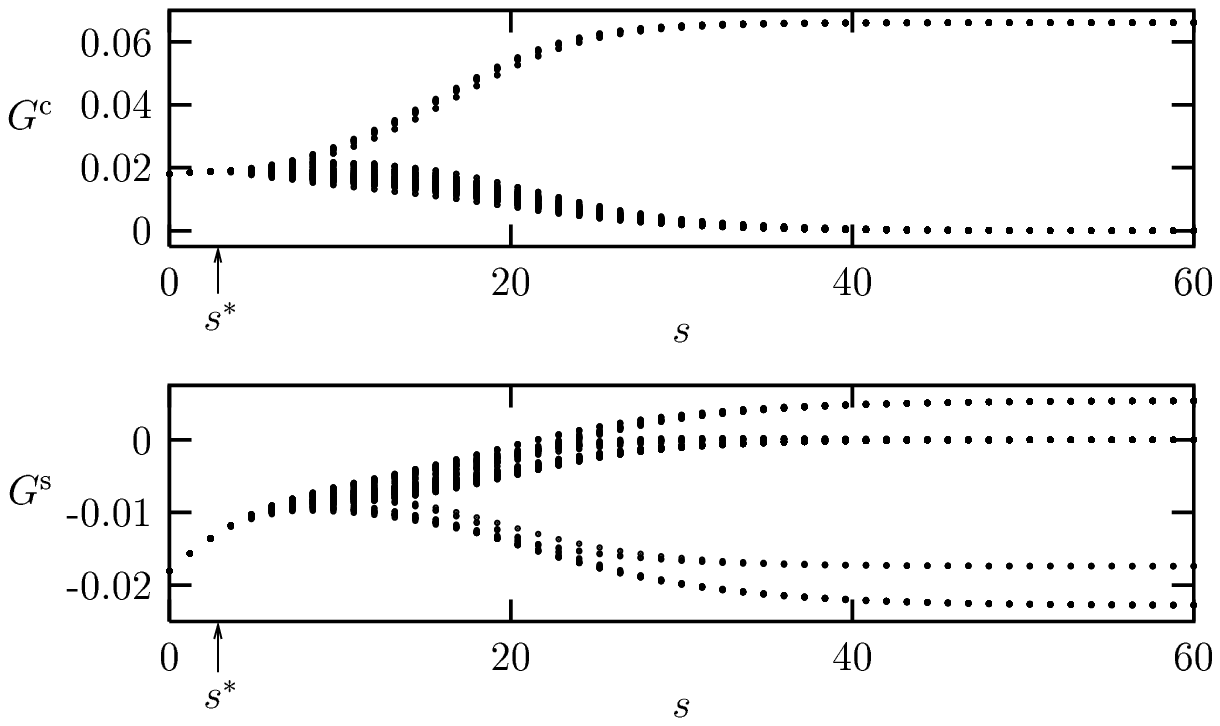}
  \includegraphics[width=8cm]{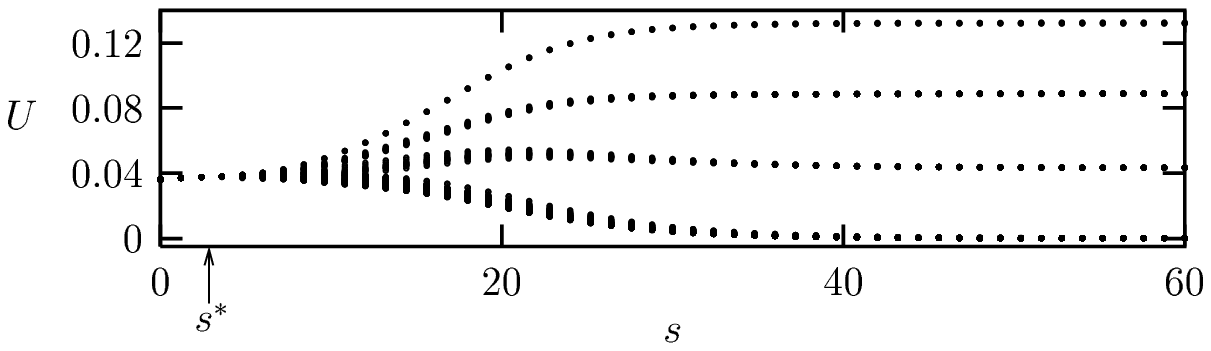}
  \caption{Flow of the three types of couplings, for $N=8$ chains, $\tperp/\tpar=0.1$, and the initial condition $\gc=0.03=-\gs$ and $U=0.06$.}
  \label{fig:flots_gcsu}
\end{figure}
\begin{figure}[t]
  \includegraphics[width=8cm]{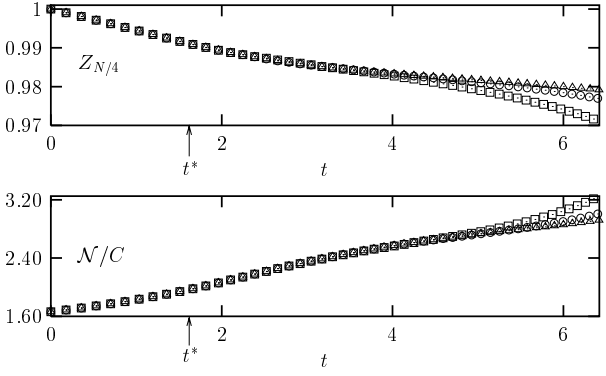}
  \caption{The top figure represents the evolution of $Z_{N/4}$ as a function of $t$ for $N$=8 (squares), 16 (circles) and 32 (triangles). The bottom figure shows how the norm varies with time $t$. In fact, in order to allow the comparison, we have divided the norm for $N=16$ and $32$ by a constant $C$ so as to make all the norms equal at time $t=0$. The squares thus represent the norm $\mathcal{N}$ for $N=8$, the circles represent $\mathcal{N}/C$ ($N=16$) with $C=(16/8)^{3/2}=2\sqrt{2}$ and the triangles represent $\mathcal{N}/C$ ($N=32$) with $C=(32/8)^{3/2}=8$.}
  \label{fig:comp_Z_um_N}
\end{figure}

A weak coupling initial condition is one for which the \fs does not get completely flat during the RG flow (\ie a non-zero value of $\Lambda^*$ exists), and for which all couplings that are irrelevant (\ie that do not exist at zero energy) do go to zero (after dividing by the norm) during the flow. As a consequence, the cross-over scale $\Lambda^*$ between the Luttinger and the Fermi liquid behavior is much larger than the typical scale for the onset of long range order. The system is therefore in a deconfined regime, in the sense that it allows for coherent transverse motion of electron-like excitations. An example of this is obtained while using initial Hubbard couplings $\gc=0.03=-\gs$ and $U=0.06$, and as usual $\tperp/\tpar=0.1$. It is not worth representing the deformation of the \fs in this case, for it is very small (for $N=8$, the effective $\tperp/\tpar$ is about 0.0995, so that the correction is of the order of half a percent). Let us however represent the flow of the couplings, on Fig.~\ref{fig:flots_gcsu}, in the $N=8$ case, and the flow of the quasiparticle weight (and of the norm of the couplings) on Fig.~\ref{fig:comp_Z_um_N}. As in the incommensurate case, the time derivative of the quasiparticle weights becomes smaller (in absolute value) in an intermediate regime, and this effect is more and more visible as $N$ gets bigger. This decrease can be understood from the flow of the norm, which shows a tendency towards a plateau behavior at intermediate scales, so that the arguments previously given in Sec.~\ref{sec:sub:first_num_std} still apply. For the comparison with Fig.~\ref{fig:flots_gcsu}, let us simply say that for $0\leqslant t\leqslant 6$ the link between $s$ and $t$ is approximately linear, and for $t=6$, one has $s=14$.

Before studying the fixed direction, let us make a remark about the scale $\Lambda^*$. Because of the presence of the Umklapps, not only the scales defined by all the $\Delta k_{\rmf,\alpha}(I,J)$ play a role, but also the scales $\Delta k^U_{\rmf,\alpha}(I,J)$. However, as the filling is one-half, the average Fermi momentum is $\pi/2$, and one can check that in this case, the biggest $\Delta k^U_{\rmf,\alpha}(I,J)$ is, as the biggest $\Delta k_{\rmf,\alpha}(I,J)$, equal to twice the difference between the biggest and the smallest Fermi momenta. $\Lambda^*$ will then be defined exactly as we did when the Umklapps were zero.

The couplings on the fixed direction have very well defined values, with many being equal to zero. For the charge and spin couplings, the only non-zero couplings are the same as previously discussed in Sec.~\ref{sec:sub:first_num_std}, namely of the type $\mathcal{A}$, $\mathcal{C}$ and $\mathcal{D}$. The condition $\macs=\mccs+\mdcs$ still seems valid, and one can furthermore check that $\mcc=0$, so that $\mac=\mdc$. For the study of the Umklapps, it is also interesting to introduce $V$ couplings, which are the equivalent of the $F$ couplings: $V_\delta(I,J)=U_{J-I-\delta}(I,J)$. An analysis of the non-zero Umklapps reveals that there are only three types of such couplings: $\mcu=U_{N/2}(I,J\neq I)$, $\mcv=V_{N/2}(I,J\neq I)$ and $\mcw=U_{N/2}(I,I)=V_{N/2}(I,I)$. Furthermore, these Umklapps also seem not to be independent, but linked by the $\mcw=\mcu+\mcv$ relation. We let the reader check that the fixed direction is found by solving the following set of coupled equations (which is just a generalization of Eq.~(\ref{eq:fd})):
\begin{widetext}
  \begin{equation}
    \label{eq:fd_um}
    \left\lbrace
      \begin{array}{l}
        N\mac=(N-1)({\mdc}^2+3{\mds}^2+\mcu^2+\mcv^2-\mcu\mcv)\\
        N\mas=4{\mas}^2 +2(N-1)\left[{\mds}^2+\mds\mdc+\frac{1}{2}(\mcu^2-\mcu\mcv)\right]\\
        N\mcc=-({\mdc}^2+3{\mds}^2+\mcu\mcv-\mcu^2-\mcv^2)\\
        N\mcs=4{\mcs}^2 +2\left[{\mds}^2-\mds\mdc+\frac{1}{2}(\mcv^2-\mcu\mcv)\right]\\
        N\mdc=(N-2)({\mdc}^2+3{\mds}^2+\mcu^2+\mcv^2-\mcu\mcv)+2(\mac-\mcc)\mdc+6(\mas-\mcs)\mds+\mcw(\mcu+\mcv)\\
        N\mds=2(N-2)\left[{\mds}^2+\mds\mdc+\frac{1}{2}(\mcu^2-\mcu\mcv)\right]+2(\mas-\mcs)\mdc+2(\mac-\mcc)\mds+4(\mas+\mcs)\mds+\mcw(\mcu-\mcv)\\
        N\mcu=2(N-2)\left[(\mdc+3\mds)\mcu-2\mds\mcv\right]+2(\mdc+\mds)\mcw+2(\mac+3\mas)\mcu-4\mas\mcv+2(\mcc-\mcs)\mcu\\
        N\mcv=2(N-2)(\mdc-\mds)\mcv+2(\mdc-\mds)\mcw+2(\mac-\mas)\mcv-4\mcs\mcu+2(\mcc+3\mcs)\mcv\\
        N\mcw=2(N-1)\left[ \mdc(\mcu+\mcv)+3\mds(\mcu-\mcv)\right]+4\mac\mcw
      \end{array}
    \right..
  \end{equation}
\end{widetext}
It is easy to solve this set of equations (with the relations between the couplings), order by order in $N$. One finds a few fixed directions, but the one of interest is the following (that we give to order 3, for the independent couplings):
\begin{eqnarray}
  &&\mcs=0+\frac{1}{4N}-\frac{1}{4N^2}+\frac{47}{96N^3}+\mco\left(\frac{1}{N^4}\right),\nonumber\\
  &&\mdc=\frac{3}{8}-\frac{3}{16N}-\frac{1}{16N^2}+\frac{33}{64N^3}+\mco\left(\frac{1}{N^4}\right),\nonumber\\
  \label{eq:coup_dir_fix_av_um}
  &&\mds=-\frac{1}{8}+\frac{1}{16N}+\frac{1}{16N^2}-\frac{11}{64N^3}+\mco\left(\frac{1}{N^4}\right),\hspace{0.9cm}\\
  &&\mcu=\frac{1}{4}-\frac{1}{8N}-\frac{7}{24N^2}+\frac{89}{96N^3}+\mco\left(\frac{1}{N^4}\right),\nonumber\\
  &&\mcv=\frac{1}{2}-\frac{1}{4N}+\frac{1}{6N^2}+\frac{13}{24N^3}+\mco\left(\frac{1}{N^4}\right).\nonumber
\end{eqnarray}
We let the reader check that even the order 0 reproduces quite accurately the values of the couplings of Fig.~\ref{fig:flots_gcsu} (of course, up to an overall normalization factor). Here again, the effective Hamiltonian contains forward interactions that are $1/N$ corrections, and interactions for which the transferred momentum is the nesting vector. The $\mathcal{D}$ couplings satisfy the relation $\mdc+3\mds=0$ (in the infinite $N$ limit), so that the part of the effective low-energy Hamiltonian containing the $\mathcal{D}$ couplings is the same (apart from a numerical factor) as the one we previously obtained (see Sec.~\ref{sec:sub:first_num_std}). It is non-zero only in the triplet channel (we again consider the particle-hole parametrization of the couplings). Let us see what form the effective Umklapp Hamiltonian takes in this singlet and triplet parametrization. From Eq.~(\ref{eq:coup_dir_fix_av_um}), we see that we have a relation between $\mcu$ and $\mcv$ which reads $\mcv=2\mcu$ (this is valid up to $\mathcal{O}(1/N^2)$ terms). The corresponding interaction involving chains $I$ and $J$ on one side of the \fs, and $I+\delta$ and $J-\delta$ on the other side, may typically be written as (using Pauli's principle): 
\begin{eqnarray}
  &&\frac{\mcu}{2NL}\sum_{I,J}~ \sum_{\tau,\tau'}~ \sum_{\rho,\rho'} \left(\mathbb{I}_{\tau,\tau'} \mathbb{I}_{\rho,\rho'}-2\mathbb{I}_{\tau,\rho'} \mathbb{I}_{\rho,\tau'}\right)\nonumber\\
  &&\times\left(\cdag_{\rmr,I+N/2,\tau} \cdag_{\rmr,J-N/2,\rho} \anc_{\rml,J,\rho'} \anc_{\rml,I,\tau'}+\mbox{ h. c.}\right).\hspace{0.5cm}
\end{eqnarray}
As $\mathbb{I}_{\tau,\tau'} \mathbb{I}_{\rho,\rho'}-2\mathbb{I}_{\tau,\rho'} \mathbb{I}_{\rho,\tau'}=-\boldsymbol{\sigma}_{\tau,\tau'}\cdot \boldsymbol{\sigma}_{\rho,\rho'}$, it is easy to rewrite the Umklapp interaction in the triplet channel only. If we define the generalized current
\begin{equation}
  \boldsymbol{J}_{\rmr\rml}(x)=\sum_I \pdag_{\rmr,I+N/2,\tau}(x) \boldsymbol{\sigma}_{\tau,\tau'}\anp_{\rml,I,\tau'}(x)=\boldsymbol{J}^\dagger_{\rml\rmr}(x),
\end{equation}
the total effective Hamiltonian takes the simple following form ($g>0$):
\begin{equation}
  \label{eq:Hint_eff_esp_reel_U}
  H_\mathrm{int}^\mathrm{eff} = -\frac{g}{N}\int_0^L \rmd x : \left[\boldsymbol{J}_{\rmr\rml}(x)+\boldsymbol{J}_{\rml\rmr}(x)\right]^2:.
\end{equation}
The low-energy physics can again be described by the fluctuations of the massless modes associated to the order parameter (which is $\boldsymbol{\xi}(x)=\langle\boldsymbol{J}_{\rmr\rml}(x)\rangle$). The difference with the non half-filled case studied in Sec.~\ref{sec:sub:first_num_std}, is that the spin-density wave will be pinned to the lattice by the Umklapps. That this is indeed what happens can be seen by computing the effective action of the gapless modes, and one finds (we drop less relevant terms):
\begin{equation}
  \label{eq:eff_action}
  S_\mathrm{eff}(\boldsymbol{n})=\frac{N}{4\pi}\int {\rm d}x{\rm d}t \,\partial_\mu \boldsymbol{n} \partial^\mu \boldsymbol{n}.
\end{equation}
with $\boldsymbol{\xi}(x)=\rho \boldsymbol{n}(x)$, $\rho$ being a positive number found by solving mean-field equations, and $\boldsymbol{n}(x)$ is a real unit vector, giving the direction of the staggered magnetization. This time, there is no gapless mode associated to the ``phason'' field (see the discussion around Eqs.~(70) and (71) of our previous paper). This is physical, for a shift $\theta(x)\to\theta(x)+\Theta$ in the ``phason'' field corresponds roughly to a uniform translation of the spin-density wave condensate. As the physics of Eq.~(\ref{eq:eff_action}) has already been discussed in our previous paper, we do not consider it further.

\begin{figure}[t]
  \includegraphics[width=8cm]{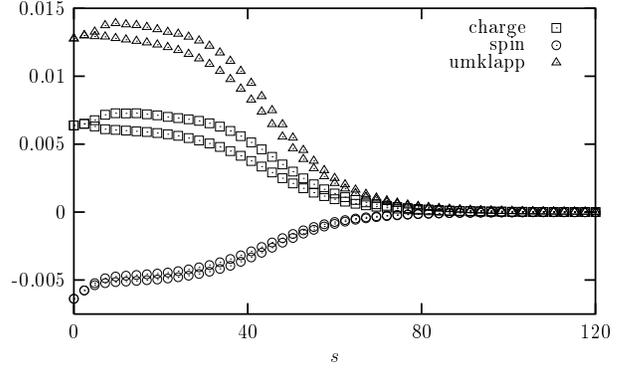}
  \caption{Evolution of the smallest and largest charge (squares), spin (circles) and Umklapp (triangles) irrelevant couplings, for $N=16$, $\tperp/\tpar=0.1$ and initial condition $U=0.02$.}
  \label{fig:irr}
\end{figure}
\begin{figure}[t]
  \includegraphics[width=8cm]{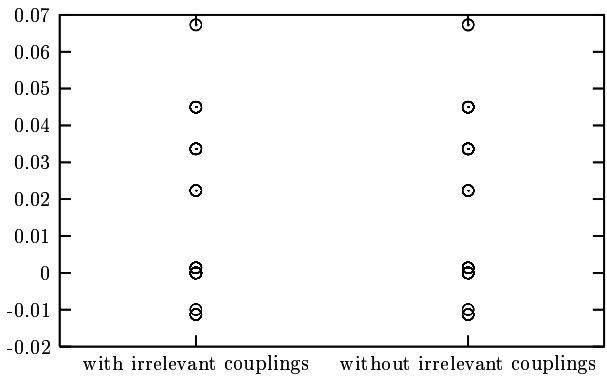}
  \caption{Values of all the couplings on the final fixed direction, for $N=16$, $\tperp/\tpar=0.1$ and weak initial Hubbard coupling $U=0.02$. On the left (respectively right), we represented the values obtained when computing the whole flow (respectively initially setting the irrelevant couplings to zero).}
  \label{fig:comp_moi_lbf}
\end{figure}

Let us now study more precisely the fate of the irrelevant couplings. We defined the initial coupling as a weak coupling if the normalized irrelevant couplings go to zero during the flow. In fact it is interesting to check that the final fixed direction that is reached is the same whether we run the complete flow, or we run the flow in which the irrelevant couplings are initially set to zero. We have done this in a system of $N=16$ chains, with $\tperp/\tpar=0.1$ and an initial Hubbard coupling $U=0.02$. The evolution of the irrelevant couplings is shown on Fig.~\ref{fig:irr}. 
In fact we have not represented all the couplings, because there are too many of them. We have decided to show only the smallest and largest charge, spin and Umklapp couplings. In order to make sure that the final fixed direction is the same as the one we would have obtained when initially setting irrelevant couplings to zero, we show the different values of the couplings on this fixed direction, in both cases, on Fig.~\ref{fig:comp_moi_lbf}.
\begin{figure}[b]
  \includegraphics[width=8cm]{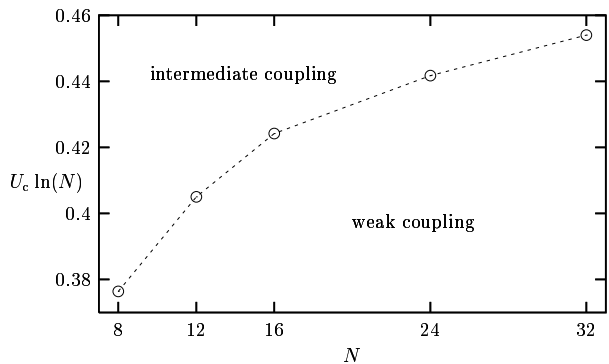}
  \caption{Representation of the numerical values of $U_\rmc \ln(N)$ as functions of $N$, where $U_\rmc$ is the value above which the irrelevant couplings do not flow to zero.}
  \label{fig:verif_critere_wc}
\end{figure}

A natural question that arises from this discussion is how small should the couplings be for being weak couplings according to the definition given above ? This question has already been answered by Lin, Balents and Fisher\cite{Lin97}. They have done so on a theoretical ground (and for a situation that is not the half-filled system, but this should not change anything), and found that the weak coupling condition reads $U \ln(N)\ll 1$ (for large $N$). Thanks to our ability to take the irrelevant couplings into account, we have tried to check this numerically. To do so, we have determined the critical coupling $U_\rmc$ for which the irrelevant couplings do not flow to zero anymore, for $N=$8, 12, 16, 24 and 32, and represented the values $U_\rmc \ln(N)$ as functions of $N$. The result is shown on Fig.~\ref{fig:verif_critere_wc}.
Because of the small values of $N$ we have used, we do not observe an horizontal line as could have been inferred from the $U \ln(N)\ll 1$ criterion. But this criterion is in fact a sufficient condition (maybe not a necessary one) to observe a weak coupling behavior, since it implies that the effective Hamiltonian hardly changes during the high-energy part of the RG flow, for scales above $\Lambda^*$.


\subsubsection{Half-filled system in intermediate coupling}

\begin{figure}[t]
  \includegraphics[width=8cm]{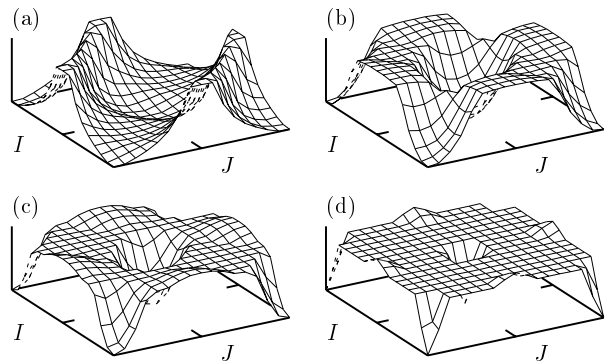}
    \caption{Evolution of the (normalized) charge forward scattering couplings $\gc_0(I,J)$, when the strength of the initial Hubbard coupling $U$ is increased ($U=0.12$ for (a), $U=0.17$ for (b), $U=0.19$ for (c) and $U=0.202$ for (d)). The number of chains is $N=16$, and $\tperp/\tpar=0.1$. The flow has been stopped at the RG time for which the biggest of the whole set of couplings is equal to 1. The charge forward scattering pictured here are the ones obtained at this time. In order to keep the figures clear, we have not put any indication on the $z$ axis. Let us simply say that the values of the couplings range from $3\cdot10^{-3}$ to $1.3\cdot10^{-2}$ in (a), and from $6.75\cdot10^{-3}$ to $7.05\cdot10^{-3}$ in (d), so that (d) is in reality much flatter than (a).}
  \label{fig:evolution_coupC_charge}
\end{figure}
When the couplings are neither strong nor weak, that is intermediate, we suspect the system will behave more and more like a 1D system, as the initial Hubbard coupling $U$ grows. Before we check that this is the case, let us clarify this notion of intermediate coupling. We have just seen at the end of the previous section (\ref{sec:sub:sub:hfwc}), that the intermediate coupling should typically be characterized by $U\simeq U_\rmc$ (remember Fig.~\ref{fig:verif_critere_wc}). In the case $N=16$ and $\tperp/\tpar=0.1$ on which we shall focus, this means $U\simeq 0.15$. We should also have $U<0.206$, value above which the system is in the confined phase. If we expect the system's behavior to change and become nearly one-dimensional for these typical values of $U$, this should mean that the effective hopping is of the same order of magnitude as the critical temperature. This will be discussed in Sec.~\ref{sec:sub:sub:phase_diagram}, when we study the phase diagram of the system. Let us simply say here that for the minimum (respectively maximum) value of the coupling $U$ we will consider, namely $U=0.12$ (respectively $U=0.202$), the effective hopping $\tperp^\mathrm{eff}=$ is about 8 times (respectively 0.9 times) the critical temperature. These values confirm the previous expectation.

\begin{figure}[t]
  \includegraphics[width=8cm]{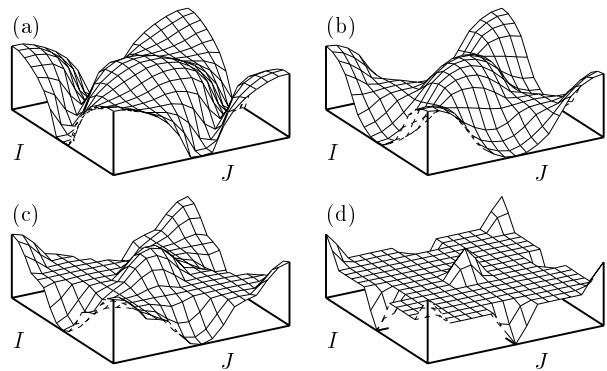}
    \caption{Same as Fig.~\ref{fig:evolution_coupC_charge}, for the spin forward scattering couplings $\gs_0(I,J)$. Here, the values of the couplings range from $-1.8\cdot10^{-3}$ to $-2\cdot10^{-4}$ in (a), and from $7.95\cdot10^{-4}$ to $7.7\cdot10^{-4}$ in (d).}
  \label{fig:evolution_coupC_spin}
\end{figure}
\begin{figure}[b]
  \includegraphics[width=8cm]{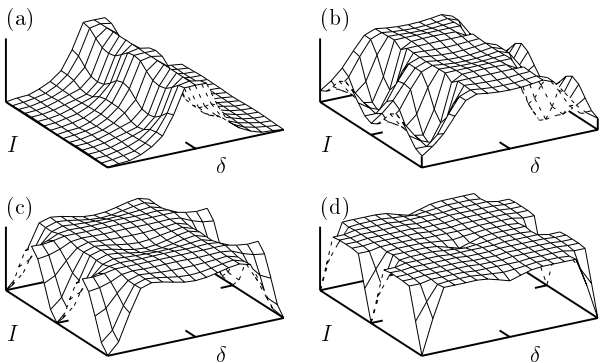}
    \caption{Same as Fig.~\ref{fig:evolution_coupC_charge}, for the Umklapp couplings $U_\delta(I,I)$. Here, the values of the couplings range from $8\cdot10^{-3}$ to $2.4\cdot10^{-2}$ in (a), and from $1.3954\cdot10^{-2}$ to $1.3961\cdot10^{-2}$ in (d).}
  \label{fig:evolution_coupC_umklapp}
\end{figure}
We have numerically studied the evolution of some special couplings, as $U$ becomes bigger. 
It is not possible to consider the couplings on the final fixed direction. Indeed, the norm is huge even before it is reached. We had neglected this problem in the strong and weak coupling regimes. In the first case, we anyway knew that the RG is not valid anymore and should be replaced by a non-perturbative analysis. In the second case, we can make the norm as small as we want when reducing the initial coupling, because in this case, the RG flow is scale invariant (the RHS of the RG equations quickly become $\nu$ independent, as $\nu$ rapidly goes to very small values). In this last case we refer the reader to our previous paper\cite{Dusuel02} for more details about the implications of this scale invariance.

We thus have chosen to stop the RG flows at the time when the biggest of all couplings (the true couplings, not the normalized ones) reaches the value 1. In this regime, the RG should be valid (of course, the two-loops contributions are not negligible when the couplings approach 1). This gives us the results shown on Figs.~\ref{fig:evolution_coupC_charge} to \ref{fig:evolution_coupC_umklapp}, which were obtained for $N=16$ and initial couplings $U=$0.12, 0.17, 0.19 and 0.202.

For the charge and spin couplings, we have represented forward scattering couplings $\gcs_0(I,J)$. In the weak coupling regime, we would have obtained $\gc_0(I,J)\sim \delta_{J,I+N/2}$ (remember the fixed direction we found in Sec.~\ref{sec:sub:sub:hfwc}). Here we also obtain peaks around $J=I+N/2$ values, but these peaks progressively disappear when $U$ grows, as is expected because the system looks more and more one-dimensional. We have chosen to represent $U_\delta(I,I)$ couplings, in the case of the Umklapps. The reason for this choice is that the biggest of all Umklapps is (numerically) found in this subset of couplings. Again, the weak coupling would give a peak $U_\delta(I,I)\sim \delta_{\delta,N/2}$, which is smeared in the case of intermediate couplings, and disappears in the strong coupling limit.


\subsubsection{Phase diagram}
\label{sec:sub:sub:phase_diagram}

\begin{figure}[t]
  \includegraphics[width=8cm]{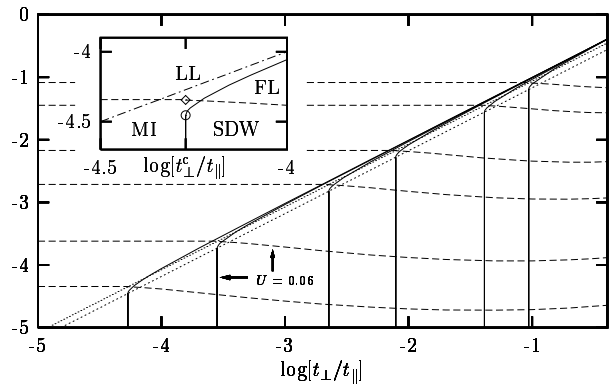}
  \caption{Phase diagram of the system, computed for $N=8$ chains. See text for a detailed description of this diagram.}
  \label{fig:phase_diag}
\end{figure}

As a conclusion of this investigation, we shall summarize the numerical results we obtained on a single figure, which is the phase diagram of the system. It is depicted in Fig.~\ref{fig:phase_diag}. 

The solid curves represent $\tperp^\mathrm{eff}/\tpar$ as a function of $\tperp/\tpar$, for $U$=0.05, 0.06 (indicated by the arrows), 0.08, 0.1, 0.15 and 0.2 (notice that both $x$ and $y$ scales are expressed in base 10 logarithms). For a given $U$, $\tperp^\mathrm{eff}$ is zero in the confined phase, for $\tperp$ smaller than a critical value $\tperp^\rmc$ (this explains the vertical lines), and it takes non-zero values as soon as $\tperp>\tperp^\rmc$. When $\tperp$ is much larger than $\tperp^\rmc$, $\tperp^\mathrm{eff}\simeq\tperp$, so that all solid curves asymptotically go to the first bisector. The dashed curves give the value of the scale at which the couplings diverge, which is the critical temperature $T_\rmc$. They are horizontal when $\tperp<\tperp^\rmc$, since in our approach the RG flows are purely one-dimensional in this regime. The upper (respectively lower) dotted curve is a straight line of slope 1 (numerically found to be 1.004), going through the points of coordinates $(\tperp^\rmc(U),T_\rmc(U,\tperp^\rmc(U))$ (respectively $(\tperp^\rmc(U),\tperp^\mathrm{eff}(\tperp^\rmc(U)^+)$), as the one represented by a diamond (respectively circle) in the inserted figure. This inserted figure is a zoom of the interesting region where both scales meet, for $U=0.05$. We have indicated the different phases (Luttinger Liquid, Fermi Liquid, Mott Insulator and Spin-Density Wave). The dash-dotted curve is the first bisector, that we did not represent in the global figure to keep it readable.

Let us now study the physical implications of the value 1 taken by the slopes of the two dotted lines. The upper one tells us that the critical value of $\tperp$ is proportional to the value of the charge gap $\Delta$ of the Mott insulating phase, and we numerically find $\tperp^\rmc\simeq 1.14 \Delta$. The lower curve gives the following relation between the value of the effective hopping for $\tperp={\tperp^\rmc}^+$, that we will denote by ${\tperp^\mathrm{eff}}^*$, and the critical value of the bare hopping: ${\tperp^\mathrm{eff}}^*\simeq0.68\tperp^\rmc$. Let us remark that this also implies: ${\tperp^\mathrm{eff}}^*\simeq 0.78\Delta$. These relations show that the confinement-deconfinement transition takes place when the bare and the effective hopping become of the order of the charge gap. These results have natural interpretations. Imagine the system is in the Mott insulating phase, with zero effective hopping. It is clear that we can only compare the bare hopping to the charge gap. If on the contrary the system is in the deconfined regime, the low-energy physics is dominated by two scales: the critical temperature and the effective hopping. When decreasing the bare hopping, we expect a phase transition to occur when these two low-energy scales become of the same order of magnitude.
These results are in quantitative agreement with those obtained by Tsuchiizu et al\cite{Tsuchiizu99_2} for a two-chain model, see for instance the insert in their Fig.~1. For values of $\tperp$ which are not too small, they indeed find a proportionality between $\Delta$ and $\tperp$ with a slope compatible with our results. For smaller values of $\tperp$, they obtain a sizeable deviation away from a linear behavior. But this difference with our conclusions comes from their choice of a fixed value for the intra-chain forward scattering, while they let the Umklapp scattering go to zero. In this regime, they observed a significant renormalization of the hopping amplitude, so that the transition is finally given by the balance between the charge gap and the renormalized hopping.


\subsubsection{Nearly half-filled system}

Up to now, we have not encountered any real difficulty when choosing the $\Lambda^*$ scale. Of course, the choice was purely pragmatic, as we simply chose the biggest of the scales $K$ appearing in the RHS of the RG flows. There was no problem in the half-filled case, because the biggest scale was the same for all the nine sorts of $K$ (remember the definitions in Eqs.~(\ref{eq:car_length_1}) to (\ref{eq:car_length_9})). Let us consider the non half-filled case, for which the mean Fermi momentum is $\overline{k}=\pi/2+\delta k$. The filling does not change the biggest $K^\pp$ and $K^\ph$, which is still the difference between the biggest and the smallest Fermi momenta, $\Delta k_\rmf^{\max}=k_\rmf^{\max}-k_\rmf^{\min}$. We let the reader check that the biggest $K^{UU}$ is now $2|\delta k|+\Delta k_\rmf^{\max}$ and the biggest $K^{GUi}$ is $|\delta k|+\Delta k_\rmf^{\max}$. Those last two scales are obviously always bigger than the first one.

What are the consequences of the existence of these three scales? At the formal level there is no real consequence. Indeed, even in the half-filled case, there were a lot of different scales, given by all the possible $K$'s, so that introducing more scales does not make much change. But, practically, we have used the pragmatic point of view that we should stop the cut-off scaling flow when the biggest scale is reached. This relied on the hypothesis that the flow of the \fs nearly stops at that scale. It is clear that if the filling is not too far from one-half, the three scales are not qualitatively different, so that we can afford stopping the flow when the biggest scale is reached. 

When the filling is quite far from one-half, things become more involved. Of course, we could simply forget about the Umklapps and perform the analysis of Sec.~\ref{sec:sub:first_num_std} devoted to the non half-filled system. But we expect that in an intermediate to strong coupling regime, the Umklapps could play a non-negligible role in the high-energy part of the flow, where there are not yet irrelevant. We would thus like to be able to take them into account. Intuitively, we expect that the \fs deformation is caused by both $U$ and $G$ couplings for a cut-off $\Lambda>\Delta k_\rmf^{\max}+(2)|\delta k|$ and by the $G$ couplings only for $\Delta k_\rmf^{\max}<\Lambda<\Delta k_\rmf^{\max}+(2)|\delta k|$. It is however not possible to implement this idea in a simple manner. Indeed, once the scale $\Delta k_\rmf^{\max}+(2)|\delta k|$ is reached, we could drop the Umklapp contribution to the flow of the \fs, but the flow of the Umklapp couplings will not stop and will depend on the shape of the fixed dressed \fs. This flow of the Umklapps will affect the flow of the $G$ couplings and as these latter still deform the \fs, we see that the flow of the Umklapps indirectly affects the flow of the \fs. As the shape of the dressed \fs comes into play {\em before} the flow of the running \fs stops, there is here no simple way to circumvent the inversion problem we have discussed in the introduction.

As a consequence, in what follows, we will restrict ourselves to situations where our pragmatic scheme works, \ie to the nearly half-filled system. We have considered a filling slightly less than 1/2, setting the chemical potential to -0.01. For this value, the difference between the initial values of $\Delta k_\rmf^{\max}$ and $\Delta k_\rmf^{\max}+(2)|\delta k|$ is about 10\%, which is reasonable. As in the half-filled case, we have observed different regimes, when changing the strength of the initial coupling namely weak, intermediate and strong coupling regimes.

\begin{figure}[t]
  \includegraphics[width=8cm]{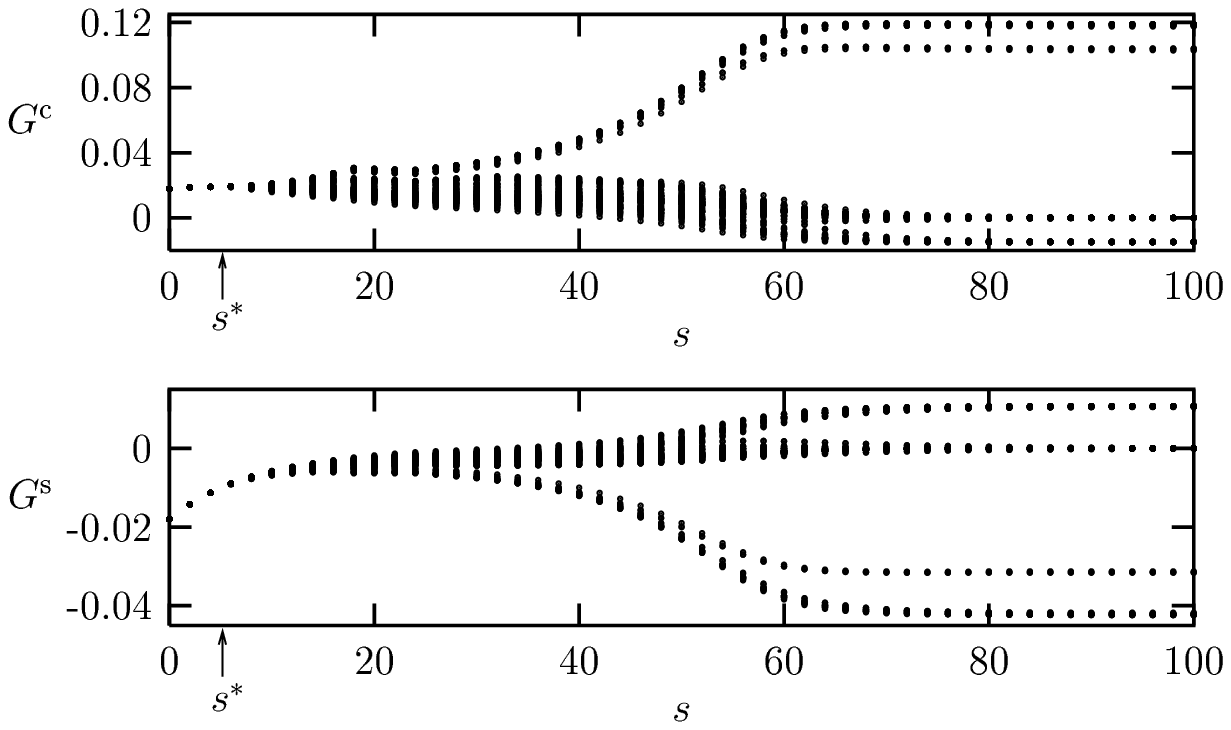}
  \includegraphics[width=8cm]{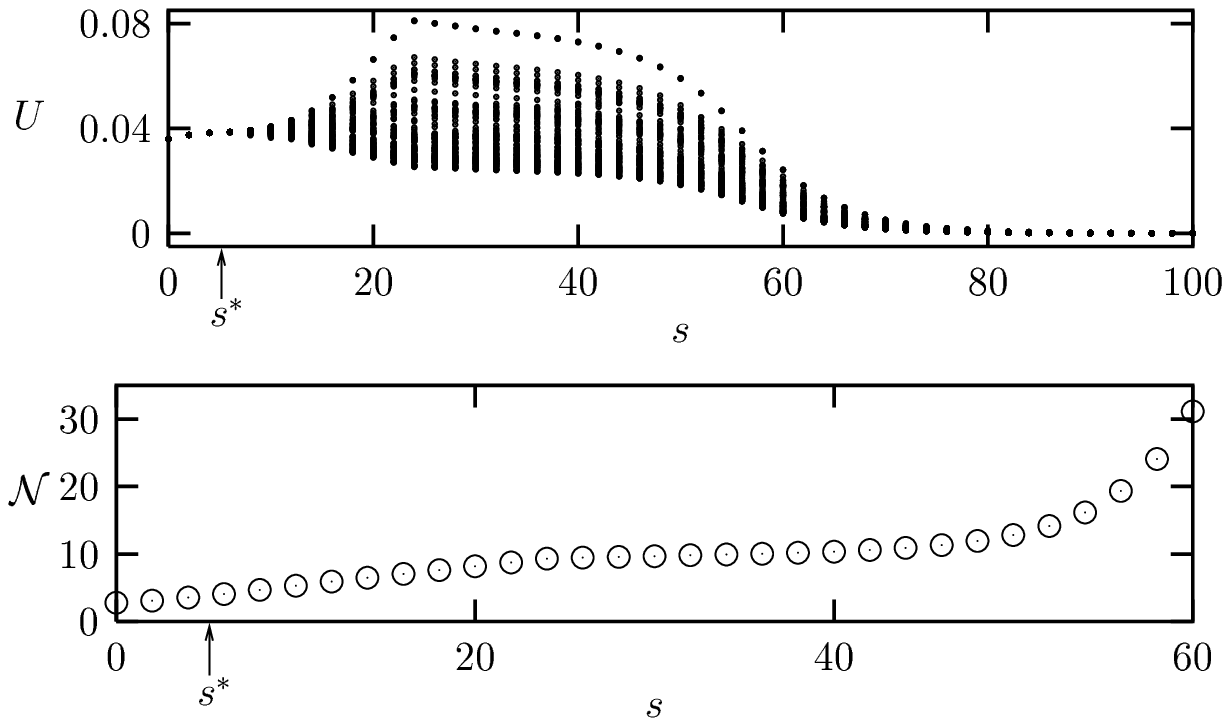}
  \caption{Flow of the three types of couplings, for $N=8$ chains, $\tperp/\tpar=0.1$, $\mu=-0.01$, and the initial condition $\gc=0.05=-\gs$ and $U=0.1$. We also have represented the flow of the norm, for $0\leqslant s\leqslant 60$.}
  \label{fig:flots_gcsu_pdr}
\end{figure}
In the strong coupling regime, as before, the system behaves as a purely 1D system, and the \fs gets completely flat. The intermediate coupling regime is the same as the one observed in the half-filled case, where the Umklapps are not suppressed, because the scale at which the phase transition takes place is bigger than the scale that measures the distance from half-filling, namely $(2)\delta k$. In the weak coupling regime, the Umklapps are irrelevant, and all vanish (for the normalized couplings) at low energies. The final fixed direction is simply the one we previously found in Sec.~\ref{sec:sub:first_num_std}, where we had set the Umklapps to zero at the beginning of the flow. The flows of the couplings are represented on Fig.~\ref{fig:flots_gcsu_pdr}, and were obtained for $N=8$ and initial coupling $U=0.1$. 
We also have represented the evolution of the norm of the couplings, because its behavior changes drastically at the precise time the Umklapps become irrelevant ($s\simeq24$). The plateau observed in the norm's evolution just after this time reveals the existence of an intermediate Fermi liquid phase. 

These flows show how our RG, taking account of the different scales of the system, is able to get rid of the irrelevant couplings, such as the Umklapps, and leads to the correct final fixed direction. This is not of a purely academic interest. Indeed, if one takes the same value of the initial couplings ($\gc=0.05=-\gs$), but set $U$ to zero from the beginning (with a flow very much like the one in Fig.~\ref{fig:flot_gcs}), the scale at which the phase transition occurs is found to be $4\cdot 10^{12}$ times smaller as the one found when the Umklapps are incorporated and vanish along the flow. This can in part be explained because neglecting the Umklapps from the beginning of the flow is a very crude approximation. We could also take the Umklapps into account in the 1D part of the flow, and then take them to zero. However, a look at Fig.~\ref{fig:flots_gcsu_pdr} shows that the Umklapps do not vanish very fast, and their influence has thus no reason to be small. Indeed, we performed this comparison, and found that this prediction is correct. In fact both methods neglecting the Umklapps at one time or another, give approximately the same critical temperature, because the Hubbard coupling $U$ is small so that the couplings do not change much in the 1D region (we found $\gc=0.074$ and $\gs=-0.038$ in the end of the 1D flow).


\section{Conclusion}

We have attempted to develop a simple physical picture to understand the forces which drive the deformations of the \fs of an interacting electron system. Using considerations from second-order time-independent perturbation theory, we showed that the shape of the dressed \fs controls the quantum zero-point motion correction to the ground-state energy. We demonstrated that a given coupling tends to deform the \fs so as to have all its four momenta precisely on the \fs, because this allows for smaller energy denominators (in the second order contribution) and thus decreases the total energy. As a consequence, we find that the \fs is deformed by irrelevant couplings, which are characterized by the impossibility of choosing all four momenta on the \fs. Because of this, the \fs deformation induced by these couplings can only occur in the high-energy regime where the kinematical constraints associated to the \fs do not play much role. Once the energy is low enough and the warping of the \fs is felt, these irrelevant couplings do not flow any more and no longer contribute to the \fs deformation. 

For the quasi 1D materials at half-filling, the Umklapp couplings belong to this category. As they undergo a strong renormalization in the high-energy regime, they have a much more drastic effect than the charge or spin couplings in the doped system. Our numerical simulation provide a description of the cross-over from the confined regime to the Fermi liquid, which is in overall good agreement with previous works.\cite{Prigodin79,Bourbonnais85,Kishine98,Tsuchiizu99_2,Biermann01} We have presented a detailed analysis of the evolution of the quasiparticle weights as a function of the typical energy scale. This confirms the existence of a Fermi liquid regime at intermediate energies, for the deconfined systems. It would be very interesting to compute the longitudinal and transverse optical conductivities, which could be done by adapting some existing methods\cite{Honerkamp01_2} to quasi 1D systems. Another problem is to investigate the nature of spin correlations in the confined regime. This can not be achieved within the present formalism since the couplings diverge at a scale associated to the charge gap which is much larger than the N\'eel temperature in the limit of small transverse hopping. This problem has been addressed by Kishine and Yonemitsu\cite{Kishine98} who used RG equations to two-loop order for the couplings. But it is not clear that the two-loop corrections provide a reliable description since the couplings do not remain small.

This limitation of our method is certainly connected to the fact that we are using a physical picture in which the fermion fields remain the elementary objects. This is valid at sufficiently high energies and thus generically adapted to the study of the \fs deformations. However in the confined regime, elementary excitations are likely to be very different from the Fermi liquid-like quasiparticles, but rather some soliton-like objects. In this case, it seems a deeper understanding of the corresponding phases should be obtained  by expanding around the exact solution for a system of uncoupled chains.\cite{Boies95,Arrigoni00,Essler02} Nevertheless this raises the important issue of the validity of an adiabatic principle for generating the ground-state at finite transverse hopping from the ground-state of uncoupled chains. At half-filling this adiabatical principle certainly holds in the confined phase where the charge gap is finite. But its validity beyond the critical value of the transverse hopping is questionable. This may explain the qualitative discrepancy between the perturbation approach of Essler and Tsvelik,\cite{Essler02} which leads to a disconnected \fs with electron and hole pockets, and the dynamical mean-field theory of Biermann et al, who have obtained a conventional \fs (see Fig.~5 of the latter work\cite{Biermann01}). Away from half-filling, this notion of adiabatic continuity is even less obvious to prove since the energy gap vanishes for uncoupled Luttinger liquids. However we believe the use of a skeleton expansion by Arrigoni\cite{Arrigoni00} likely provides a way to circumvent this potential difficulty.


\section*{ACKNOWLEDGMENTS}

We would like to thank Fern\~{a}o Vistulo de Abreu for providing the initial impetus to this work. We have also benefitted from many valuable remarks by Dra\v{z}en Zanchi and Benedikt Binz. Finally we are grateful to Bertrand Delamotte and Dominique Mouhanna for sharing their experience with various aspects of the renormalization group.


\appendix
\section{Equivalence between minimizing the total energy and stability criteria for the dressed single particle propagator}
\label{app:equi}

\subsection{Two-chain system at first order}
\label{sec:sub:sub:pert_ex}

We wish to check that the \se computed by using a free excited state, with a trial Fermi sea which should be determined only at the end of the calculation, gives the correct results of Sec.~\ref{sec:sub:sub:min_energy}. The sole difference with the calculation of 
Appendix~\ref{sec:sub:sub:std_pert} below is that we shall consider here the following free propagators:
\begin{eqnarray}
  \widetilde{G}^{(0)}_{\rmr,I}(k,\omega)=\hspace{5.8cm}\nonumber\\
  \label{eq:free_propR_tilde}
  \frac{1}{\omega-\left[\mu^{(0)}+v_\rmf^{(0)}(k-k_{\rmf,I}^{(0)})\right]+\rmi\eta\sgn(k-k_{\rmf,I})},\hspace{0.4cm}\\
  \widetilde{G}^{(0)}_{\rml,I}(k,\omega)=\hspace{5.8cm}\nonumber\\
  \label{eq:free_propL_tilde}
  \frac{1}{\omega-\left[\mu^{(0)}-v_\rmf^{(0)}(k+k_{\rmf,I}^{(0)})\right]-\rmi\eta\sgn(k+k_{\rmf,I})}.\hspace{0.4cm}
\end{eqnarray}
These formulas differ with (\ref{eq:free_propR}) and (\ref{eq:free_propL}) below
only in the imaginary parts. We let the reader check (some more details about 
one-loop \se calculations can be found in Appendix~\ref{sec:sub:sub:std_pert} 
devoted to the standard perturbation theory) that the \ses read:
\begin{eqnarray}
  &&\Sigma_{\rmr,0}^{(1)}(k,\omega)=A\frac{\Lambda_0+(k_{\rmf,0}-k_{\rmf,0}^{(0)})}{2\pi}\nonumber\\
  &&\hspace{2.5cm}+C\frac{\Lambda_0-(k_{\rmf,0}-k_{\rmf,0}^{(0)})}{2\pi},\\
  &&\Sigma_{\rmr,\pi}^{(1)}(k,\omega)=B\frac{\Lambda_0-(k_{\rmf,0}-k_{\rmf,0}^{(0)})}{2\pi}\nonumber\\
  &&\hspace{2.5cm}+C\frac{\Lambda_0+(k_{\rmf,0}-k_{\rmf,0}^{(0)})}{2\pi}.
\end{eqnarray}
It is sufficient to compare these results with (\ref{eq:en_1_1part0}) and (\ref{eq:en_1_1partpi}) to understand that we will {\em exactly} find (\ref{eq:dk0_vrai}) and (\ref{eq:dmu_vrai}). The energy minimization method or this self-consistent computation of the \se carried formally to first order in interaction do generate the same higher order terms. These appear since corresponding contributions are sensitive to the shape of the trial \fs. As shown in Appendix \ref{app:diff_re}, this feedback effect is missing in the standard perturbative approach.

\subsection{Formal proof for a finite size system}

Let us choose a trial Fermi surface, with the corresponding occupation
numbers $n(k) \in \{0,1\}$ and $\Phi(k)=1-2n(k)$. This generates a free
particle state (Slater determinant) which may be used as a starting point
for perturbative expansions of the total energy $E(\{n\})$ and the
single particle propagator $G_{\Phi}(k,\omega)$. Suppose we add one
particle to the system, so that the total momentum is increased by
$k_{0}$. This is achieved by using $n'(k)=n(k)+\delta_{k,k_{0}}$ instead 
of $n(k)$, assuming that $n(k_{0})=0$. This induces a change in the total
energy of the system $\Delta E =  E(\{n'\})-E(\{n\})$. For a finite size
system, it is easy to connect this energy shift to the single particle
propagator $G_{\Phi}(k_{0},\omega)$. Indeed, we have the well-known
spectral decomposition:
\begin{eqnarray}
&&G_{\Phi}(k_{0},\omega)=\sum_{\alpha} \frac{\left|\bra{N+1,k_{0},\alpha}\cdag_{k_{0}}\ket{N}\right|^{2}} {\omega - E(N+1,k_{0},\alpha)+E(N)+\rmi\eta}\nonumber\\
&&\hspace{0.5cm}+ \sum_{\beta} \frac{\left|\bra{N-1,-k_{0},\beta}\anc_{k_{0}}\ket{N}\right|^{2}} {\omega + E(N-1,-k_{0},\beta)-E(N)-\rmi\eta}.
\end{eqnarray}
Here, $\ket{N}$ is the eigenstate with $N$ particles obtained perturbatively
from the free-particle state with distribution $n(k)$, and the kets
$\ket{M,k,\alpha}$ denote eigenstates with $M$ particles and a total momentum
$k$ with respect to the total momentum of state $\ket{N}$. The total energy
of these states is of course $E(M,k,\alpha)$. So the energy shift
$\Delta E$ is one of the poles of $G_{\Phi}(k_{0},\omega)$ seen as
a function of $\omega$. Denoting the typical interaction strength by $V$,
it is sensible to assume that for a finite-size system, energy differences
like $E(N+1,k_{0},\alpha)-E(N)$ (and  $\Delta E$ in particular)
can be expanded as power series in $V$. In the non-interacting case,
$\Delta E=\ve_{0}(k_{0})$, and $G^{(0)}_{\Phi}(k_{0},\omega)^{-1}
=\omega- \ve_{0}(k_{0})+\rmi\eta$. Therefore, $\Delta E$ is the pole
of $G_{\Phi}(k_{0},\omega)$ (as a function of $\omega$) which goes smoothly
towards $\ve_{0}(k_{0})$ as $V$ goes to zero.  
Writing $G_{\Phi}(k_{0},\omega)^{-1}=\omega- \ve_{0}(k_{0})
-\Sigma_{\Phi}(k_{0},\omega)$, and given the fact that 
$\Sigma_{\Phi}(k_{0},\omega)$ has a well-defined power series expansion
in $V$ which vanishes as $V$ goes to zero, we may conclude that
$\Delta E$ may be obtained as a formal power series in $V$ from the
solution of the following equation for $\omega$:
\begin{equation}
\omega- \ve_{0}(k_{0})-\Re \Sigma_{\Phi}(k_{0},\omega)=0.
\end{equation}
Indeed, let us denote by $\omega (k_{0})$ the solution of this
equation which goes to $\ve_{0}(k_{0})$ as $V$ goes to zero.
Then, we have:
\begin{equation}
\Delta E = \omega (k_{0}).
\end{equation}
Now, if the trial state $\ket{N}$ obtained from $n(k)$ minimizes the total
energy $E(\{n\})$, it means that removing a particle at $k_{1}$ 
on the Fermi surface associated to the distribution $n(k)$ and adding
another particle at $k_{2}$ also on the Fermi surface does not change the
total energy (up to corrections which are negligible for very large systems).
This yields $\omega(k_{1})=\omega(k_{2})$, implying that quasiparticle
energies are constant (equal the the dressed chemical potential $\mu$)
on the Fermi surface associated to $n(k)$. This is exactly condition i)
 (see Eq.~(\ref{eq:conditioon_re_fs})) for the dressed propagator discussed in Sec.~\ref{sec:sub:gen_cons}. Assuming condition
i) holds, condition ii) on the imaginary part follows from standard
phase-space arguments and analyticity considerations developed 
already long ago by Luttinger\cite{Luttinger61} or Langer\cite{Langer61}. The main idea is to use an expression
for the self-energy in terms of skeleton graphs. Condition i) suggests
that the full one-particle spectral function (which determines completely
the internal lines of these graphs) is qualitatively similar to the
one of a Fermi liquid with the Fermi surface obtained from $n(k)$. 


\subsection{Extension to an infinite system}

Applying the previous argument to infinite systems requires some care.
In fact, we have to prove that the coefficients of perturbative series
in powers of $V$ for $\Delta E$ or $\omega(k_{0})$ have a well-defined
infinite volume limit. Our experience with other systems including models
for an unstable state coupled to a continuum suggests that such a limit
does not exist in general. However, perturbation theory in powers of
the interaction strength for fermion systems with local two-body
interactions is likely to be a favorable case for which this limit may be
safely taken. For instance, the perturbative expansion of the ground-state
energy involves connected Feynman graphs with no external lines, which 
contributions are easily shown to be proportional to the volume. 
Similarly, standard techniques based on the Luttinger-Ward energy
functional (see for instance the text by Nozi\`eres,\cite{Nozieres_anglais} pages
222 to 229) show that:
\begin{eqnarray}
  \Delta E &=& E(\{n'\})-E(\{n\})\nonumber\\
  &=& \ve_{0}(k_{0}) + \int_{-\infty}^{\infty}
  \frac{\rmd\omega}{2\pi \rmi} \log \Bigg[\frac{\omega -\ve_{0}(k_{0})-\rmi\eta}
  {\omega -\ve_{0}(k_{0})+\rmi\eta}\\
  &&\hspace{2cm}\times \frac{\omega -\ve_{0}(k_{0})-\Sigma_{\Phi}(k_{0},\omega)+\rmi\eta}
  {\omega -\ve_{0}(k_{0})-\Sigma_{\Phi}(k_{0},\omega)-\rmi\eta}\Bigg].\nonumber
\end{eqnarray}
up to terms which vanish in the thermodynamical limit.
If $k_{0}$ is close enough to the dressed Fermi surface so that the inverse
life time of the corresponding ``quasiparticle'' is small compared to
$\eta$, it is easy to show that $\Delta E = \omega(k_{0})$.
This shows that the series expansion of $\omega(k_{0})$ has a well-defined
infinite volume limit. This fact is a priori non trivial since any
perturbative algorithm for $\omega(k_{0})$ involves partial derivatives
at any order for $\Sigma_{\Phi}(k_{0},\omega)$ with respect to
$\omega$, taken at $\omega = \ve_{0}(k_{0})$. Although 
$\Sigma_{\Phi}(k_{0},\omega)$ has a good thermodynamical limit, 
some difficulties arise while considering derivatives with respect to
$\omega$. Indeed, their expressions for finite size systems involve
sums of rational functions of $\omega$ with multiple poles, and these
are not easily converted into converging integrals in the infinite
volume limit. But the above connection between $\Delta E$ (which has a
thermodynamical limit) and $\omega(k_{0})$ shows that all the wild 
terms which are expected to appear in a perturbative expression of
$\omega(k_{0})$, eventually cancel.


\section{Difficulties with the real part of $\Sigma$
in the traditional perturbation scheme}
\label{app:diff_re}

\subsection{Two-chain system at first order}
\label{sec:sub:sub:std_pert}

Let us begin by the calculation of the \se in the usual case where one starts from the free ground-state. The free propagators are simply given by:
\begin{eqnarray}
  G^{(0)}_{\rmr,I}(k,\omega)=\hspace{5.8cm}\nonumber\\
  \label{eq:free_propR}
  \frac{1}{\omega-\left[\mu^{(0)}+v_\rmf^{(0)}(k-k_{\rmf,I}^{(0)})\right]+\rmi\eta\sgn(k-k_{\rmf,I}^{(0)})},\hspace{0.4cm}\\
  G^{(0)}_{\rml,I}(k,\omega)=\hspace{5.8cm}\nonumber\\
  \label{eq:free_propL}
  \frac{1}{\omega-\left[\mu^{(0)}-v_\rmf^{(0)}(k+k_{\rmf,I}^{(0)})\right]-\rmi\eta\sgn(k+k_{\rmf,I}^{(0)})}.\hspace{0.4cm}
\end{eqnarray}
We will restrict ourselves to the study of the right propagators, because the left ones can be analyzed in an analogous way. The first order correction to the right propagators is given by the tadpole graph represented on Fig.~\ref{fig:tadpole_spinless}, where the solid (respectively dashed) lines represent right (respectively left) propagators, and where the black dot denotes one of the couplings. 
The \se for the (R,0) fermions is given by two terms: either the interaction is $A$, in which case the left propagator in the loop is on branch 0, or the interaction is $C$, and the left propagator is on branch $\pi$. It is a simple matter to evaluate the tadpole, and to show that in the thermodynamic limit the \se given by the $A$ interaction reads:
\begin{equation}
  \Sigma_{\rmr,0;A}^{(1)}=A \int\frac{\rmd q}{2\pi} n_{\rml,0}^{(0)}(q),
\end{equation}
where $n_{\rml,0}^{(0)}(q)$ is the particle distribution on branch (L,0), \ie it is 1 if $q\geqslant -k_{\rmf,0}$ and 0 otherwise. Of course, exactly as in the energy minimization scheme described in Sec.~\ref{sec:sub:2chaines_ordre1}, we get infinite results because our linearized dispersion relations have been extended to include infinitely many states. We will thus here too regularize these divergences by putting an ultra-violet cut-off $\Lambda_0$ on the momenta, around the four {\em free} Fermi momenta (remember Fig.~\ref{fig:cut-off} for one band).
It is then easy to show that $\Sigma_{\rmr,0;A}^{(1)}=A\Lambda_0/(2\pi)$. We let the reader check that the final results for the \ses of right fermions are:
\begin{eqnarray}
  \Sigma_{\rmr,0}^{(1)}(k,\omega)&=&(A+C)\frac{\Lambda_0}{2\pi},\\
  \Sigma_{\rmr,\pi}^{(1)}(k,\omega)&=&(B+C)\frac{\Lambda_0}{2\pi}.
\end{eqnarray}
The renormalized chemical potential $\mu$ and Fermi momenta $k_{\rmf,0(\pi)}$ can now be deduced from the condition that the inverse propagators vanishes for $\omega=\mu$ and $k=k_{\rmf,0}$ or $k=k_{\rmf,\pi}$, and from the conservation of the number of particles. This last condition is nothing but the Luttinger theorem. We thus have to solve for the following system of three equations for three unknown quantities:
\begin{eqnarray}
  &&\mu-\left[\mu^{(0)}+v_\rmf^{(0)}(k_{\rmf,0}-k_{\rmf,0}^{(0)})\right]-(A+C)\frac{\Lambda_0}{2\pi}=0,\hspace{0.8cm}\\
  &&\mu-\left[\mu^{(0)}+v_\rmf^{(0)}(k_{\rmf,\pi}-k_{\rmf,\pi}^{(0)})\right]-(B+C)\frac{\Lambda_0}{2\pi}=0,\hspace{0.8cm}\\
  &&k_{\rmf,0}+k_{\rmf,\pi}=k_{\rmf,0}^{(0)}+k_{\rmf,\pi}^{(0)}.
\end{eqnarray}
The chemical potential is found by summing the first two equations and making use of the third one. Then one gets the difference between interacting and free Fermi momenta at one loop:
\begin{eqnarray}
  \label{eq:dmu_faux}
  &&\mu^{(1)}-\mu^{(0)}=(A+B+2C)\frac{\Lambda_0}{4\pi},\\
  \label{eq:dk0_faux}
  &&k_{\rmf,0}^{(1)}-k_{\rmf,0}^{(0)}=(B-A)\frac{\Lambda_0}{4\pi v_\rmf^{(0)}}.
\end{eqnarray}
These are the results given by the standard perturbation theory.
This last result is to be compared with (\ref{eq:dk0_vrai}). To first order in the couplings, both results are equal. But (\ref{eq:dk0_vrai}) contains next order contributions that are not present in (\ref{eq:dk0_faux}). This happens although both computations assume the same physics, namely the validity of the Hartree approximation. As has been shown in Appendix ~\ref{sec:sub:sub:pert_ex}, 
consistency between both viewpoints is recovered only if the electron \se is computed with free propagators corresponding to the {\em dressed} \fs. A similar conclusion also holds for the chemical potential shift, as a comparison between Eqs.~(\ref{eq:dmu_vrai})
and ~(\ref{eq:dmu_faux}) readily shows.

 
\subsection{Formal calculation to second order}

\begin{figure}[b]
  \includegraphics[width=7cm]{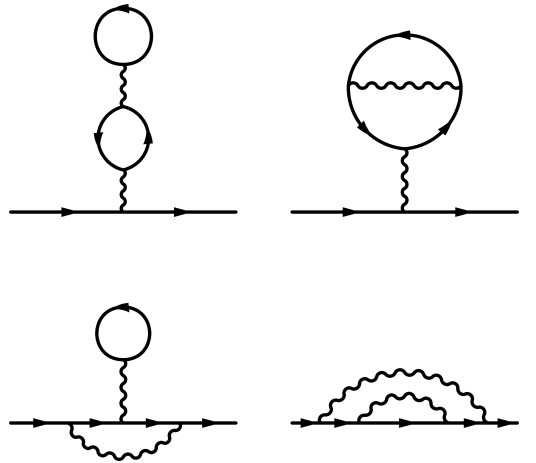}
    \caption{The four anomalous Kohn-Luttinger graphs contributing to the \se at two loops.}
  \label{fig:KL_diag}
\end{figure}

Let us consider a system of interacting spinless Fermions in $d=3$
dimensions. In fact, the actual value of $d$ does not have much influence
in the following discussion, the main point is that $d \geqslant 2$ so the
Fermi surface  is in general a smooth manifold of codimension one in \vk-space.
We take the following Hamiltonian:
\begin{eqnarray}
  &&H = \int \frac{\rmd^{3}\vk}{(2\pi)^{3}}\ve(\vk)\cdag(\vk)c(\vk)\\
  &&\hspace{0.8cm}+\frac{V}{2}\int \frac{\rmd^{3}\vk}{(2\pi)^{3}}\int \frac{\rmd^{3}\vkp}{(2\pi)^{3}}
  \int \frac{\rmd^{3}\vq}{(2\pi)^{3}}f(\vk,\vkp,\vq)\nonumber\\
  &&\hspace{3cm}\times\cdag(\vk+\vq)\cdag(\vkp-\vq)c(\vkp)c(\vk).\nonumber
\end{eqnarray}
We assume the Fermi surface for $V=0$ is connected and that each half-line
starting from the origin in \vk-space intersects it only once. For any
unit vector \vu, we thus define a positive number $k_{\rmf,0}(\vu)$ such that
$k_{\rmf,0}(\vu)\vu$ belongs to the Fermi surface. The Fermi sea is then the
set of \vk \hspace{0ex} points such that $\vk = k\vu$ with \vu \hspace{0ex}
unit vector and
$0 \leqslant k \leqslant  k_{\rmf,0}(\vu)$. The total particle number is assumed to be
fixed, independently of the coupling strength $V$. We now consider 
eigenstates obtained by adiabatic switching of the interaction $V$ on
free particle states with a deformed Fermi surface $\vu \mapsto k_{\rmf}(\vu)$.
Denoting $k_{\rmf}(\vu)- k_{\rmf,0}(\vu)=\delta k_{\rmf}(\vu)$, the constraint
on the total particle number reads:
\begin{eqnarray}
  &&\int \rmd^{2}\vu \Big[k^{2}_{\rmf,0}(\vu)\delta k_{\rmf}(\vu)\\
  &&\hspace{2cm}+k_{\rmf,0}(\vu)\delta k^{2}_{\rmf}(\vu)+\frac{1}{3}\delta k^{3}_{\rmf}(\vu)\Big]=0,\nonumber
\end{eqnarray}
where $\rmd^{2}\vu$ is the usual area element on the unit sphere, for instance
$\rmd^{2}\vu=\sin\theta\, \rmd\theta \rmd\phi$ in spherical coordinates. We now wish to
choose $\delta k_{\rmf}(\vu)$ in order to minimize the total energy of the
corresponding eigenstate, while keeping a constant particle number.
To second order in $V$, this total energy is given by:
\begin{eqnarray}
  &&E(\{k_{\rmf}\}) = \int \frac{\rmd^{3}\vk}{(2\pi)^{3}}n(\vk)\ve(\vk)\\
  &&\hspace{0.5cm}+ \frac{V}{2}\int \frac{\rmd^{3}\vk}{(2\pi)^{3}}\int \frac{\rmd^{3}\vkp}{(2\pi)^{3}}
  n(\vk)n(\vkp)g(\vk,\vkp,0)\nonumber\\
  && \hspace{0.5cm}+ \frac{V^{2}}{4}\int \frac{\rmd^{3}\vk}{(2\pi)^{3}}
  \int \frac{\rmd^{3}\vkp}{(2\pi)^{3}}\int \frac{\rmd^{3}\vq}{(2\pi)^{3}} g(\vk,\vkp,\vq)^{2}\nonumber\\
  &&\hspace{1.5cm}\times\frac{n(\vk)n(\vkp)\big(1-n(\vk+\vq)\big)\big(1-n(\vkp-\vq)\big)}
  {\ve(\vk)+\ve(\vkp)-\ve(\vk+\vq)-\ve(\vkp-\vq)}
  \nonumber\\
  &&\hspace{0.5cm}+\mathcal{O}(V^{3}).\nonumber
\end{eqnarray}
Here, $n(k\vu)=1$ if $k$ is smaller than $k_{\rmf}(\vu)$
and $n(k\vu)=0$ otherwise. We have also defined $g(\vk,\vkp,\vq)\equiv
f(\vk,\vkp,\vq)-f(\vk,\vkp,\vkp-\vk-\vq)$. After some algebra, we find that
$\delta k_{\rmf}(\vu)=V\delta k_{\rmf,1}(\vu)+V^{2}\delta k_{\rmf,2}(\vu)+
\mathcal{O}(V^{3})$, and $\mu = \mu_{0}+ 
V\mu_{1}+V^{2}\mu_{2}+\mathcal{O}(V^{3})$, 
where $\delta k_{\rmf,1}(\vu)$, $\delta k_{\rmf,2}(\vu)$, $\mu_{1}$ and $\mu_{2}$
are given by the following coupled linear equations:
\begin{equation}
  v_{\rmf}(\vu)\delta k_{\rmf,1}(\vu)+\Sigma^{(1)}\big(\vkfz(\vu)\big)-\mu_{1} = 0,
\end{equation}
\begin{equation}
  \int \rmd^{2}\vu k^{2}_{\rmf,0}(\vu)\delta k_{\rmf,1}(\vu) = 0,
\end{equation}
\begin{eqnarray}
  &&v_{\rmf}(\vu)\delta k_{\rmf,2}(\vu)+\Re\Sigma^{(2)}\Big(\vkfz(\vu),\ve\big(\vkfz(\vu)\big)\Big)\nonumber\\
  &&\hspace{0.3cm}+\vu\cdot\nabla_{\vk}\Sigma^{(1)}\big(\vkfz(\vu)\big)+\frac{1}{2}v_{\rmf}'(\vu) \delta k^{2}_{\rmf,1}(\vu)\\
  &&\hspace{0.6cm}+\Delta\Sigma^{(1)}\big(\vkfz(\vu)\big)-\mu_{2} = 0, \mbox{ and}\nonumber
\end{eqnarray}
\begin{equation}
  \int \rmd^{2}\vu \Big[k^{2}_{\rmf,0}(\vu)\delta k_{\rmf,2}(\vu)+
  k_{\rmf,0}(\vu)\delta k^{2}_{\rmf,1}(\vu)\Big] = 0.
\end{equation}
In these expressions, $v_{\rmf}(\vu)$ and $v_{\rmf}'(\vu)$ denote the first
and second derivatives of the function 
$x \mapsto \ve\big(\big(k_{\rmf,0}(\vu)+x\big)\vu\big)$, taken at $x=0$.
$\Sigma^{(1)}(\vk)$ and $\Sigma^{(2)}(\vk,\omega)$ are the 
self-energies computed
to first and second order in $V$, using the standard algorithm:
\begin{equation}
  \Sigma^{(1)}(\vk) = \int \frac{\rmd^{3}\vkp}{(2\pi)^{3}} n^{(0)}(\vkp)
  g(\vk,\vkp,0),\mbox{ and}
\end{equation}
\begin{widetext}
  \begin{eqnarray}
    &&\Re\Sigma^{(2)}(\vk,\omega) = \frac{1}{2}\int \frac{\rmd^{3}\vk}{(2\pi)^{3}}  
    \int \frac{\rmd^{3}\vkp}{(2\pi)^{3}} \int \frac{\rmd^{3}\vq}{(2\pi)^{3}}
    \;g(\vk,\vkp,\vq)^{2}\\
    &&\hspace{4cm}\times\frac{n^{(0)}(\vkp)\big(1-n^{(0)}(\vk+\vq)\big)\big(1-n^{(0)}(\vkp-\vq)\big)+
      \big(1-n^{(0)}(\vkp)\big)n^{(0)}(\vk+\vq)n^{(0)}(\vkp-\vq)}
    {\omega+\ve(\vkp)-\ve(\vk+\vq)-\ve(\vkp-\vq)}.\nonumber
  \end{eqnarray}
\end{widetext}
The most interesting quantity in these formulae is $\Delta\Sigma^{(1)}(\vk)$.
It is the change in the first-order (with respect to $V$) self-energy
due to the fact that the Fermi surface has changed by an amount 
$\delta k_{\rmf,1}$. More precisely, we have:
\begin{eqnarray}
  &&\Delta\Sigma^{(1)}(\vk)= \nonumber\\
  &&\hspace{0.5cm}\int \frac{\rmd^{2}\vup}{(2\pi)^{3}} 
  g\big(\vk,\vkfz(\vup),0\big) k_{\rmf,0}^{2}(\vup)\delta k_{\rmf,1}(\vup).\hspace{1cm}
\end{eqnarray}
It turns out this term is {\em not} recovered in the naive perturbation
algorithm. This latter procedure is based on solving for $k_{\rmf}(\vu)$ in the equations:
\begin{equation}
\ve\big(\vkf(\vu)\big)+\Re\Sigma\big(\vkf(\vu),\mu\big) = \mu,
\end{equation}
where $\Sigma(\vk,\omega)$ is computed with the free propagators associated
to the non-interacting Fermi surface. As before, $\mu$ is chosen
to keep a constant total particle number. This second approach yields
the same set of equations as before, except that the term
$\Delta\Sigma^{(1)}(\vkfz(\vu))$ is missing in the first equation for
$\delta k_{\rmf,2}(\vu)$. This shows that the naive algorithm is not able to 
keep track of the first-order Fermi surface deformation while evaluating the
Hartree-Fock corrections to second order. Intuitively, these effects are 
expected to be associated to the four second order graphs for $\Sigma$
shown on Fig.~\ref{fig:KL_diag}.
However, these graphs are anomalous according to
Kohn and Luttinger, and their contribution vanishes in the $T=0$ perturbation
theory scheme. We believe this illustrates the crucial problem with naive 
perturbation theory. As we go to higher orders in $V$, lower order graphs
for $\Sigma$ are modified by the changes already induced on the occupation
numbers of single particle states. One would expect to capture these
changes thanks to self-energy insertions in the internal lines of the
lower-order graphs for $\Sigma$. But the example of anomalous graphs 
shows this does not work so well in general.

As discussed in section \ref{sec:sub:use_ct} the natural cure for this problem is to fix the 
dressed Fermi surface, thanks to counterterms which gradually modify the
single particle dispersion of the original Hamiltonian, as $V$ is increased.
In this approach, we choose the dressed Fermi surface $\vu \mapsto \vkf(\vu)$
and the dressed dispersion relation $\ve(\vk)$. We therefore
compute the self-energy $\widetilde\Sigma(\vk,\omega)$ with respect to this
dressed Fermi surface. Denoting by $\mu=\ve(\vkf(\vu))$ the dressed
chemical potential, the counterterms $\Sigma_{CT}(\vu)$ are defined by:
\begin{equation}
\widetilde\Sigma(\vkf(\vu),\mu)+\Sigma_{CT}(\vu) = 0.
\end{equation}
In this third approach, the ``bare'' Fermi surface $\vu \mapsto \vkfz(\vu)$
becomes a function of $V$ and $\{k_{\rmf}\}$. It is obtained from:
\begin{equation}
\ve(\vkfz(\vu))+\Sigma_{CT}(\vu) = \mu_{0},
\end{equation}
where, as always, $\mu_{0}$ is chosen in order to conserve the total 
particle number. We therefore have to solve:
\begin{equation}
\ve(\vkf(\vu))-\ve(\vkfz(\vu))+\widetilde\Sigma(\vkf(\vu),\mu) = 
\mu-\mu_{0}.
\end{equation}
Since $\widetilde\Sigma$ is computed with free propagators whose singularities
lie on the dressed Fermi surface, it is easy to check that this yields the same
expressions for $\delta k_{\rmf,1}$ and $\delta k_{\rmf,2}$ as the energy minimizing
procedure, in complete agreement with the general conclusions of Appendix \ref{app:equi}.


\section{Difficulties with the imaginary part of $\Sigma$
in the traditional perturbation scheme}
\label{app:diff_im}

Let us consider the traditional perturbation scheme around the unperturbed
Fermi surface. With the same notation as before, this corresponds to the
choice of free propagator: 
$G^{(0)}(k,\omega)^{-1}=\omega- \ve_{0}(k)+ \rmi\eta \sgn
(\ve_{0}(k)-\mu_{0})$. In the discussion, we shall use
the spectral densities $\rho_{p,h}(k,\omega)$
for excited states involving $p$ particles
and $h$ holes, with a total momentum $k$ and a total energy $\omega$.
We have:
\begin{widetext}
  \begin{eqnarray}
    &&\rho_{p,h}(k,\omega)=\prod_{i=1}^{p}\int \frac{\rmd k_{i}}{(2\pi)^d}\theta
    (\ve_{0}(k_{i})-\mu_{0})\prod_{j=1}^{h}\int \frac{\rmd k_{j}'}{(2\pi)^d}\theta
    (\mu_{0}-\ve_{0}(k_{j}'))\\
    &&\hspace{6cm}(2\pi)^{d}\delta(k-\sum_{i=1}^{p}k_{i}
    +\sum_{j=1}^{h}k_{j}')
    \times(2\pi)\delta(\omega-\sum_{i=1}^{p}\ve_{0}(k_{i})
    +\sum_{j=1}^{h}\ve_{0}(k_{j}')).\nonumber
  \end{eqnarray}
\end{widetext}
Let us consider the simple second order diagram for
$\Sigma(k,\omega)$ shown on Fig.~\ref{fig:sunrise_spinless} (the sunrise diagram).
It is simple to check that the imaginary part of this diagram 
is proportional to $\rho_{2,1}(k,\omega)-\rho_{1,2}(-k,-\omega)$.
As is well known since Landau, this quantity vanishes at the bare chemical
potential $\mu_{0}$, and behaves as $(\omega-\mu_{0})^{2}$ in magnitude
for $\omega$ close to $\mu_{0}$. The effect of Fermi surface deformation
on $\Sigma(k,\omega)$ arises via the replacement of bare propagators
by sequences of these propagators separated by lower order self-energy
insertions. The main point we wish to emphasize here is that there is
no simple way to predict the influence of these insertions on the
frequency dependence of $\Im \Sigma(k,\omega)$. For some graphs, and 
some patterns of insertions, the resulting $\Im \Sigma(k,\omega)$ will
continue to vanish at $\omega=\mu_{0}$, whereas some other combinations
will produce a finite contribution to $\Im \Sigma(k,\omega)$ at
$\omega=\mu_{0}$. Therefore, the traditional scheme does not allow for a good
control of the analytical structure of the self-energy.

Let us show this on a typical example. The simplest interesting situation
is obtained for the sunrise graph, for \se insertions which are
assumed not to depend on $k$ nor on $\omega$, since it is then easy to
perform the frequency integrals. In the more general case of an arbitrary
frequency dependence for the insertions, the natural procedure would be
to Taylor-expand them in the vicinity of $\mu_{0}$. The strongest effect
is obtained for the constant term in these expansions, and this leads 
to our toy example. For a total number $n$ of constant insertions,
we get a contribution to  $\Im \Sigma(k,\omega)$ proportional to
$\rho_{2,1}^{(n)}(k,\omega)-(-1)^{n}\rho_{1,2}^{(n)}(-k,-\omega)$,
where $\rho_{p,h}^{(n)}(k,\omega)$ stands for the $n$-th partial derivative
of  $\rho_{p,h}(k,\omega)$ with respect to $\omega$. Using the fact that
$\rho_{2,1}(k,\omega)$ and $\rho_{1,2}(-k,-\omega)$ behave as 
$(\omega-\mu_{0})^{2}$ for $\omega$ close to $\mu_{0}$, we notice that a
single self-energy insertion in the sunrise graph preserves the property
that $\Im \Sigma(k,\omega)$ vanishes for $\omega=\mu_{0}$. More generally,
this implies that $\Im \Sigma(k,\omega=\mu_{0})=0$ up to third order in 
perturbation theory. However, a Fermi surface deformation already occurs
usually for the simplest Hartree and Fock graphs, which are first order in
the coupling strength. In the standard perturbation scheme, the dressed
chemical potential and the dressed Fermi surface are determined by solving the
infinite set of equations:
\begin{equation}
\mu- \ve_{0}(k_{F})-\Re \Sigma(k_{F},\mu)=0,
\end{equation}
with the constraint that the total volume of the Fermi surface 
does not change as interactions are switched on. We therefore see that,
in this scheme, there is no reason for which $\Im \Sigma(k_\rmf,\omega=\mu)$
should vanish. This is rather unsatisfactory on physical grounds, since it
would imply a finite life-time for particle-like excitations 
lying just on the dressed Fermi surface.


\section{Field-theoretical RG}
\label{app:field_th}

This Appendix is devoted to a detailed derivation of the field-theoretical RG
equations which we have gathered in Appendix \ref{app:flow_eq}. The two main
considerations we wish to stress here are: i) the choice of external
momenta in the renormalization prescriptions, which has to be adapted to the 
\fs shape, and ii) the use of the logarithmic approximation.

\subsection{Motivation and general idea}
\label{sec:sub:sub:motivation_gal_idea}

In the usual ``field-theoretical'' RG, the high-energy Hamiltonian is given and fixed, and one parametrizes the theory by low-energy values of proper Green's functions (for instance the interaction vertices) at a typical scale $\nu$. Requiring that all these theories at different energy scales should correspond to one and the same high-energy theory yields RG flows, when $\nu$ is varied. This approach is physically natural, because it is based on the calculation of low-energy observables. Furthermore, as we will see, it allows for a study of crossovers between high and low-energy regimes.

One of its limitations is that it requires renormalizable interactions (\ie the existence of a continuous limit). But it is not really a severe drawback, since non-renormalizable interactions are expected to be irrelevant (by power counting) in the low-energy limit, which is the most interesting to us. Note that the renormalizability constraint disappears in RG schemes based on Wilson's idea of gradual mode elimination. Several groups have recently implemented Wilson's approach to the RG, expressed via the Polchinski equation,\cite{Zanchi96,Halboth00} or its one-particle irreducible version.\cite{Jungnickel96,Honerkamp01} Although these equations are exact, they are quite complicated, since effective interactions involving an arbitrary number of particles are generated along the RG flow. Any numerical computation therefore requires drastic truncations in the effective action. By contrast, the field-theory approach involves only a much smaller set of effective or running couplings, which is a good feature for practical implementations.


\subsection{Renormalization of the interactions}
\label{sec:sub:sub:ren_int}

First of all, we have to define the renormalized couplings. The two corresponding Green's functions (in real space for the direction parallel to the chains) are:
\begin{eqnarray}
  &&G^{I\!\!I}(X_4,X_3,X_2,X_1)=\nonumber\\
  &&\hspace{1cm}-\bra{0} \rT\Big[ \anp_{\rmr,I+\delta,\tau}(X_4) \anp_{\rml,J-\delta,\rho}(X_3)\\
  &&\hspace{3cm}\times\pdag_{\rml,J,\rho'}(X_2) \pdag_{\rmr,I,\tau'}(X_1)\Big] \ket{0},\nonumber\\
  &&U^{I\!\!I}(X_4,X_3,X_2,X_1)=\nonumber\\
  &&\hspace{1cm}-\bra{0} \rT\Big[ \anp_{\rmr,I+\delta,\tau}(X_4) \anp_{\rmr,J-\delta,\rho}(X_3)\\
  &&\hspace{3cm}\times\pdag_{\rml,J,\rho'}(X_2) \pdag_{\rml,I,\tau'}(X_1)\Big] \ket{0}.\nonumber
\end{eqnarray}
In the above equations, $\ket{0}$ is the interacting ground-state, T is the time ordering operator, and $X$ is a shorthand notation for $(t,x)$. The renormalized couplings are the values of the amputated one-particle irreducible parts of these Green's functions, divided by $\rmi$. In fact, the charge and spin couplings are the coefficients obtained from the Fourier transform of $G^{I\!\!I}$, factor of $\mathbb{I}_{\tau,\tau'} \mathbb{I}_{\rho,\rho'}$ and $\boldsymbol{\sigma}_{\tau,\tau'} \cdot \boldsymbol{\sigma}_{\rho,\rho'}$ respectively. The Umklapp coupling $U_\delta(I,J)$ is defined in the same way from $U^{I\!\!I}$, as the coefficient of $\mathbb{I}_{\tau,\tau'} \mathbb{I}_{\rho,\rho'}$ (the one in front of $\mathbb{I}_{\rho,\tau'} \mathbb{I}_{\tau,\rho'}$ being $-U_{J-I-\delta}(I,J)$).

The set of external frequencies is chosen to be the same for all types of couplings. We have decided to take:
\begin{equation}
  \label{eq:energies_pattes_externes}
  \omega_1=\nus,\quad \omega_2=\nus,\quad \omega_3=\tnus,
\end{equation}
and by energy conservation we have of course $\omega_4=\omega_1+\omega_2-\omega_3=-\nus$. $\nu$ is the typical energy scale of the interaction process, and is the quantity to be varied to get RG flows.

It is a bit more difficult to choose the external momenta, because of the warping of the \fs. Our choice has been dictated by a few natural requirements. First the symmetries of the \fs should be respected, as the right-left symmetry, the up-down symmetry (\ie $k_y\leftrightarrow -k_y$, in terms of the original transverse momenta). Interactions processes for which it is possible to choose all external momenta on the \fs, should be computed for this special choice, because it would otherwise mean the introduction of a spurious energy scale. Let us first consider $\gcs_\delta(I,J)$. It is possible to choose $k_1=k_{\rmf,I}$, $k_2=-k_{\rmf,J}$, $k_3=-k_{\rmf,J-\delta}$, and $k_4=k_{\rmf,I+\delta}$, only if momentum conservation $k_{\rmf,I}-k_{\rmf,J}=k_{\rmf,I+\delta}-k_{\rmf,J-\delta}$ is respected. In general, this will not be possible, for we will have $\Delta k_\delta(I,J)=(k_{\rmf,I+\delta}+k_{\rmf,J})-(k_{\rmf,I}+k_{\rmf,J-\delta})\neq 0$. Notice that up to a minus sign and a factor of 2, $\Delta k_\delta(I,J)$ is simply the generalization of $\Delta k_\rmf$ in the two-chain model. It is then natural to split this quantity equally among the four momenta. One can check that the following choice fulfills all the conditions we have mentioned:
\begin{eqnarray}
  \label{eq:choix_4impulsions}
  &&k_1=k_{\rmf,I}+\frac{\Delta k_\delta(I,J)}{4},\nonumber\\
  &&k_2=-\left(k_{\rmf,J}-\frac{\Delta k_\delta(I,J)}{4}\right),\\ 
  &&k_3=-\left(k_{\rmf,J-\delta}+\frac{\Delta k_\delta(I,J)}{4}\right),\mbox{ and }\nonumber\\
  &&k_4=k_{\rmf,I+\delta}-\frac{\Delta k_\delta(I,J)}{4}.\nonumber
\end{eqnarray}

For the Umklapps, the choice of external momenta is dictated by the same requirements. The equivalent of $\Delta k_\delta(I,J)$ is now $\Delta k^U_\delta(I,J)=2\pi-(k_{\rmf,I}+k_{\rmf,J}+k_{\rmf,I+\delta}+k_{\rmf,J-\delta})$, and the natural choice of momenta reads:
\begin{eqnarray}
  \label{eq:choix_4impulsions_U}
  &&k_1=-\left(k_{\rmf,I}+\frac{\Delta k^U_\delta(I,J)}{4}\right),\nonumber\\
  &&k_2=-\left(k_{\rmf,J}+\frac{\Delta k^U_\delta(I,J)}{4}\right),\\ 
  &&k_3=k_{\rmf,J-\delta}+\frac{\Delta k^U_\delta(I,J)}{4},\mbox{ and }\nonumber\\
  &&k_4=k_{\rmf,I+\delta}+\frac{\Delta k^U_\delta(I,J)}{4}.\nonumber
\end{eqnarray}

The next task is to draw all possible Feynman diagrams, and compute them. One can then establish the field theoretical RG flows, requiring the couplings measured at two different scales should correspond to the same high-energy theory. The couplings' flow equations are given in Appendix \ref{app:flow_eq_coup}, and are obtained in the one-loop approximation. As argued in Sec.~\ref{sec:csfs}, the most interesting effects connected to \fs deformation appear at the two-loop level for the single-electron propagator. For the sake of simplicity, we shall use a hybrid scheme, involving a one-loop approximation for the couplings and a two-loop approximation for the electronic \se. In the case where all the couplings remain weak, it is reasonable to keep only the dominant term in the corresponding flow equation. If by contrast couplings have a tendency to grow and become large at low energies, experience from the Kondo problem suggests adding the subleading terms to the couplings' flow does not provide a better physical picture. For the Kondo problem, the two-loop approximation predicts an intermediate coupling fixed point,\cite{Fowler71,Abrikosov70} whereas the low-energy physics corresponds to an infinite coupling fixed point.\cite{Anderson70,Wilson75}

We shall not give any technical detail on the derivation of these couplings' flow equations which is standard.\cite{Solyom79_dans_articles} The main new feature is the use of special sets of external momenta, described in Eqs.~(\ref{eq:choix_4impulsions}) and (\ref{eq:choix_4impulsions_U}) above. However, it is worth focusing on the $f$ function that appears in these equations (the $f$ function is defined by $f(t=\ln(\Lambda_0/\nu),\delta)=1$ if $\nu\geqslant |\delta|$ and 0 otherwise, see Appendix \ref{app:flow_eq_coup}). In fact, the ``true'' RG equations do not involve this function, but rather:
\begin{equation}
  \widetilde{f}(t=\ln(\Lambda_0/\nu),\delta)=\frac{1}{2}\left( \frac{\nu}{\nu-\delta+\rmi\eta} + \frac{\nu}{\nu+\delta-\rmi\eta} \right).
\end{equation}
It is obvious that these $\widetilde{f}$ diverge for $\nu=|\delta |$, so that some couplings will diverge or vanish singularly at the scales given in (\ref{eq:car_length_1}) to (\ref{eq:car_length_9}). Though this physically signals the crossing of the characteristic scales, it is practically unpleasant for the numerical simulations. Furthermore, if we did not work at zero temperature, the energy scale given by the temperature $T$ would suppress these divergences. Notice that the presence of $\rmi\eta$ factors also implies that the couplings will not remain real. 

\begin{figure}[t]
  \includegraphics[width=8cm]{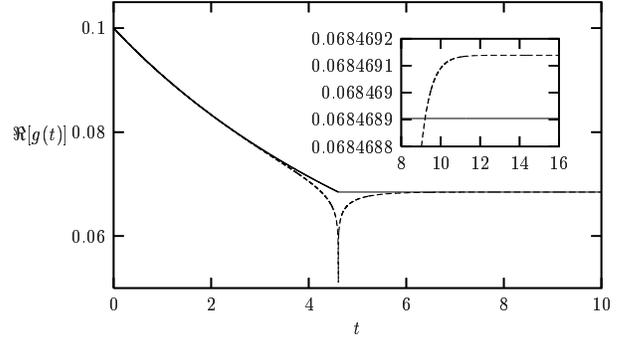}
  \caption{The dashed curve is the real part of the coupling $g$ satisfying (\ref{eq:test_1coup}), with initial condition $g(0)=0.1$. Here $\delta/\Lambda_0=10^{-2}$, so that the coupling goes to zero singularly for $t\simeq 4.6$. The solid line represents the solution of the approximate equation where $\widetilde{f}$ is replaced by $f$. The inserted plot is a zoom on the end of the flows.}
  \label{fig:sol_div_pos_regularisee}
\end{figure}

For all these reasons, it is thus natural to try to find a way to get rid of these singular behaviors. This can be achieved by replacing $\widetilde{f}$ by a function that extends its asymptotic behaviors (for $\nu\gg|\delta|$ and $\nu\ll|\delta|$) up to $\nu=|\delta|$. This is exactly what the function $f$ does. The quality of the approximation can be checked on a very simple flow equation, for which one knows the exact solution:
\begin{equation}
  \label{eq:test_1coup}
  \partial_\nu g(\nu)=\frac{1}{2}\left( \frac{1}{\nu-\delta+\rmi\eta} + \frac{1}{\nu+\delta-\rmi\eta} \right) g^2(\nu).
\end{equation}
This is simply a RPA-like flow, for only one coupling $g$. We will not study this in detail, but we show the good quality of the approximate solution for a positive initial coupling, on Fig.~\ref{fig:sol_div_pos_regularisee}

The approximation we use is in fact nothing but a logarithmic approximation. Indeed, $\widetilde{f}$ appears in the flow equations, after we have differentiated $\ln[(\nu-\delta+\rmi\eta)(\nu+\delta-\rmi\eta)/(2v_\rmf\Lambda_0)^2]$ factors (with respect to the scale $\nu$), coming from the Feynman graphs' logarithmic divergences. Changing $\widetilde{f}$ in $f$ just amounts to replacing this logarithm by the approximation $2\ln[\mathrm{Max}(\nu,|\delta|)/(2v_\rmf\Lambda_0)]$.


\subsection{Renormalization of the propagator}

\begin{figure}[t]
  \includegraphics[width=8cm]{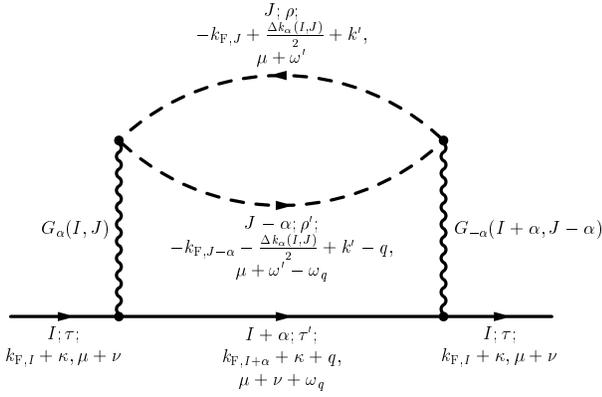}
  \caption{Sunrise diagram involving two $G$ interactions.}
  \label{fig:se_gg_sunrise}
\end{figure}

For reasons explained at the end of Sec.~\ref{sec:sub:sub:gen_th_an}, the one-loop \se correction does not have much influence on the \fs deformation for the quasi 1D Hubbard systems considered in this paper. Therefore, we will only focus on the two-loop sunrise diagram (the Kohn-Luttinger diagram being equal to zero). We only show the Feynman diagram here, with all the information about the internal lines, for the ``$G^2$'' contribution, on Fig.~\ref{fig:se_gg_sunrise}. 
There is of course a ``$U^2$'' contribution, and the spin algebra has to be taken into account. In order to simplify the expressions, we have computed the \se at $k=k_{\rmf,I}$, and not for a general momentum, since we have decided not to take Fermi velocity renormalization into account. We give the expression of the \se in Appendix \ref{app:flow_eq_prop}. From this, the first thing to do is to compute the counterterms, so that the inverse propagators vanish on the dressed \fs. Focusing on second order terms only, this amounts to require:
\begin{eqnarray}
  &&\forall I\in\{1,\ldots,N\},\nonumber\\
  &&-\Sigma_{\rmr,I}^{(2)}(k=k_{\rmf,I},\omega=\mu)-\delta\mu^{(2)}+v_\rmf\delta k_I^{(2)}=0,\hspace{0.8cm}
\end{eqnarray}
together with the conservation of the total particle number. $\delta\mu^{(2)}$ is found by summing all these equations: $\delta\mu^{(2)}=-\left[\sum_I \Sigma_{\rmr,I}^{(2)}(k=k_{\rmf,I},\omega=\mu)\right]/N$. We let the reader check that it is in fact sufficient to take the Umklapp contribution in this last equation, because the $G$ contribution sums to zero. This is true because of the following properties: $\Delta k_\alpha(I,J)=-\Delta k_{-\alpha}(J,I)$, $\gcs_\alpha(I,J)=\gcs_{-\alpha}(J,I)$ and $\gcs_{-\alpha}(I+\alpha,J-\alpha)=\gcs_\alpha(J-\alpha,I+\alpha)$, so that in the sum over $I$, $J$ and $\alpha$, the terms $(I,J,\alpha)$ and $(J,I,-\alpha)$ will cancel each other.

Once the chemical potential is known, the Fermi momenta counterterms can be found:
\begin{eqnarray}
  \label{eq:fs_counterterm}
  &&\delta k_I^{(2)}=\frac{1}{v_\rmf} \Big[ \Sigma_{\rmr,I}^{(2)}(k=k_{\rmf,I},\omega=\mu)\\
  &&\hspace{2.5cm}-\frac{1}{N}\sum_I \Sigma_{\rmr,I}^{(2)}(k=k_{\rmf,I},\omega=\mu) \Big].\nonumber
\end{eqnarray}
In order to save space, we will not give the full expressions of these counterterms.

From all this, we can deduce a dressed propagator, as we did in Eq.~(\ref{eq:dressed_prop_2c}) for the two-chain model. As in Eq.~(\ref{eq:dressed_prop_2c}), the result is still divergent when $\Lambda_0$ is sent to infinity. This simply means that the counterterms are not sufficient. Something more is needed, and as is well known, this is wave function renormalization. The renormalized (R,I) propagator is defined as usual by:
\begin{equation}
  {G_{\rmr,I}^{(\mathcal{R})}}^{-1}(k,\omega)=Z_{\rmr,I} G^{-1}_{\rmr,I}(k,\omega),
\end{equation}
and the wave function renormalization factor $Z_{\rmr,I}$ is found by imposing the following renormalization prescription:
\begin{eqnarray}
  &&{G_{\rmr,I}^{(\mathcal{R})}}^{-1}(k=k_{\rmf,I},\omega=\mu+\nu)=\nonumber\\
  \label{eq:prescr_renorm_prop}
  &&\hspace{1cm}Z_{\rmr,I} G^{-1}_{\rmr,I}(k=k_{\rmf,I},\omega=\mu+\nu)=\nu.\hspace{0.8cm}
\end{eqnarray}
We have not written which variables $Z_{\rmr,I}$ depends on, in order to make the equations lighter, but it should be clear that it is a function of the couplings at scale $\nu$, of the dressed Fermi momenta, of $\nu$ and of $\Lambda_0$. The calculation of the renormalized propagator is achieved thanks to the standard observation that Eq.~(\ref{eq:prescr_renorm_prop}) implies:
\begin{eqnarray}
  \label{eq:lien_prop_2echelles}
  &&{G_{\rmr,I}^{(\mathcal{R})}}^{-1}(k,\omega;g;\nu,\Lambda_0)=\\
  &&\hspace{1cm}\vp_{\rmr,I}(g;\nu,\nu',\Lambda_0)\, {G_{\rmr,I}^{(\mathcal{R})}}^{-1}(k,\omega;{g}';\nu',\Lambda_0), \mbox{ with }\nonumber\\
  &&\vp_{\rmr,I}(g;\nu,\nu',\Lambda_0)=\frac{Z_{\rmr,I}(g;\nu,\Lambda_0)}{Z_{\rmr,I}({g}';\nu',\Lambda_0)}\nonumber\\
  \label{eq:definition_phi}
  &&\hspace{2.4cm}=\frac{Z_{\rmr,I}(g;\nu,\Lambda_0)}{Z_{\rmr,I}(\overline{g}(g;\nu,\nu',\Lambda_0);\nu',\Lambda_0)}.
\end{eqnarray}
In these equations, $g$ is a shorthand notation for all the couplings, at scale $\nu$, and $g'$ is the same at scale $\nu'$. In the last equation, the function $\overline{g}$ is relating the value of the couplings at two different scales by $g'=\overline{g}(g;\nu,\nu',\Lambda_0)$. Finally, the RG flow equations for the $\vp$ functions are found by differentiating the multiplicative relation $\vp_{\rmr,I}(g;\nu,\nu'',\Lambda_0)=\vp_{\rmr,I}(g;\nu,\nu',\Lambda_0) \vp_{\rmr,I}({g}';\nu',\nu'',\Lambda_0)$, and one obtains:
\begin{eqnarray}
  \label{eq:eq_flot_phi_gale}
  &&\frac{\partial}{\partial \nu'}\vp_{\rmr,I}(g;\nu,\nu',\Lambda_0)=\vp_{\rmr,I}(g;\nu,\nu',\Lambda_0)\\
  &&\hspace{1cm}\times\left.\left[ \frac{\partial}{\partial \nu''} \vp_{\rmr,I}(\overline{g}(g^\beta;\nu,\nu',\Lambda_0);\nu',\nu'',\Lambda_0)\right]\right|_{\nu''=\nu'}.\nonumber
\end{eqnarray}
Notice that the importance of the $\vp$ functions lay in the close link between these and the renormalized propagator, that one can deduce from Eq.(\ref{eq:lien_prop_2echelles}) and the renormalization prescription of Eq.~(\ref{eq:prescr_renorm_prop}):
\begin{equation}
  \label{eq:expression_prop_renorm}
  {G_{\rmr,\rml}^{(\mathcal{R})}}^{-1}(k=k_\rmf,\omega;g^\alpha;\nu,\Lambda_0)=\omega\, \vp_{\rmr,\rml}(g^\alpha;\nu,\omega,\Lambda_0).
\end{equation}
For the $N$ chains we are interested in, the flow equations of $\vp_{\rmr,I}$ are given in Appendix \ref{app:flow_eq_prop}, Eq.~(\ref{eq:flot_phi}). The flow equations of $\vp_{\rml,I}$ can be checked to be exactly the same (this is due to the Left-Right symmetry of the system we study).


\section{RG flow equations}
\label{app:flow_eq}

\subsection{Flows of the couplings}
\label{app:flow_eq_coup}

The flow equations for the couplings that are given below are the 
field-theoretical RG equations, obtained after the general analysis of Appendix
\ref{app:field_th} has been performed. In these, the RG time $t$ is given by
$t=\ln(\Lambda_0/\nu)$ ($\Lambda_0$ being the ultra-violet cut-off and $\nu$ the typical energy scale of the interaction), and the running couplings are low-energy interactions $G(\nu)$.

The high-energy flow equations of the couplings in the cut-off scaling scheme 
are easily deduced from the ones in the field theoretical version. 
They are in fact the same, with the sole difference that the $f$ functions 
appearing in the flows are to be replaced by 1 
(which is the value these $f$ functions take at high energies). 
Notice however that the quantities entering the flow equations now 
have different physical meanings. The cut-off scaling time is 
$t=\ln(\Lambda_0/\Lambda)$ ($\Lambda$ being the running cut-off), and the 
couplings are running bare couplings $G_\rmb(\Lambda)$.

The field theoretical RG flow equations for the charge, spin and Umklapp 
couplings are given below:
\begin{widetext}
  \begin{eqnarray}
    &&\partial_t \gc_\delta(I,J)=\frac{1}{N}\sum_\alpha\nonumber\\
    &&\hspace{0.5cm}\Biggl\lbrace f\left(t,2v_\rmf K^\ph_{\alpha;\delta}(I,J)\right)\Big[\gc_\alpha(I,J+\alpha-\delta) \gc_{\delta-\alpha}(I+\alpha,J)+3 \gs_\alpha(I,J+\alpha-\delta) \gs_{\delta-\alpha}(I+\alpha,J)\Big]\nonumber\\
    \label{eq:eqRGNchaines_c}
    &&\hspace{0.5cm}-f\left(t,2v_\rmf K^\pp_{\alpha;\delta}(I,J)\right)\Big[ \gc_\alpha(I,J) \gc_{\delta-\alpha}(I+\alpha,J-\alpha)-3 \gs_\alpha(I,J) \gs_{\delta-\alpha}(I+\alpha,J-\alpha)\Big]\\
    &&\hspace{0.5cm}+f\left(t,2v_\rmf K^{UU}_{\alpha;\delta}(I,J)\right)\Big[ U_\alpha(I,J+\alpha-\delta) U_{\delta-\alpha}(I+\alpha,J)+ U_{J-I-\delta}(I,J+\alpha-\delta) U_{J-I-\delta}(I+\alpha,J) \nonumber\\
    &&\hspace{4cm}-\frac{1}{2} U_{\delta-\alpha}(I+\alpha,J) U_{J-I-\delta}(I,J+\alpha-\delta)-\frac{1}{2} U_{J-I-\delta}(I+\alpha,J) U_\alpha(I,J+\alpha-\delta) \Big] \Biggr\rbrace.\nonumber
  \end{eqnarray}

  \begin{eqnarray}
    &&\partial_t \gs_\delta(I,J)=\frac{1}{N}\sum_\alpha\nonumber\\
    &&\hspace{0.5cm}\Biggl\lbrace f\left(t,2v_\rmf K^\ph_{\alpha;\delta}(I,J)\right)\Big[2 \gs_\alpha(I,J+\alpha-\delta) \gs_{\delta-\alpha}(I+\alpha,J)\nonumber+ \gs_\alpha(I,J+\alpha-\delta) \gc_{\delta-\alpha}(I+\alpha,J) \nonumber\\
    &&\hspace{4cm}+\gc_\alpha(I,J+\alpha-\delta) \gs_{\delta-\alpha}(I+\alpha,J)\Big]\nonumber\\
    \label{eq:eqRGNchaines_s}
    &&\hspace{0.5cm}+f\left(t,2v_\rmf K^\pp_{\alpha;\delta}(I,J)\right)\Big[2 \gs_\alpha(I,J) \gs_{\delta-\alpha}(I+\alpha,J-\alpha) - \gs_\alpha(I,J) \gc_{\delta-\alpha}(I+\alpha,J-\alpha)\\
    &&\hspace{4cm}-\gc_\alpha(I,J) \gs_{\delta-\alpha}(I+\alpha,J-\alpha)\Big]\nonumber\\
    &&\hspace{0.5cm}+f\left(t,2v_\rmf K^{UU}_{\alpha;\delta}(I,J)\right)\Big[ U_{J-I-\delta}(I,J+\alpha-\delta) U_{J-I-\delta}(I+\alpha,J)-\frac{1}{2} U_{\delta-\alpha}(I+\alpha,J) U_{J-I-\delta}(I,J+\alpha-\delta)\nonumber\\
    &&\hspace{4cm} -\frac{1}{2} U_{J-I-\delta}(I+\alpha,J) U_\alpha(I,J+\alpha-\delta) \Big] \Biggr\rbrace.\nonumber
  \end{eqnarray}

  \begin{eqnarray}
    &&\partial_t U_\delta(I,J) = \frac{1}{N}\sum_\alpha\nonumber\\
    &&\hspace{0.5cm}\Biggl\lbrace f\left(t,2v_\rmf K^{GU1}_{\alpha;\delta}(I,J)\right)\big[ \gc_{-\alpha}(J+\alpha-\delta,I)-\gs_{-\alpha}(J+\alpha-\delta,I) \big] U_{\delta-\alpha}(I+\alpha,J) \nonumber\\
    &&\hspace{0.5cm}+f\left(t,2v_\rmf K^{GU2}_{\alpha;\delta}(I,J)\right)\big[ \gc_{\delta-\alpha}(I+\alpha,J)-\gs_{\delta-\alpha}(I+\alpha,J) \big] U_\alpha(I,J+\alpha-\delta) \nonumber\\
    \label{eq:eqRGNchaines_u}
    &&\hspace{0.5cm}-f\left(t,2v_\rmf K^{GU3}_{\alpha;\delta}(I,J)\right)\big[2\gs_{I-\alpha}(\alpha+\delta,I) U_{J-\alpha-\delta}(\alpha,J)\big]\\
    &&\hspace{0.5cm}-f\left(t,2v_\rmf K^{GU4}_{\alpha;\delta}(I,J)\right) \big[2\gs_{J-\alpha-\delta}(\alpha,J) U_{I-\alpha}(\alpha+\delta,I)\big] \nonumber\\
    &&\hspace{0.5cm}+f\left(t,2v_\rmf K^{GU5}_{\alpha;\delta}(I,J)\right)\big[ \gc_{-\alpha}(I+\delta+\alpha,I)+3\gs_{-\alpha}(I+\delta+\alpha,I)\big] U_\delta(I+\alpha,J) \nonumber\\
    &&\hspace{0.5cm}+f\left(t,2v_\rmf K^{GU6}_{\alpha;\delta}(I,J)\right)\big[ \gc_{J-\alpha}(\alpha-\delta,J)+3\gs_{J-\alpha}(\alpha-\delta,J)\big] U_\delta(I,\alpha) \Biggr\rbrace.\nonumber
  \end{eqnarray}
\end{widetext}

In the above three flow equations, we have used the following notations:
\begin{eqnarray}
  \label{eq:car_length_1}
  K^\pp_{\alpha;\delta}(I,J)&\!\!=\!\!&\frac{1}{4}\left[ \Delta k_\delta(I,J)-2\Delta k_\alpha(I,J)\right]\\
  K^\ph_{\alpha;\delta}(I,J)&\!\!=\!\!&\frac{1}{4}\left[ \Delta k_\delta(I,J)-2\Delta k_\alpha(I,J+\alpha-\delta)\right]\\
  K^{UU}_{\alpha;\delta}(I,J)&\!\!=\!\!&\frac{1}{4}\left[\Delta k_\delta(I,J)-2\Delta k^U_\alpha(I,J+\alpha-\delta)\right]\quad\quad
\end{eqnarray}
\begin{eqnarray}
  K^{GU1}_{\alpha;\delta}(I,J)&\!\!=\!\!&\frac{1}{4}\left[ \Delta k^U_\delta(I,J)-2\Delta k_\alpha(I,J+\alpha-\delta)\right]\\
  K^{GU2}_{\alpha;\delta}(I,J)&\!\!=\!\!&\frac{1}{4}\left[ \Delta k^U_\delta(I,J)-2\Delta k_{\alpha-\delta}(J,I+\alpha)\right]\\
  K^{GU3}_{\alpha;\delta}(I,J)&\!\!=\!\!&\frac{1}{4}\left[ \Delta k^U_\delta(I,J)-2\Delta k_{\alpha-I}(I,\alpha+\delta)\right]\quad\quad
\end{eqnarray}
\begin{eqnarray}
  K^{GU4}_{\alpha;\delta}(I,J)&\!\!=\!\!&\frac{1}{4}\left[ \Delta k^U_\delta(I,J)-2\Delta k_{\alpha+\delta-J}(J,\alpha)\right]\\
  K^{GU5}_{\alpha;\delta}(I,J)&\!\!=\!\!&\frac{1}{4}\left[ \Delta k^U_\delta(I,J)-2\Delta k_\alpha(I,I+\alpha+\delta)\right]\quad\quad\\
  \label{eq:car_length_9}
  K^{GU6}_{\alpha;\delta}(I,J)&\!\!=\!\!&\frac{1}{4}\left[ \Delta k^U_\delta(I,J)-2\Delta k_{\alpha-J}(J,\alpha-\delta)\right]
\end{eqnarray}
We refer the reader to Sec.~\ref{sec:sub:sub:ren_int} for the definitions of $\Delta k_\delta(I,J)$ and $\Delta k^U_\delta(I,J)$. The $f$ function is defined as follows: $f(t=\ln(\Lambda_0/\nu),\delta)=1$ if $\nu\geqslant |\delta|$ and 0 otherwise.


\subsection{Renormalization of the propagator}
\label{app:flow_eq_prop}

The two-loop \se has the following expression:
\begin{widetext}
  \begin{eqnarray}
    &&\Sigma_{\rmr,I}^{(2)}(k=k_{\rmf,I},\omega=\mu+\nu)=\frac{1}{4N^2}\sum_{J,\alpha}\nonumber\\
    \label{eq:se_Nchaines}
    &&\hspace{1.5cm}\Biggl\lbrace 2\left[ \gc_\alpha(I,J) \gc_{-\alpha}(I+\alpha,J-\alpha) + 3 \gs_\alpha(I,J) \gs_{-\alpha}(I+\alpha,J-\alpha) \right]\\
    &&\hspace{6cm}\times\Big(\nu+v_\rmf \Delta k_\alpha(I,J) \Big) \ln\left[ \frac{\big| \nu^2-\big(v_\rmf \Delta k_\alpha(I,J)\big)^2\big|}{(2v_\rmf\Lambda_0)^2}\right]\nonumber\\
    &&\hspace{1.5cm}+U_\alpha(I,J) \big[ 2U_\alpha(I,J)-U_{J-I-\alpha}(I,J) \big]\Big( \nu+v_\rmf \Delta k^U_\alpha(I,J) \Big) \ln\left[ \frac{\big| \nu^2-\big(v_\rmf \Delta k^U_\alpha(I,J)\big)^2\big|}{(2v_\rmf\Lambda_0)^2}\right]\Biggr\rbrace.\nonumber
  \end{eqnarray}
\end{widetext}

The $\vp$ functions relating renormalized propagators at two different scales satisfy the general Eq.(\ref{eq:eq_flot_phi_gale}), which in the case of $N$ chains reads:
\begin{widetext}
  \begin{eqnarray}
    &&\partial_t \ln(\vp_I)=\frac{1}{2N^2}\sum_{J,\alpha}\nonumber\\
    &&\hspace{1.5cm}\Biggl\lbrace 2\left[ \gc_\alpha(I,J) \gc_{-\alpha}(I+\alpha,J-\alpha) + 3 \gs_\alpha(I,J) \gs_{-\alpha}(I+\alpha,J-\alpha) \right]\nonumber\\
    &&\hspace{3cm}\times\left[ \left(1+\frac{v_\rmf \Delta k_\alpha(I,J)}{\nu}\right)f\big(\nu,v_\rmf \Delta k_\alpha(I,J)\big)-\frac{v_\rmf \Delta k_\alpha(I,J)}{\nu}l\big(\nu,v_\rmf \Delta k_\alpha(I,J)\big)\right]\nonumber\\
    \label{eq:flot_phi}
    &&\hspace{1.5cm}+U_\alpha(I,J) \big[ 2U_\alpha(I,J)-U_{J-I-\alpha}(I,J) \big]\\
    &&\hspace{3cm}\times\left[ \left(1+\frac{v_\rmf \Delta k^U_\alpha(I,J)}{\nu}\right)f\big(\nu,v_\rmf \Delta k^U_\alpha(I,J)\big)-\frac{v_\rmf \Delta k^U_\alpha(I,J)}{\nu}l\big(\nu,v_\rmf \Delta k^U_\alpha(I,J)\big)\right]\Biggr\rbrace,\nonumber
  \end{eqnarray}
\end{widetext}
where the function $l$ is defined by $l(\nu,\delta)=\ln(|\nu/\delta|)$ if $\nu\geqslant |\delta|$ and $=0$ otherwise. This function is, as $f$, a logarithmic approximation of a more complex function, diverging at scale $|\delta|$. From the definition of $\vp_I$ (see Eq.~\ref{eq:definition_phi}), it is clear that if the initial high-energy quasiparticle weight is equal to one, then one simply has : $Z_I(t)=1/\vp_I(t)$.

The flow equation for the running \fs in the cut-off scaling scheme is also obtained thanks to the \se, and we find:
\begin{widetext}
  \begin{eqnarray}
    \partial_t k_{\rmf,I}^{(0)}&=&\frac{1}{2N^2}\sum_{J,\alpha} \Biggl\lbrace 2\Delta k_\alpha^{(0)}(I,J) \Big[ \gc_{\rmb,\alpha}(I,J) \gc_{\rmb,-\alpha}(I+\alpha,J-\alpha) + 3 \gs_{\rmb,\alpha}(I,J) \gs_{\rmb,-\alpha}(I+\alpha,J-\alpha) \Big]\nonumber\\
    \label{eq:flot_fs}
    &&\hspace{1.5cm}+ {\Delta k^U_\alpha}^{(0)}(I,J) \, U_\alpha(I,J) \big[ 2U_\alpha(I,J)-U_{J-I-\alpha}(I,J) \big]\\
    &&\hspace{5cm}-\frac{1}{N}\sum_I {\Delta k^U_\alpha}^{(0)}(I,J) \, U_\alpha(I,J) \big[ 2U_\alpha(I,J)-U_{J-I-\alpha}(I,J) \big]\Biggr\rbrace.\nonumber
  \end{eqnarray}
\end{widetext}


\end{document}